\documentclass[iop,apj,numberedappendix]{emulateapj}

\slugcomment{Received ; accepted ; published }
\slugcomment{Submitted to The Astrophysical Journal.} 
 
\shorttitle{The ASL neutrino treatment}
\shortauthors{Perego, Cabez\'on and K\"appeli}

\usepackage{graphics}
\usepackage{natbib}
\usepackage{amsmath}

\newcommand{\spmom}{\mathbf{p},\mathbf{x}}
\newcommand{\spc}{\mathbf{x}}
\newcommand{\msun}{{\rm M}_{\odot}}
\newcommand{\nub}{\bar{\nu}}
\newcommand{\nue}{\nu_e}
\newcommand{\nueb}{\bar{\nu}_e}

\newcommand{\numt}{\nu_{\mu,\tau}}
\newcommand{\numtb}{\bar{\nu}_{\mu,\tau}}
\newcommand{\boltz}{{\texttt{BOLTZTRAN}}~}
\newcommand{\agile}{{\textit{Agile}}~}
\newcommand{\aboltz}{\textit{Agile}-\texttt{BOLTZTRAN}~}
\newcommand{\aasl}{\textit{Agile}-ASL~}
\newcommand{\ye}{$Y_{\rm e}$}
\newcommand{\pasa}{Publication of the ASA}
\newcommand{\nar}{New Astronomy Review}

\newcounter{savfig}

\usepackage{color}
\usepackage[normalem]{ulem}  % \sout{old text} for strikeout
\usepackage{xspace}
\definecolor{green2}{rgb}{0,0.8,0}
\definecolor{orange}{rgb}{1,0.5,0}

\begin{document}

\title{An advanced leakage scheme for neutrino treatment in astrophysical simulations}

\author{A. Perego\altaffilmark{1},
        R. M. Cabez\'{o}n\altaffilmark{2} and
        R. K\"{a}ppeli\altaffilmark{3}
} 

% affiliations
\altaffiltext{1}{Institut f\"ur Kernphysik, 
Technische Universit\"at Darmstadt, 
Schlossgartenstra{\ss}e 2, 
D-64289 Darmstadt, 
Germany}
\altaffiltext{2}{Physics Department,
University of Basel,
Klingelbergstrasse 82,
CH-4056 Basel,
Switzerland}
\altaffiltext{3}{Seminar for applied Mathematics,
ETH Z\"urich,
R\"amistrasse 101,
8092 Z\"urich,
Switzerland}
\email{albino.perego@physik.tu-darmstadt.de}

\begin{abstract}
We present an Advanced Spectral Leakage (ASL) scheme to model neutrinos in the context 
of core-collapse supernovae and compact binary mergers.
Based on previous gray leakage schemes, the ASL scheme computes the neutrino
cooling rates by interpolating local production and diffusion rates (relevant in optically thin and
thick regimes, respectively), separately for discretized values of the neutrino energy. 
Neutrino trapped components are also modeled, based on equilibrium and timescale arguments. 
The better accuracy achieved by the spectral treatment allows
a more reliable computation of neutrino heating rates in optically thin conditions. 
The scheme has been calibrated and tested against Boltzmann transport in the context of Newtonian
spherically symmetric models of core-collapse supernovae.
ASL shows a very good qualitative and a partial quantitative agreement, for key quantities from 
collapse to a few hundreds of milliseconds after core bounce. We have proved the adaptability and 
flexibility of our ASL scheme coupling it to an axisymmetric Eulerian and to a 
three-dimensional SPH code to simulate core-collapse. 
Therefore, the neutrino treatment presented here is ideal for large parameter-space explorations, 
parametric studies, high-resolution tests, code developments, and long-term modeling of 
asymmetric configurations, where more detailed neutrino treatments are not available or 
currently computationally too expensive.
\end{abstract}

\keywords{neutrinos, radiative transfer, hydrodynamics, star: neutron, stars: supernovae: general}

\section{Introduction}
\label{sec: intro}

Neutrinos are elusive, weakly interacting particles. Due to their small cross-sections 
with ordinary matter, which make their detection 
on the Earth so challenging, they represent a very efficient way for astrophysical dense 
and hot plasma to radiate energy away.
In particular, they are expected to be copiously emitted in stellar explosive scenarios,
including core-collapse supernovae (CCSNe) \citep[e.g.,][for recent reviews]{Janka:2012,Burrows2013,Foglizzo.etal:2015}
and compact binary mergers \citep[e.g., ][for recent reviews]{Shibata.Taniguchi:2011,Faber.Rasio:2012,Rosswog:2015a}.
The modeling of such systems is extremely stimulating, due to the large variety of involved scales,
and to the complex and rich physics required. The treatment of neutrinos, in particular their transport
from optically thick to optically thin regions, is among the most crucial and difficult parts to model.
This is even more evident for intrinsically multidimensional problems, where the solution of
the Boltzmann transport equation would result in a genuine seven-dimensional problem \citep[e.g.,][]{Lindquist:1966}.

The large variety of questions and possible initial conditions, together with the parallel increase of
computational power, have motivated the development of several neutrino treatments, which differ 
in complexity and accuracy.
The solution of the complete Boltzmann equation for 
neutrino radiation has been performed for spherically symmetric simulations of CCSNe 
\citep{Mezzacappa.Bruenn:1993a,Mezzacappa.Bruenn:1993b,Mezzacappa.Bruenn:1993c,Liebendorfer2004,Sumiyoshi2005}.
Solutions in 2D \citep[e.g.,][]{Livne.etal:2004,Ott.etal:2008,Brandt.etal:2011} and, recently, in 3D 
\citep{Sumiyoshi.Yamada:2012,Sumyioshi.etal:2015}, 
neglecting velocity dependent terms, in the context of collapsing stellar cores, 
have also been presented.
Another sophisticated approach to the neutrino transport problem is represented by the 
so called Moment schemes. In these schemes, the explicit angular dependence in the neutrino momentum space 
is removed by integrating the distribution function and introducing momenta (for example, the energy density 
is the zeroth momentum, while the linear momentum density is the first momentum).
The Boltzmann transport equation is replaced by time evolution equations for the different momenta.
Among them, we recall the (multi-group) flux limited diffusion schemes, 
(MG)FLD, where only the 0th moment is considered (\citet{Arnett:1977,Bowers.Wilson:1982,Bruenn1985,Swesty.Myra:2009,Zhang.etal:2013};
for related applications see, e.g.,
\citet{Fryer:1999,Dessart.etal:2006,Burrows2007b,Dessart2009,Yakunin.etal:2010,Bruenn.etal:2013,Dolence.etal:2015}), 
and the M1 schemes, where both the 0th and 1st momenta are taken into account (see, e.g., 
\citet{Pons.etal:2000,Kuroda.etal:2012,O'Connor.Ott:2013,Obergaulinger.etal:2014,
O'Connor:2015,Just.etal:2015,Foucart.etal:2015a}).
The closure relation in M1 schemes is usually provided by an analytic expression.
It is possible to design moment schemes where the closure relation is not analytic, but
it is given by a variable Eddington tensor
\citep[e.g.,][]{Burrows.etal:2000,Rampp2002,Thompson2003,Buras.etal:2006a,Buras.etal:2006b,Mueller.etal:2010,Tamborra.etal:2013}). 
This latter solution is close to the solution of the full Boltzmann equation.
Another noteworthy approximate transport scheme is the Isotropic Diffusion Source Approximation 
(IDSA, \citet{Liebendorfer2009}; see \citet{Suwa.etal:2011,Takiwaki.etal:2014,Nakamura.etal:2014,Pan.etal:2015,Suwa.etal:2015} for
some recent applications), where the distribution function is separated into a trapped and a free-streaming component. 
Also Monte Carlo methods have been developed in the context of neutrino transport 
\citep[see, for example, ][]{Janka.Hillebrandt:1989a,Janka.Hillebrandt:1989b,Janka:1992,Abdikamalov2012,Richers.etal:2015}. 
More approximate treatments include gray transport schemes 
\citep[][]{Scheck.etal:2006}, gray neutrino leakage schemes 
\citep[][]{Ruffert1996,Rosswog2003} and light-bulb schemes 
\citep[][]{Murphy.Burrows:2008,Hanke2012,Fernandez:2012,Fernandez.Metzger:2013,Couch.Ott:2013,Handy.etal:2014}.

% hisotry and applications of leakage scheme
In this paper, we present an improved and more sophisticated version of the classical leakage scheme.
Leakage schemes have a long history in computational astrophysics:
due to their reduced computational cost and to their flexibility, 
they were, after FLD schemes, among the first developed neutrino treatments 
for spherically symmetric core-collapse models
\citep{VanRiper1981,Bludman1982,Cooperstein.etal:1986}, and for the first three-dimensional compact binary merger 
simulations \citep{Ruffert1996,Ruffert1999,Rosswog2003}.
More recently, gray leakage schemes have been widely applied to study, for example:
i) neutrino and gravitational wave emissions from rotating and/or magnetized collapsing stellar cores, in axisymmetry
\citep{Kotake2005,Kotake2012} or in three dimensions \citep{Scheidegger2010a,Takiwaki2011};
ii) supernova explosions caused by the magneto-rotational mechanism \citep[e.g.,][]{Suwa2007,Takiwaki2009,Winteler.etal:2012};
iii) the influence of the pions and hyperons in the nuclear equation of state (EoS) of stellar cores collapsing into a black hole 
\citep{Peres2013}, 
iv) the impact of asphericity in the progenitor model for core-collapse simulations \citep{Couch.Ott:2013},
v) gravitational waves and neutrino emission from Newtonian simulations
of compact binary mergers \citep{Rosswog:2013}.
General relativistic extensions of the gray scheme have also been developed \citep{Sekiguchi2010,Oconnor2010,Galeazzi2013}.
They have been used, for example, to simulate:
i) spherically symmetric models of supernovae and black hole formation 
from massive stellar cores \citep{OConnor2011};
ii) three-dimensional general relativistic core-collapse models \citep{Ott2013},
including also magnetic fields \citep{Moesta.etal:2014} ; 
iii) gravitational waves and neutrino emission from general relativistic simulations
of compact binary mergers \citep{Kiuchi2012,Deaton2013,Sekiguchi.etal:2015}; 
iv) the collapse of rotating stellar core to a black hole surrounded by an 
accretion disc \citep{Sekiguchi2011a}.

% what's new
Despite such a broad application field, detailed comparisons between the results obtained by a
leakage scheme and more sophisticated neutrino transports are difficult to find. 
For example, \cite{Dessart2009,Foucart.etal:2015a,Foucart.etal:2015b} compared 
the luminosities obtained by a gray leakage scheme with the ones obtained by 
MGFLD or moment schemes, during and after a compact binary merger.
Moreover, \cite{Oconnor2010} and \cite{Sekiguchi2010} provide temporal profiles of the neutrino
luminosities and mean energies occurring after core bounce in CCSNe
simulations, together with a few radial profiles of some relevant quantities (e.g., entropy or
electron fraction). These results can be compared with reference results in the literature 
(e.g., \cite{Liebendorfer2005a}). 
They found a qualitative good agreement, even if quantitative differences were present. 
This confirms the idea that leakage schemes capture the dominant aspects
of neutrino cooling, even if the evolution of the neutrino field in the opaque region
and the inclusion of consistent absorption terms in the optically thin region remain challenging.

In the Advanced Spectral Leakage (ASL) treatment we conjugate the usual positive 
aspects associated with leakage schemes (mainly, the reduced computational cost and flexibility), 
with an improved accuracy, obtained using a spectral approach (i.e., solving different leakage schemes, 
for different energy bins), modeling a neutrino trapped component in the optically thick region,
and including a consistent absorption term obtained from the spectral cooling rates.
The development of the ASL treatment has been performed in the framework of spherically symmetric models of CCSNe,
where the new treatment has been compared against a detailed Boltzmann neutrino transport.
\cite{Perego.etal:2015} used it, in combination with the IDSA for electron flavor neutrinos, 
to model heavy flavor neutrinos in spherically symmetric, 
artificially induced explosions of CCSNe.
However, one of the goals of this approximate scheme 
is the application to multidimensional models, with reduced computational costs.
Applications of the ASL scheme in multidimensional astrophysical simulations have been already performed 
in the past few years. \citet{Winteler.etal:2012} simulated a magnetically-driven CCSN explosion of a 
15~$\msun$ progenitor star, using the three-dimensional, Cartesian MHD code FISH \citep{Kappeli2011} coupled
with a previous version of the ASL scheme to model the neutrino cooling.
\citet{Perego.etal:2014b} studied the neutrino-driven wind that emerges from the remnant of a binary
neutron star merger, in presence of a long-lived massive neutron star. They used
the FISH code coupled with the ASL scheme. The neutrino heating rates were based on 
neutrino densities in optically thin regions, computed by a ray-tracing algorithm.

% layout of the paper
In Section \ref{sec: the new ASL}, we provide a detailed presentation of the ASL scheme and of its terms.
In Section \ref{sec: tests and validity}, we test the new scheme in the context of spherically symmetric models of
CCSNe, comparing the results obtained by the new treatment with the solution of the 
Boltzmann transport equation. We also briefly explore the impact of the variation of the few free parameters 
present in the scheme.
In Section \ref{sec: multiD models}, we show the flexibility and the versatility of the ASL scheme by 
implementing it in two different multidimensional codes, a grid code and an SPH code, modeling CCSNe.
Finally, in Section \ref{sec: conclusions}, we summarize and discuss our results.

\section{The ASL treatment}
\label{sec: the new ASL}

\subsection{Neutrino description and interactions}
The ASL scheme is an approximate neutrino treatment designed for neutrinos and antineutrinos of all flavors.
While electron neutrinos ($\nue$) and antineutrinos ($\nueb$) are considered separately,
$\mu$ and $\tau$ neutrinos, as well as their antiparticles, are treated as a single neutrino species ($\numt$).
For each of the three independent species,
we perform a spectral treatment, i.e. we distinguish between neutrinos with different energies.

The interaction between matter and neutrinos is provided by weak interaction processes.
For each production, absorption, or scattering reaction, we compute the corresponding spectral emissivity $j_{\nu}(E,\spc)$, \
absorptivity $\chi_{\nu,{\rm ab}}(E,\spc)$, or scattering rate $\chi_{\nu,{\rm sc}}(E,\spc)$, respectively, 
as they are defined in the Boltzmann transport equation, for a neutrino energy $E$ and a position $\mathbf{x}$
(in this section we consider a fixed time $t$).
The total emissivity and absorptivity are computed as the sum over all considered 
neutrino processes.
While the emissivity provides the local rates of neutrino production, the absorptivity and the 
scattering rates are the sources of the local neutrino opacity. 
The local opacity can be expressed in terms of
the \textit{total} mean free path $\lambda_{\nu, {\rm tot}}$,
\begin{equation}
 \lambda_{\nu,{\rm tot}}(E,\spc) = 
 c \left( \sum\limits_{r} \chi_{\nu,{\rm ab},r}(E,\spc) + \sum\limits_{s} \chi_{\nu,{\rm sc},s}(E,\spc) \right)^{-1}
 \label{eqn: lambda tot}
\end{equation}
where $c$  is the speed of light, and the indexes $r$ and $s$ run over all the 
considered absorption and scattering reactions, respectively.
Besides the total mean free path, in which all reactions are treated equally, we define also an 
{\it energy} mean free path, $\lambda_{\nu,{\rm en}}$.
The latter represents the mean free path over which neutrinos can effectively exchange energy with the fluid.
To compute it, we perform the geometrical mean between the 
total mean free path and the mean free path only due to highly inelastic processes, i.e. all the 
absorption processes and the scattering processes where the energy of the incoming and out-coming neutrinos
is expected to differ significantly \footnote{Elastic 
scattering processes enter the definition of $\lambda_{\nu,{\rm en}}$ via $\lambda_{\nu,{\rm tot}}$. 
Even if they do not allow direct energy exchange between neutrinos and matter, 
they still provide opacity and increase the probability of inelastic processes to locally happen.} 
(see, for example, \cite{Shapiro1986} or \cite{Raffelt2001} for analogous expressions):
\begin{equation}
\lambda_{\nu,{\rm en}}(E,\spc) = \left( c^{-1} \sum\limits_{s'} \chi_{\nu,{\rm inel},s'}(E,\spc) \right)^{-1/2} \, \\
        \left( \lambda_{\nu,{\rm tot}}(E,\spc) \right)^{1/2},
         \label{eqn: lambda eff}
\end{equation}
where we have restricted the sum only over inelastic processes, abbreviated by {\it inel} and labeled by $s'$.

%Modern leakage schemes rely on the {\it optical depth}.
The neutrino optical depth $\tau_{\nu}$ is defined as the path integral of 
the inverse neutrino mean free path, $\lambda_{\nu}^{-1}$ , calculated on a typical radiation path, 
$\gamma$, connecting any point $\spc$ of the system with its edge:
\begin{equation}
\label{eqn: tau def}
 \tau_{\nu,\gamma}(E,\spc) = \int_{\gamma:\mathbf{x} \rightarrow + \infty} \frac{1}{\lambda_{\nu}(E,\mathbf{x'}(s))} \, {\rm d}s .
\end{equation}
From a physical point of view, it is a measure of the accumulated opacity of matter to radiation along an escape 
path: it counts the number of interactions that, on average, a radiation particle, emitted at a certain point, 
experiences before leaving the system.
The two mean free paths introduced above can be used to compute two different optical depths,
a total optical depth, $\tau_{\nu,{\rm tot}}(E,\spc)$, 
and an energy optical depth, $\tau_{\nu,{\rm en}}(E,\spc)$. 
We note that $\tau_{\nu,{\rm en}} \leq \tau_{\nu,{\rm tot}}$, by definition.
In the case of spherically symmetric models, the optical depth retains the spherical symmetry and 
$\tau_{\nu}$ can be simply calculated along radial paths:
\begin{equation}
\tau_{\nu}(E,R) = \int_R^{+ \infty} \frac{1}{\lambda_{\nu}(E,r)} \, {\rm d}r.
\end{equation}
Otherwise, for more general geometries, multidimensional algorithms are required to
compute the optical depth along paths that minimize the number of neutrino-matter interactions
(e.g., \cite{Perego.etal:2014a}).

According to the values of $\tau_{\nu,{\rm tot}}$ and $\tau_{\nu,{\rm en}}$
for the most relevant neutrino energies, 
several different regimes can be distinguished:
\begin{enumerate}
\item $\tau_{\nu,{\rm tot}} \gg 1$ and $\tau_{\nu,{\rm en}} \gtrsim 1$,
the \textit{equilibrium-diffusive} regime.
The radiation field is in thermal and weak equilibrium 
with the surrounding matter, and neutrinos can be considered as a trapped Fermi 
gas inside the fluid, behaving like a fluid component itself. Under these assumptions, 
the distribution functions describing the neutrino gas can be expressed as
$ f_{\nu}(\spmom) = f_{\nu}^{\rm tr}(\spmom) + \delta f_{\nu}(\spmom) $,
where $f_{\nu}^{\rm tr}$ is the trapped component and $\delta f_{\nu}$ is a small deviation 
from equilibrium, that we neglect \citep{Cooperstein.etal:1986,Cooperstein.etal:1987a,Cooperstein1988}.
We assume further that the trapped component has no explicit angular dependence in the momentum space, 
$f_{\nu}^{\rm tr}(E,\spc)$ (for a more detailed discussion of the decomposition 
of the neutrino distribution function in a (isotropic) trapped component and a free streaming one
see \cite{Liebendorfer2009}). 
The integration over the neutrino phase space of $f_{\nu}^{\rm tr}$ and $f_{\nu}^{\rm tr} \, E$ gives information 
about the particle and the energy contents of the neutrino gas:
\begin{equation}
\label{eq: ynu def}
Y_{\nu}(\spc) = \frac{4 \pi}{(hc)^3} \frac{m_{\rm b}}{\rho(\spc)} \int f_{\nu}^{\rm tr}(E,\spc) E^2 dE  \, ,
\end{equation}
\begin{equation}
\label{eq: znu def}
Z_{\nu}(\spc) = \frac{4 \pi}{(hc)^3} \frac{m_{\rm b}}{\rho(\spc)} \int f_{\nu}^{\rm tr}(E,\spc) E^3 dE \, ,
\end{equation}
where $m_{\rm b}$ is the baryon mass and $h$ the Planck constant.

\item $\tau_{\nu,{\rm tot}} \gg 1$, but $\tau_{\nu,{\rm en}} \lesssim 1$, the {\it diffusive} regime.
Neutrinos still diffuse, but they are not necessarily in thermal equilibrium with the surrounding plasma.
\item $\tau_{\nu,{\rm tot}} \sim 1$, the {\it semi-transparent} regime. The solution of the Boltzmann transport problem would be here ideal to
model the radiation transport with accuracy. The surfaces defined by the conditions $\tau_{\nu} = 2/3$ are called {\it neutrino surfaces} 
(or {\it neutrinospheres}, in spherically symmetric models). In the case of $\tau_{\nu,{\rm tot}}$, they are considered as the last-interaction 
surfaces, before neutrinos can stream away freely. For $\tau_{\nu,{\rm en}}$, they represent the 
surface at which neutrinos decouple thermally from matter.
\item $\tau_{\nu,{\rm tot}} \lesssim 1$, the {\it free streaming} regime. In this regime, neutrinos that are locally produced can
stream out freely, almost with no interaction with matter. At the same time, a fraction of the large neutrino fluxes 
coming from the neutrino surfaces can be here re-absorbed by matter.
\end{enumerate}

\subsection{The coupling with hydrodynamics}

In the following we consider a Newtonian hydrodynamical system at time $t$, 
which is evolving in time with a time-step $\Delta t$.
To be more general, we consider a three-dimensional domain and, unless it is explicitly said, no
symmetries are assumed.
The system is described by its density $\rho$, temperature
$T$, electron fraction \ye, and velocity field $\mathbf{v}$.
The trapped neutrino components are defined by $Y_{\nu}$ and $Z_{\nu}$.
All these quantities are given at every position $\mathbf{x}$.
We define the vector of conserved variables, $U$, and the corresponding flux tensor, $F$, as:
\begin{equation}
\label{eqn: variables and fluxes}
U=\left( \begin{array}{c}
\rho \\
\rho  \mathbf{v}\\
\rho \left( e+\frac{v^2}{2}\right) \\
\rho Y_{{\rm e}}\\
\rho Y_{\nu}\\
\left( \rho Z_{\nu}\right) ^{\frac{3}{4}}
\end{array}\right), \, F=\left( \begin{array}{c}
\mathbf{v} \rho \\
\mathbf{v} \rho \mathbf{v}+\mathbb{I}P\\
\mathbf{v}\rho \left( e+\frac{v^{2}}{2}+\frac{P}{\rho }\right) \\
\mathbf{v}\rho Y_{{\rm e}}\\
\mathbf{v}\rho Y_{\nu}\\
\mathbf{v}\left( \rho Z_{\nu}\right) ^{\frac{3}{4}}
\end{array}\right) ,
\end{equation}
where the specific internal energy $e$ and the fluid pressure $P$ are provided 
by an EoS as functions of $(\rho, T, Y_{\rm e})$, while $v$ is the modulus 
of the fluid velocity. 
The evolution of the system is determined by:
\begin{equation}
\label{eqn:hydro equations}
\frac{\partial}{\partial t} U + \nabla \cdot  F = g_{\rm grav} + g_{\nu} \, ,
\end{equation}
where $g_{\rm grav}$ is the gravitational source term, depending on the gravitational potential $\phi$, 
and $g_{\nu}$ is the neutrino source term.
The latter is related to the variation of the specific internal energy, $\dot{e}$, of the electron fraction, $\dot{Y}_{\rm e}$,
of the neutrino trapped components, $\dot{Y}_\nu$ and $\dot{Z}_\nu$, and of the fluid velocity, $\dot{\mathbf{v}}$, 
provided by neutrinos:
\begin{equation}
 g_{\rm grav} = \left( 
\begin{array}{c}
0 \\ 
- \rho \, \nabla \phi \\  
- \rho \mathbf{v} \cdot \nabla \phi \\
0  \\
0  \\
0
\end{array}
\right)
,\, g_{\nu} = \left( \begin{array}{c}
0 \\ 
\rho \dot{\mathbf{v}} \\ 
\rho \dot{e} + \rho \mathbf{v} \cdot \dot{\mathbf{v}}  \\ 
\rho \dot{Y_{\rm e}} \\ 
\rho \dot{Y}_{\nu} \\ 
\frac{3}{4} \, \rho^{3/4} \frac{\dot{Z}_{\nu}}{{Z}_{\nu}} 
\end{array}
\right).
\label{eq: source terms}
\end{equation}
The goal of the ASL scheme is the estimation of the neutrino source term, $g_{\nu}$, from the present values
of the thermodynamical state $U$ of the system.
Before proceeding, it is important to notice that any leakage scheme models the local net loss of leptons 
in form of neutrinos, i.e. the variations of the total (trapped) lepton number, $\dot{Y}_{\rm l}$, and 
of the specific total internal energy, $\dot{u}$. These quantities are related with the variation rates appearing in
Equation~(\ref{eq: source terms}) by
\begin{equation}
 \dot{Y}_{\rm l} = \dot{Y}_{\rm e} + \dot{Y}_{\nue} - \dot{Y}_{\nueb}
\end{equation}
and
\begin{equation}
 \dot{u} = \dot{e} + \frac{1}{m_{\rm b}}\left( \dot{Z}_{\nue} + \dot{Z}_{\nueb} 
 + 4 \, \dot{Z}_{\numt} \right).
\end{equation}
Note that the contributions to $\dot{Y}_{\rm l}$ given by $\numt$ and $\numtb$ cancel in our approach.
Concerning the neutrino stress, we distinguish $\dot{\mathbf{v}}$ into two contributions:
\begin{equation}
 \dot{\mathbf{v}} = \left( \dot{\mathbf{v}} \right)_{\tau_{\nu} > 1} + \left( \dot{\mathbf{v}} \right)_{\tau_{\nu} \lesssim 1} \, ,
\end{equation}
i.e., one related with the trapped neutrinos, $\left( \dot{\mathbf{v}} \right)_{\tau_{\nu} > 1}$, and one 
with the absorption of radiation in optically thin conditions, $\left( \dot{\mathbf{v}} \right)_{\tau_{\nu} \lesssim 1}$.

In Section \ref{subsec: trapped part} we show how we compute the rates for the trapped components 
$\dot{Y}_{\nu}$ and $\dot{Z}_{\nu}$, and the neutrino stress in trapped conditions $(\dot{\mathbf{v}})_{\tau_{\nu}>1}$.
The calculation of the leakage rates, $\dot{u}$ and $\dot{Y}_{\rm l}$, and of the neutrino stress in free-streaming 
conditions, $(\dot{\mathbf{v}})_{\tau_{\nu} < 1}$, is exposed in Section \ref{subsec: effective part}.
In both sections, we omit the explicit dependence on time and position in the equations, apart when the 
neutrino energy is involved. In that case, we only omit the temporal dependence.

\subsection{Trapped component rates}
\label{subsec: trapped part}
The variation rates for the trapped components are computed as:
\begin{eqnarray}
 \dot{Y}_\nu = \frac{\tilde{Y}_{\nu,t+\Delta t} - Y_\nu}{\Delta t} \, ,\\
 \dot{Z}_\nu = \frac{\tilde{Z}_{\nu,t+\Delta t} - Z_\nu}{\Delta t} \, ,
\end{eqnarray}
where $\tilde{Y}_{\nu,,t+\Delta t}$ and $\tilde{Z}_{\nu,,t+\Delta t}$ are guesses of the
trapped components at $t+\Delta t$, only due to neutrino processes.
Their computation is done following these steps.
i) Reconstruct approximated trapped components of the neutrino distribution functions, $f_{\nu}^{\rm tr}$, 
at the current time $t$ based on $Y_{\nu}$. This is done assuming that
\begin{equation}
 f^{\rm tr}_{\nu}(E,\spc) = \gamma(\spc) \left( f_{\nu}(E,\spc) \right)_{\rm eq} \left( 1 - e^{ - \tau_{\nu,{\rm en}}(E,\spc)} \right).
 \label{eqn: f_tr reconstruction}
\end{equation}
The exponential cut-off ensures that $f_{\nu}^{\rm tr}$ is significantly different from 0 only when 
neutrino trapping conditions are fulfilled with respect to the energy optical depth, i.e. 
$\tau_{\rm en} \gtrsim 1$, and that it is proportional to $(f^{\rm tr})_{\rm eq}$ for 
$\tau_{\rm en} \gg 1$.
The local parameter $\gamma(\spc)$ is fixed by Equation~(\ref{eq: ynu def})
\footnote{We notice that the usage of $Y_{\nu}$ to reconstruct $f^{\rm tr}_{\nu}$ does not 
ensure the exact reconstruction of $Z_{\nu}$. 
On one hand we have tested that the usage of $Z_{\nu}$ instead of $Y_{\nu}$
does not change our results significantly. On the other hand, 
the usage of the original (i.e., not reconstructed) values of $Z_{\nu}$ in the 
computation of $\dot{Z}_{\nu}$ still ensures energy conservation.}.
The equilibrium distribution functions are assumed to be Fermi-Dirac distribution functions 
of a neutrino gas in thermal and weak equilibrium with matter
\begin{equation}
(f_\nu)_{\rm eq}(E,\spc) = \frac{1}{e^{(E/T_{\nu}(\spc)-\eta_{\nu}(\spc))}+1},
\end{equation}
where $T_{\nu}$ is assumed to be equal to the matter temperature (true for $\tau_{\nu,{\rm en}}\gtrsim 1$) 
and $\eta_{\nu}$ is the neutrino degeneracy parameter. 
For $\nue$ and $\nueb$, we use the equilibrium degeneracy parameter:
\begin{equation}
 \eta_{\nue} = \left( \mu_{e} - \mu_n + \mu_p \right) / T = - \, \eta_{\nueb} \, ,
\end{equation}
where $\mu_p$, $\mu_n$ and $\mu_e$ are the relativistic chemical potentials of neutrons, protons and electrons, respectively.
For $T \gtrsim 0.5 \, {\rm MeV}$, we assume $ \mu_{e^+} + \mu_{e^-} = 0$.
If $\numt$ are produced by electron-positron annihilation, this relation 
suggests that neutrinos and antineutrinos of these flavors have approximately opposite
chemical potentials. However, the substantial equivalence between $\numt$ and 
$\numtb$\footnote{This approximation is valid
as long as the temperature is not high enough to produce a significant amount of muons 
and antimuons ($T \ll m_{\mu}\approx 105.7 \, {\rm MeV}$),
and nuclear effects distinguishing $\nu$ and $\nub$ 
(like weak magnetism) are neglected.}
implies also the equivalence between their chemical potentials. 
Thus, the degeneracy parameter for $\numt$ is set to 0.

\noindent ii) Evolve $f_{\nu}^{\rm tr}$ according to timescale arguments between $t$ and $t+\Delta t$.
This is done considering the neutrino production and diffusion as
two competing processes:
\begin{equation}
 \frac{{\rm d} f^{\rm tr}_{\nu}}{{\rm d} t} = \dot{f}^{\rm tr}_{{\nu},{\rm prod}}
        + \dot{f}^{\rm tr}_{{\nu},{\rm diff}}
 \label{eqn: dist func update}
\end{equation}
with:
\begin{equation}
\dot{f}^{\rm tr}_{{\nu},{\rm prod}} = 
\frac{((f_{\nu})_{\rm eq}-f^{\rm tr}_{\nu})}{\max( t_{{\nu},{\rm prod}},\Delta t)}
\, \exp{ \left(  { - \frac{t_{{\nu},{\rm prod}}}{t_{{\nu},{\rm diff}}}} \right)}
\label{eqn: fdot prod}
\end{equation}
and
\begin{equation}
\dot{f}^{\rm tr}_{{\nu},{\rm diff}} = - \frac{ f^{\rm tr}_{{\nu} }}{\max({t_{{\nu},{\rm diff}}, \Delta t})} \, 
\exp{  \left( - \frac{ {  t_{{\nu},{\rm diff}}}}{ {t_{{\nu},{\rm prod}}}} \right) }.
\label{eqn: fdot diff}
\end{equation}
In Equations~(\ref{eqn: fdot prod}) and (\ref{eqn: fdot diff}), all the quantities are evaluated at position 
$\mathbf{x}$ and for a certain neutrino energy $E$. 
In Equation~(\ref{eqn: fdot prod}), the term before the exponential
ensures that the distribution function reaches the equilibrium value whenever 
the production timescale is small enough compared to the time-step $\Delta t$.
Similarly, the first part of Equation~(\ref{eqn: fdot diff}) causes the distribution 
function to go to 0, if the diffusion 
timescale is small enough compared with $\Delta t$.
The exponential factors in both expressions are a switch between the diffusive 
and the free-streaming regime. 
The production timescale, $t_{\nu,{\rm prod}}$, is set by the neutrino emissivity:
\begin{equation}
t_{{\nu},{\rm prod}}(E,\spc) = \frac{1}{j_{\nu}(E,\spc)}.
\label{eqn: tprod}
\end{equation}
Following \cite{Rosswog2003} (cf. \cite{Ruffert1996}) we define the diffusion timescale, 
$t_{\nu,{\rm diff}}$, as
\begin{equation}
t_{\nu,{\rm diff}}(E,\spc) =  \frac{\Delta x_{\nu}(E,\spc)}{c}\tau _{\nu,{\rm tot}}(E,\spc)
\label{eqn: diffusion timescale}
\end{equation}
The quantity $\Delta x_{\nu}$ can be understood as the effective width of a layer drained by the diffusion flux 
and it is calculated as
\begin{equation}
  \Delta x_{\nu}(E,\spc) = \alpha_{\rm diff} \, \tau _{\nu,{\rm tot}}(E,\spc) \, \lambda _{\nu,{\rm tot}}(E,\spc).
  \label{eqn: delta x}
\end{equation}
Usually, $\alpha_{\rm diff} \sim 3$ (e.g., \cite{Mihalas1984}, \cite{Ruffert1996} and \cite{Rosswog2003}).
While for large optical depths Equation~(\ref{eqn: diffusion timescale}) provides an estimate of a proper 
diffusion timescale, in the optically thin regime its value decreases significantly, due to 
its quadratic dependence on the optical depth.

\noindent iii) Obtain $\tilde{Y}_{\nu,t+\Delta t}$ and $\tilde{Z}_{\nu,t+\Delta t}$ from $f_{\nu}^{\rm tr}(t+\Delta t,E,\spc)$, 
based on Equations~(\ref{eq: ynu def}) and (\ref{eq: znu def}).

Trapped neutrinos provide a source of stress for the fluid.
This stress is determined by the gradient of the pressure of the neutrino gas
\begin{equation}
  \label{eqn: vdot opaque regime}
  \left( \dot{\mathbf{v}} \right)_{\tau_{\nu} > 1} = - \frac{\nabla {P_{\nu,{\rm tot}}}}{\rho},
\end{equation}
where the neutrino pressure is evaluated based on the energy content of
the neutrino gas,
\begin{equation}
P_{\nu,{\rm tot}} = \sum_\nu \: P_{\nu} = \frac{1}{3} \frac{\rho}{m_b} (Z_{\nue} + Z_{\nueb} + 4 \, Z_{\numt} ) \, .
\label{pnu1}
\end{equation}

\subsection{Emission rates}
\label{subsec: effective part}
The rates for the total lepton number and total specific internal energy are obtained as the net balance 
between the emission rates ($R^0_{\nu}$ for the particles and $R^1_{\nu}$ for the energy) 
and the absorption rates in optically thin conditions
($H^0_{\nu}$ for the particles and $H^1_{\nu}$ for the energy):
\begin{equation}
\dot{Y}_{\rm l} = - \, m_b \, \left( R^0_{\nu_e} - 
R^0_{\nub_e} - H^0_{\nu_e} + H^0_{\nub_e} \right),
\label{eqn: yldot}
\end{equation}
and
\begin{equation}
\dot{u} =  
    - \left( R^1_{\nue} + R^1_{\nueb} + 4 R^1_{\numt} \right) 
    + \left( H^1_{\nue} + H^1_{\nueb} \right).
\label{eqn: udot}
\end{equation}
In Equations~(\ref{eqn: yldot}) and (\ref{eqn: udot}), $\numt$ and $\numtb$ do not provide 
any net contribution to the lepton number and do not contribute
to the absorption in optically thin conditions.
The emission and absorption rates are obtained from 
spectral emission ($r_{\nu}$) and absorption ($h_{\nu}$) rates, according to:
\begin{equation}
R^k_{\nu}(\spc) =  \int_0^{+\infty} {r_{{\nu}}(E,\spc)}\, E^{2+k} \, dE , 
 \label{eqn: particle cooling rate}
\end{equation}
\begin{equation}
H^k_{\nu}(\spc) =  \int_0^{+\infty} h_{\nu}(E,\spc) \, E^{2+k} \, dE  ,   
\label{eqn: particle heating rate}
\end{equation}
with $k=0,1$.

The emission rates are computed as smooth interpolation between the production
rates,
\begin{equation}
\label{eqn: r_prod definition}
r_{\nu,{\rm prod}}(E,\spc) =  \frac{4 \pi}{(hc)^3} \,
\frac{j_{\nu}(E,\spc)}{\rho(\spc)} \, ,
\end{equation}
and the diffusion rates,
\begin{equation}
\label{eqn: spectral r_diff definition}
r_{\nu,{\rm diff}}(E,\spc) = \frac{4 \pi}{(hc)^3} \, \frac{1}{\rho(\spc)} \,  
\frac{(f_{\nu})_{\rm eq}(E,\spc) \, }{t_{\nu,{\rm diff}}(E,\spc)} \, .
\end{equation}
The former are expected to be dominant in optically thin conditions,
while the latter in the opaque region.
The interpolation formula is provided by
\begin{eqnarray}
 \label{eqn: r_eff last version}
   r_{{\nu}}(E,\spc) = \left( 1 - \alpha_{\nu,{\rm blk}} \right) \tilde{r}_{\nu}(E,\spc) \times \nonumber \\ 
   \frac{1}{\Psi_{\nu}\left( \spc \right)} \exp{\left( -\tau_{\nu,{\rm en}}\left(E,\spc\right)/ \tau_{\rm cut} \right)} ,
\end{eqnarray}
where 
\begin{equation}
 \label{eqn: r_eff old version}
   \tilde{r}_{{\nu}}(E,\spc) = \frac{ r_{{\nu},{\rm prod}}(E,\spc) \times r_{{\nu},{\rm diff}}(E,\spc)} 
    { r_{{\nu},{\rm prod}}(E,\spc) + r_{{\nu},{\rm diff}}(E,\spc) }
\end{equation}
is the interpolation expression used for the emission rates in gray leakage schemes
\citep[cf, for example, ][]{Ruffert1996, Rosswog2003}. 
$\alpha_{\nu,{\rm blk}}$ and $\tau_{\rm cut}$ are parameters, and $\Psi_{\nu}$ is a local normalization
factor:
\begin{equation}
 \Psi_{\nu}(\spc) = \frac{\int_0^{+\infty} \tilde{r}_{\nu}(E,\spc) e^{-\tau_{\nu,{\rm en}}\left( E,\spc \right)/ \tau_{\rm cut}} \, E^2 {\rm d}E  } 
 {\int_0^{+\infty} \tilde{r}_{\nu}(E,\spc) \, E^2 {\rm d}E }
 \label{eqn: normalization factor N}
\end{equation}
Then, for $\alpha_{\nu,{\rm blk}}=0$ and $\tau_{\rm cut} \rightarrow + \infty$, 
we obtain $r_{\nu} = \tilde{r}_{\nu}$.
Finite, non-zero values of these constants introduce two important improvements in the ASL scheme:\\
i) in the case where $\alpha_{\rm blk}=0$, the total amount of emitted $\nue$'s and $\nueb$'s 
is usually overestimated.
When a large fraction of the emission rates is produced in the semi-transparent and
optically thin conditions, the interpolation favors the usage of the production rates,
calculated as integral over the whole solid angle of isotropic emission rates, Equation~(\ref{eqn: r_prod definition}).
However, a significant fraction of those neutrinos are emitted toward the optically thick region.
Moreover, the emission rates can be significantly reduced by Pauli's blocking factors, provided by
the large amount of free streaming neutrinos emitted at the neutrino surface or locally produced.
To efficiently take into account these effects, we have introduced the factor 
$\left( 1 - \alpha_{\nu,{\rm blk}} \right)$ in Equation~(\ref{eqn: r_eff last version}).
$\alpha_{\nu,{\rm blk}}$ is a free parameter of the model and it is expected to be $\sim 0.5$ for
$\nue$ and $\nueb$ in strongly accreting systems. On the other hand, since the emission 
of $\numt$ from optically thin conditions is usually negligible, we use $\alpha_{{\rm blk},\numt} \sim 0$.\\
ii) The rates obtained with Equation~(\ref{eqn: r_eff old version}) retain spectral information 
of the local thermodynamical properties of matter. 
Nevertheless during the diffusion process, high energy neutrinos
coming from optically thick regions thermalize to lower energies, 
at least as long as $\tau_{\nu,{\rm en}} \gg 1$.
Thus, the spectrum emerging from the neutrinospheres is softer than the one provided by 
Equation~(\ref{eqn: r_eff old version}). 
To mimic this transition, we have included the term $ \exp(-\tau_{\nu,{\rm en}}/\tau_{\rm cut})/\Psi_{\nu}$
in Equation~(\ref{eqn: r_eff last version}). The definition of $\Psi_{\nu}$, Equation~(\ref{eqn: normalization factor N}), 
ensures that the number of neutrinos emitted is preserved, while their 
final spectrum is modeled according to the energy optical depth. Since a few inelastic interactions 
are necessary to thermalize the spectrum, we expect 
$\tau_{\rm cut} \sim \mathcal{O}(10)$.

\subsubsection{Absorption rates}

Eventually, neutrinos emitted at the neutrino surface and above it stream away in the optically thin region,
with a non-negligible probability to be re-absorbed by the fluid. 
In the ASL scheme, we include an estimate of this non-local absorption rates, $h_{\nu}(E,\spc)$, 
based on the computation of the neutrino densities outside the neutrino surfaces.
The spectral absorption rates for $\nue$ and $\nueb$ in optically thin conditions are calculated as:
\begin{eqnarray}
 h_{\nu}(E,\spc) & = & \frac{1}{\rho(\spc)} \, n_{\nu,\tau_{\nu} \lesssim 1}(E,\spc) 
 \, \times \nonumber \\
 & & \chi_{\nu,{\rm ab}}(E,\spc) \; \mathcal{F}_{e}(E,\spc) \; \mathcal{H}(E,\spc).
 \label{eqn: heating term outside}
\end{eqnarray}
$\mathcal{F}_{e,\nu}$ is the Pauli blocking factor for electrons or positrons in the final state:
\begin{equation}
\label{eqn: Pauli-blocking electron}
 \mathcal{F}_{e^\mp} = \left( 1 - \frac{1}{\exp{\left( (E \pm Q \mp \mu_e)/T \right) }+1}  \right)
\end{equation}
where we have assumed that the electron or positron produced by the absorption of an 
electron neutrino or antineutrino, respectively, has an energy equal to the energy of the incoming neutrino, 
corrected by the mass difference $Q= \left( m_n - m_p \right) c^2 \approx 1.293 \, {\rm MeV}$. 
In Equation~(\ref{eqn: heating term outside}), $\mathcal{H}(E,\spc) \equiv \exp(- \tau_{\nu,{\rm tot}}(E,\spc))$ is 
an exponential term that ensures the application of the heating rates only in the optically
thin region.
The quantity $ n_{\nu,\tau_{\nu} \lesssim 1} $ is the spectral 
neutrino density, defined such as the total neutrino density outside the neutrino surface, 
$ N_{\nu,\tau_{\nu} \lesssim 1}$, is:
\begin{equation}
 N_{\nu,\tau_{\nu} \lesssim 1}(\spc) = \int_{0}^{+ \infty}  \: n_{\nu,\tau_{\nu} \lesssim 1} (E,\spc) \, E^2 \: {\rm d} E .
 \label{eqn: integrated neutrino density}
\end{equation}

One of the limits of any leakage scheme is that it does not model the spatial and angular distribution 
of the emission outside the neutrino surface. 
Thus, the calculation of $h_{\nu}$ requires first a separate evaluation of 
$ n_{\nu,\tau_{\nu}\lesssim 1} $.
This task depends strongly on the nature of the problem and on the symmetry of the system.
For spherically symmetric models, the preferential propagation direction is the radial one, 
and the neutrino densities are related to the spherically symmetric neutrino spectral particle 
luminosities, $l_{\nu}$, by
\begin{equation}
 n_{\nu,\tau_{\nu}\lesssim 1} (E,R) = \frac{ l_{\nu}(E,R)}{4 \pi R^2 c \, \mu_{\nu}(E,R)}
 \label{eqn: neutrino density 1D}
\end{equation}
where $R$ is the radial coordinate (e.g., \cite{Janka2001}).
This conversion between the neutrino flux and density involves the (spectral) flux factor 
$ \mu_{\nu}(E,R)$. It represents the average of the cosine of the propagation angle for the free-streaming neutrinos. 
Far from the neutrinospheres, the distribution functions are expected to peak 
in the forward direction, meaning $\mu_{\nu}(R \gg R_{\nu}) \sim 1$. Close to the neutrinospheres, 
assuming that radiation is emitted isotropically above the plane tangential to the neutrinospheres, 
$\mu_{\nu}(R \sim R_{\nu}) \sim 1/2$. Following \cite{Liebendorfer2009}, we use an analytic
approximation for $R> R_{\nu}(E)$:
\begin{equation}
 \mu_{\nu}(E,R) = \frac{1}{2} \left( 1+\sqrt{1- \left( \frac{R_{\nu}(E)}{\max(R,R_{\nu}(E))} \right)^2} \right) \, .
\end{equation}
The quantity $l_{\nu}$ is computed at each radius and for each neutrino energy as a solution of the 
differential equation
\begin{eqnarray}
  \frac{{\rm d}l_{\nu}(E,R)}{{\rm d}R} & = & \, 4 \pi R^2 \, \rho(R) \, r_{\nu}(E,R) - \nonumber \\ 
  & & \frac{\chi_{\rm ab}(E,R)}{c} \, \mathcal{H}(E,R) \, l_{\nu}(E,R).
  \label{eqn: neutrino luminosity equation}
\end{eqnarray}

The neutrinos absorbed in free-streaming conditions deposit momentum in the fluid. In the case
of spherically symmetric models, the related stress is computed as:
\begin{equation}
  \label{eqn: vdot transparent regime}
  \left( \dot{\mathbf{v}} \right)_{\tau_{\nu} \lesssim 1} (R) = \frac{1}{c} 
  \int_0^{+\infty} h(E,R) \, \mu_{\nu}(E,R) \, E^3 \: {\rm d}E \, .
\end{equation}

For systems with an approximate spherical symmetry (like collapsing stellar cores), the procedure
described above to compute $n_{\nu}$ and $\left( \dot{\mathbf{v}} \right)_{\tau_{\nu} \lesssim 1} $ can be applied 
in a ray-by-ray fashion (i.e., along radial paths starting from the center of the system).
For more general geometries, ray-tracing algorithms can be designed (e.g., \cite{Perego.etal:2014b}).

\section{Calibration and validations}
\label{sec: tests and validity}

We implement the ASL scheme, as it is described in Section~\ref{sec: the new ASL}, in the implicit, spherically symmetric hydrodynamics code 
\agile \citep[e.g.,][and references therein]{Liebendoerfer.etal:2002a,Fischer.etal:2010}.
For the calculation of the neutrino emissivities, absorptivities and scattering rates, we include 
a minimal set of neutrino-matter reactions containing the most relevant ones.
For the production and the absorption of electron flavor neutrinos,
\begin{eqnarray}
\label{eqn: neutrino reaction 1}
e^- + p & \leftrightarrow & n + \nu_e,  \\
e^- + (A,Z) & \leftrightarrow & (A,Z-1) + \nu_e, \\
e^+ + n & \leftrightarrow & p + \nub_e.
\end{eqnarray}
where $e^-$ and $e^+$ refer to electrons and positrons, while $n$, $p$ and $(A,Z)$ to neutrons, protons and
nuclei with mass number $A$ and atomic number $Z$, respectively.
Pair processes, like electron-positron annihilation and neutrino bremsstrahlung from free nucleons 
(generically referred as $N$), are expected to be secondary sources for $\nue$ and $\nueb$, but primary for $\numt$:
\begin{eqnarray}
\label{eqn: neutrino reaction 2}
e^- + e^+   & \leftrightarrow     & \nu + \nub,  \label{eqn: ep annihilation}\\
N + N & \leftrightarrow     & N + N + \nu + \nub. \label{eqn: bremsstrahlung}
\end{eqnarray}
All the reactions listed above are considered as inelastic.
Major sources of opacity for all neutrino species are provided by scattering
on nucleons and nuclei:
\begin{eqnarray}
N + \nu & \rightarrow & N + \nu,  \\
(A,Z) + \nu & \rightarrow & (A,Z) + \nu.
\label{eqn: neutrino reaction 3}
\end{eqnarray}
These scattering reactions are considered as elastic in the computation of the total and energy mean free paths.
All these weak interactions are implemented according to \cite{Bruenn1985}, apart from neutrino bremsstrahlung 
\citep{Hannestad1998} and pair production \citep{Bruenn1985,Mezzacappa.Messer:1999}, whose implementation
is described in Appendix~\ref{appA}.
The opacity provided by the scattering of neutrinos on electrons and positrons has not yet been implemented. 
It has been shown that this process is relevant to thermalize neutrinos during the collapse of the core 
(e.g., \cite{Mezzacappa.Bruenn:1993b}) and in the cooling phase of the proto-neutron star (PNS) 
in exploding models, seconds after core bounce \citep{Fischer2012}. 
Even if the effect of this reaction on the total and energy mean free path is
not considered, the thermalization effect provided by it in the 
optically thick regime is partially taken into account in our scheme by enforcing the usage of equilibrium 
Fermi-Dirac distribution functions to model the neutrino trapped component.
The impact of this effective treatment in the different parts of a core-collapse simulation is discussed 
later in more detail.

\subsection{15~$\msun$ progenitor}
\label{subsec: 15msun test}

% effective nature of the scheme: validation and verification procedure
The ASL scheme is an effective treatment for neutrino-radiation hydrodynamics. Therefore,
it requires to be tested and compared against a reference solution, not only to
check its validity and accuracy, but also to set the free parameters that appear in the scheme.
To perform this test, we choose the case of spherically symmetric core-collapse models.
We start with a zero age main sequence (ZAMS) 15~$\msun$ progenitor model, obtained by \cite{Woosley2002}.
We follow the collapse of the core and the first 300 milliseconds after core bounce.
We include $\approx 2.05~\msun$ from the initial progenitor, distributed over 103 radial zones on the adaptive Lagrangian 
grid of \agile. The corresponding initial outer radius is $\approx 7500 \, {\rm km}$ far from the origin. 
As reference solution, we use the results obtained by the \boltz code
\citep[][and references therein]{Mezzacappa.Bruenn:1993a,Mezzacappa.Bruenn:1993b,Mezzacappa.Bruenn:1993c,Mezzacappa.Messer:1999,Liebendorfer2004,
Fischer2012}, also coupled with the \agile code.
\boltz solves the Boltzmann equation using the method of discrete ordinates with
a Gauss-Legendre quadrature. \boltz incorporates all the 
neutrino reactions listed in Equations~(\ref{eqn: neutrino reaction 1})-(\ref{eqn: neutrino reaction 3}).
For consistency with our ASL implementation, we rely on the neutrino reactions implementation
reported in \cite{Liebendorfer2005a}. Recently, \cite{Lentz.etal:2012a} has shown the impact
of modern neutrino rates in \aboltz runs.
We postpone the implementation of more accurate reaction rates in the ASL to a future step. 
Furthermore, we notice that, even if neutrino-electron scattering is implemented in 
the \boltz version we are using, for consistency our reference runs do not 
include it, if not stated otherwise. We perform our tests and compute our reference solutions 
assuming Newtonian gravity
to be able to compare later with different multidimensional Newtonian hydrodynamical schemes.
In all our \aasl runs, as well as in the \aboltz ones, 
we use the Lattimer-Swesty EoS \citep{Lattimer1991}, 
with nuclear compressibility $K=220 \, {\rm MeV}$.
Neutrino energies are discretized by 20 geometrically increasing energy groups in the 
range $3 \, {\rm MeV} \leq E_{\nu} \leq 300 \, {\rm MeV}$.
In the \aboltz runs, the neutrino propagation angle is discretized 
by 6 angular bins.

% Parameter choosen
\subsubsection{Parameter choice}

\begin{table}[ht]
\begin{center}
\caption{ \label{tab:parameter and runs}}
\begin{tabular}{cccc} 
    \tableline
    Name & $\alpha_{\rm diff}$  & $\alpha_{\rm blk}$ & $\tau_{\rm cut}$   \\
    \tableline
    \hline
    \multicolumn{4}{c}{Calibration sets} \\
      CAL & $ \left\{ 3,4,5,6 \right\} $   & $ \left\{0.4,0.45,0.5,0.55,0.6 \right\} $ & $ \left\{ 10,15,20,25 \right\} $ \\
    \tableline
    \multicolumn{4}{c}{Standard set} \\
          STD   & $3 + 2 X_h$ & 0.55 & 20 \\
    \tableline
    \multicolumn{4}{c}{Parameter variation study}  \\
          AD\_2   & 2           & 0.55 & 20 \\
          AD\_5   & 5           & 0.55 & 20 \\
          AB\_45  & $3 + 2 X_h$ & 0.45 & 20 \\
          AB\_65  & $3 + 2 X_h$ & 0.65 & 20 \\
          TC\_7   & $3 + 2 X_h$ & 0.55 & 7  \\
          TC\_54  & $3 + 2 X_h$ & 0.55 & 54 \\
    \tableline
\end{tabular}
\tablecomments{Table with a summary of the values of the parameters used in the \aasl runs.
In the calibration runs (CAL), the three parameters are varied independently. The standard set (STD) is also
used in the multidimensional runs.}
\end{center}
\end{table}

\begin{figure*}
\begin{center}
\begin{minipage}{0.48 \linewidth}
\centering
\includegraphics[width = 0.7 \linewidth,angle=-90]{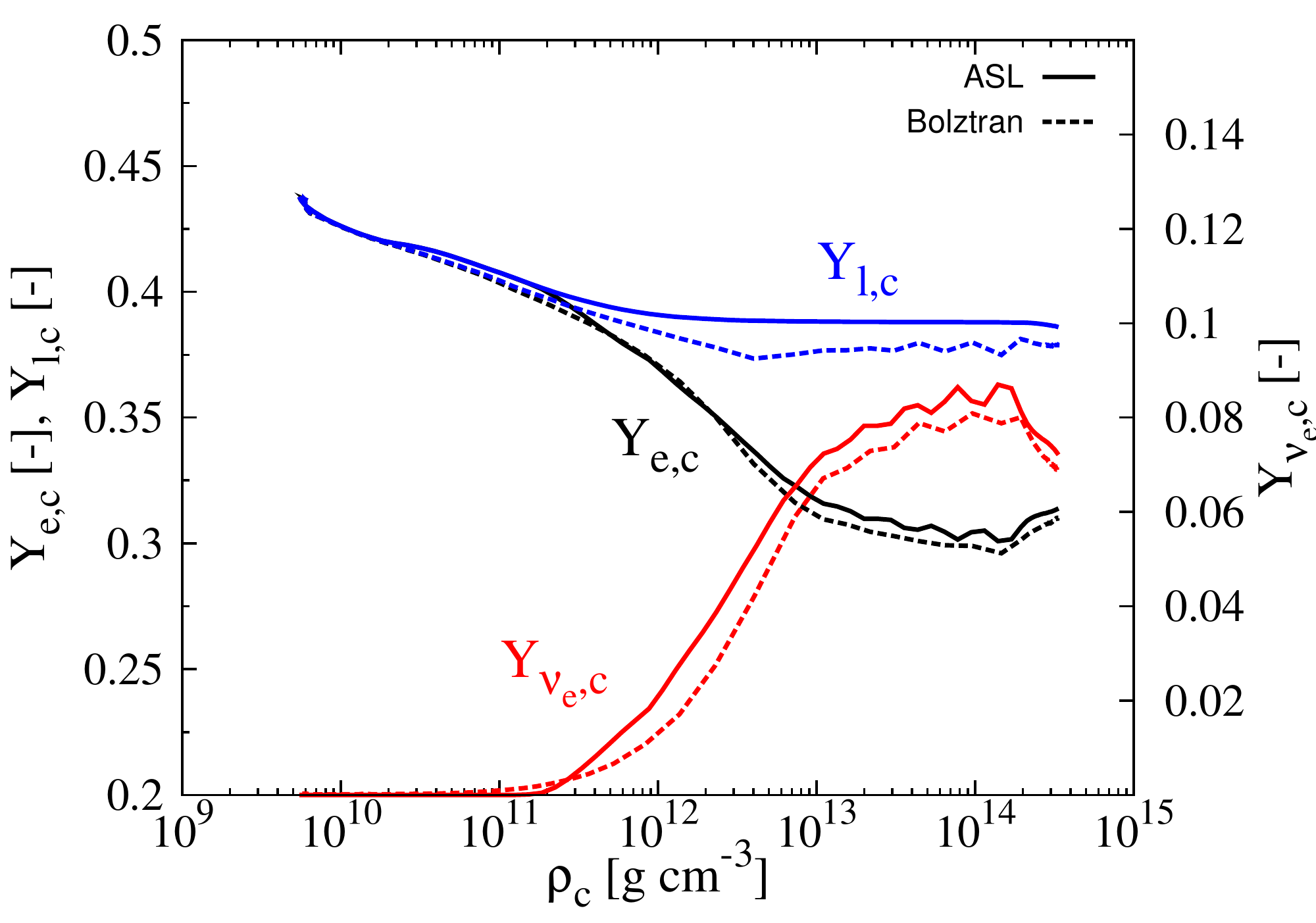}
\end{minipage}
\begin{minipage}{0.48 \linewidth}
\centering
\includegraphics[width = 0.7 \linewidth,angle=-90]{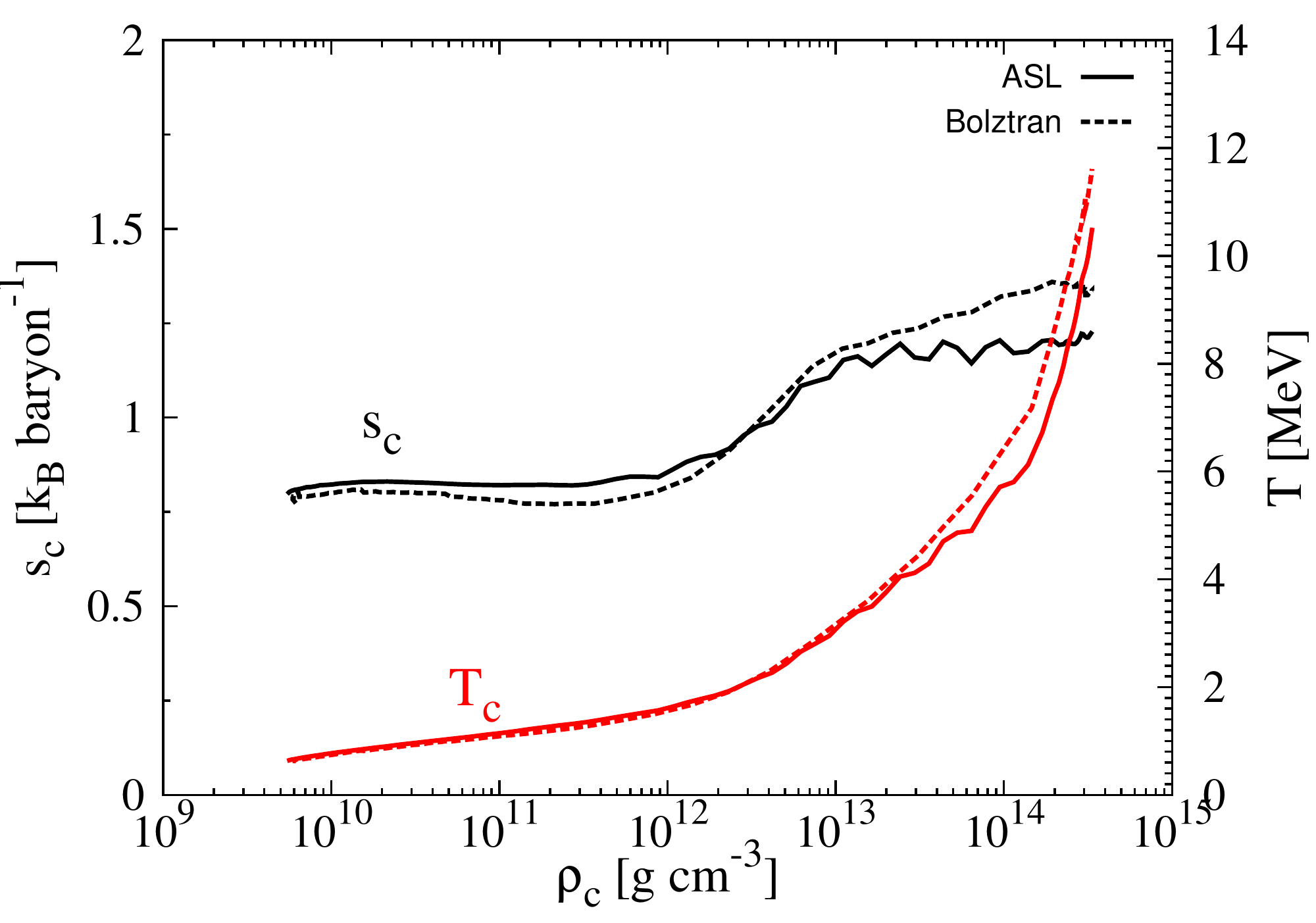} 
\end{minipage}
\end{center}
\caption{Left: Central electron fraction (black lines), electron neutrino fraction (red lines) and lepton fraction (blue lines) evolution during 
the collapse of the 15~$\msun$ model, as a function of the central density. The solid lines refer to the run obtained with the ASL scheme, 
while the dashed to the run obtained with \boltz.
Right: Same as for the left panel, but for central entropy (black lines) and temperature (red lines), as a function of the central density during the collapse. 
}
\label{fig1}
\end{figure*}

\begin{figure*}
\centering
\includegraphics[width = 0.28 \linewidth,angle=-90]{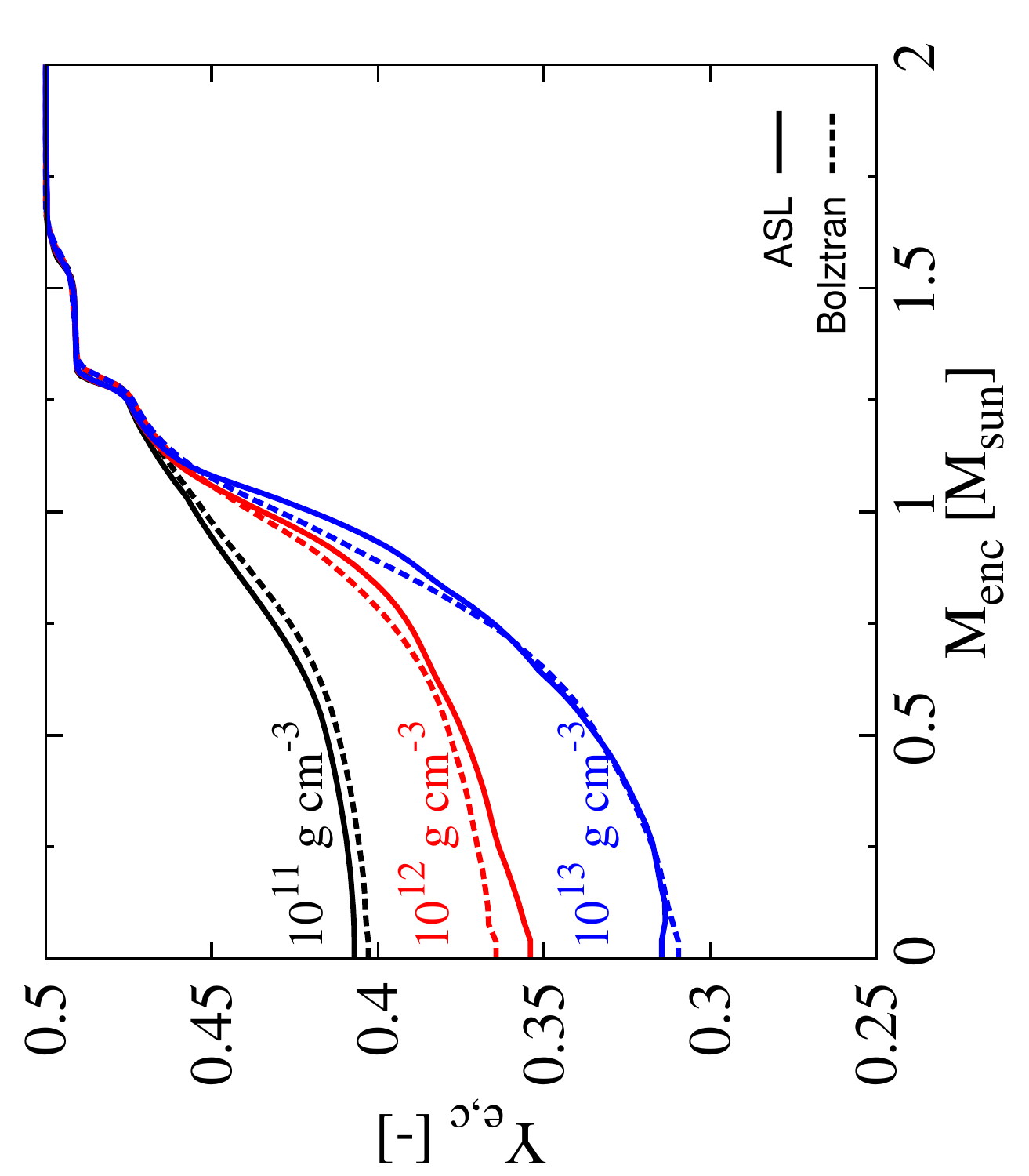}
\includegraphics[width = 0.28 \linewidth,angle=-90]{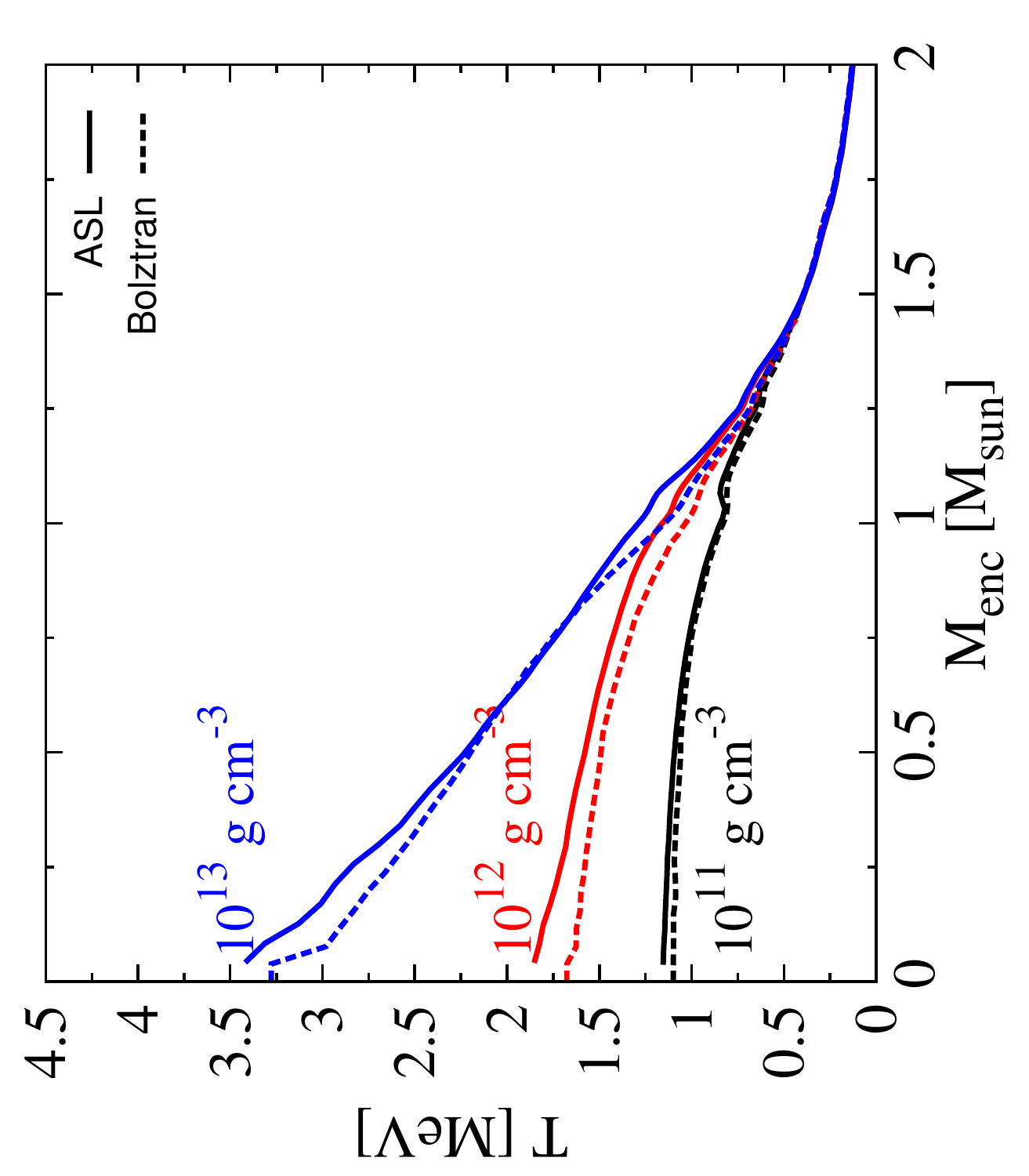}
\includegraphics[width = 0.28 \linewidth,angle=-90]{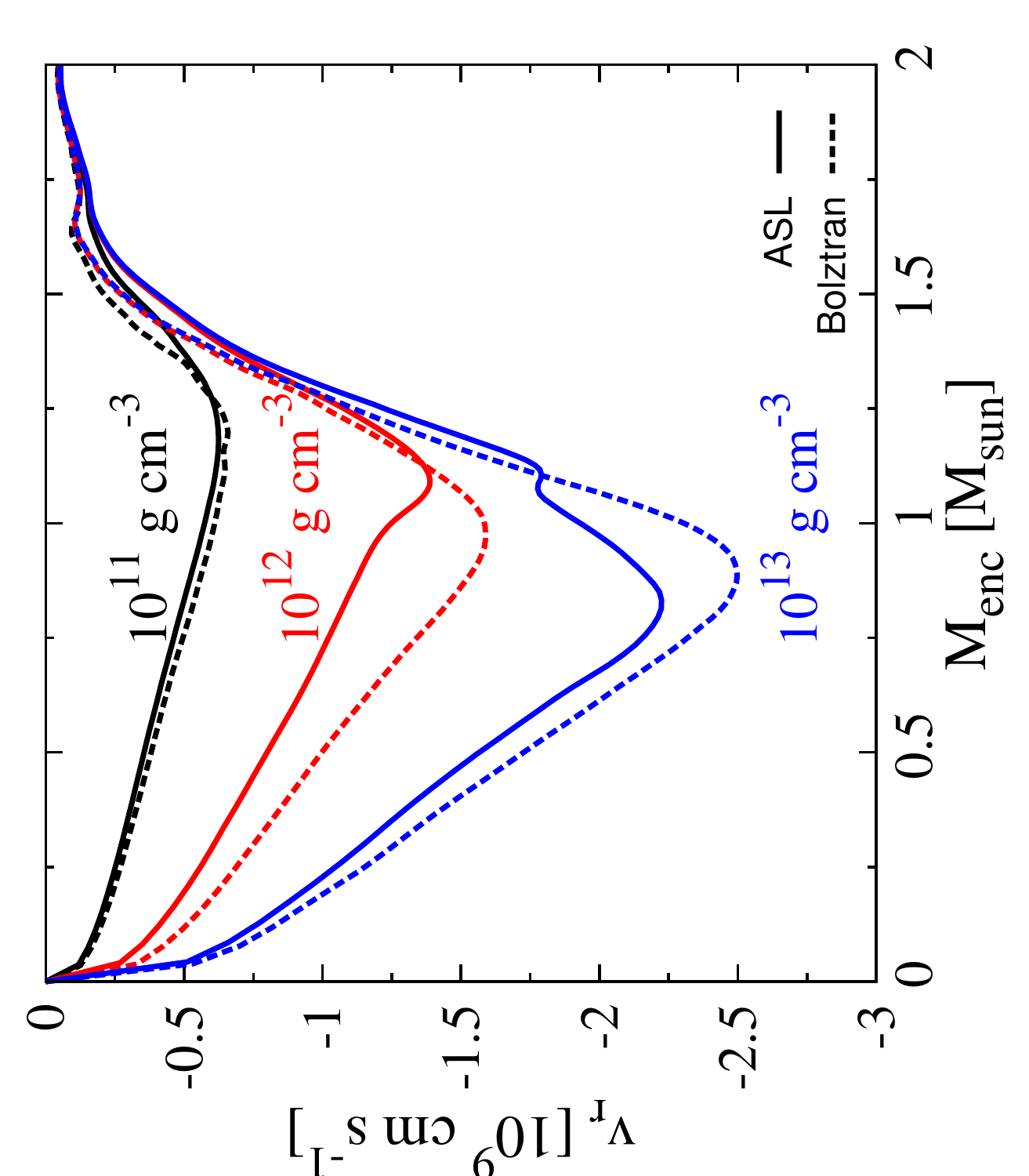}
\caption{Profiles of the electron fractions (left), temperature (middle) and radial velocity (right) as a function of the enclosed mass profile for
the 15~$\msun$ model, at three different times labeled by their central densities (black: $10^{11} \, {\rm g \, cm^{-3}}$, 
red: $10^{12} \, {\rm g \, cm^{-3}}$ and blue: $10^{13} \, {\rm g \, cm^{-3}}$). The solid lines represent the run obtained with the ASL scheme, the dashed 
the run obtained with \boltz.}
\label{fig2}
\end{figure*}

The validation and the calibration of the ASL scheme are done by comparing directly the temporal and the radial
profiles of some relevant quantities obtained with \aasl with our reference solutions.
We do not define a quantitative criterion to compare the different results, because our simulations span
a broad range of conditions and this prevents the possibility to select a single quantity as indicator.
Instead, we search for the parameter set that, overall, best matches the most important features of a
CCSN model between collapse and a few hundreds of milliseconds after core bounce. 
In particular, during the collapse we focus on central quantities (e.g., density, entropy and electron fraction).
Before and after bounce, we monitor the profiles of matter density, $\rho$, matter entropy per baryon, $s$, electron fraction, 
\ye, and radial velocity, $v_r$. Regarding the neutrino quantities, we investigate the radial profile of $Y_{\nu}$,
together with the luminosities, $L_{\nu}$, and the root mean squared (RMS) neutrino energies, $E_{\rm rms} =  \sqrt{\langle E^2_{\nu} \rangle}$,
both measured at 300 km from the center.
The different versions of \aasl differ by the usage of a distinct set of parameters,
$(\alpha_{\rm diff}, \alpha_{\rm blk}, \tau_{\rm cut})$. The ranges we have explored are
reported in the first row of Table~\ref{tab:parameter and runs}.

The values of the parameters that provide the overall best agreement in all the monitored quantities, during
the entire simulation, are:
\begin{itemize}
\item $ \alpha_{\rm diff} = 3 \, \left( 1 + 2 \,  X_h / 3 \right)$, 
where $X_{h}$ is the mass fraction of heavy nuclei. 
This peculiar dependence is a simple interpolation between two limiting behaviors: 
$\alpha_{\rm diff} \approx 5$ in the unshocked regions (more relevant in the collapse phase) and 
$\alpha_{\rm diff} \approx 3$ in the shocked ones (more relevant in the post bounce phase).
The physical interpretation of this difference is linked with the fact that in the unshocked region $\nue$ and $\nueb$ opacities depend more on
quasi-elastic scattering on nucleons and nuclei than on inelastic absorption processes ($\tau_{\nu,{\rm en}} \ll \tau_{\nu,{\rm tot}}$);
while in the shocked region both are equally important ($\tau_{\nu,{\rm en}} \lesssim \tau_{\nu,{\rm tot}}$).
When isoenergetic scattering dominates, high energy neutrinos diffuse changing more slowly 
their energy and interacting more with matter, compared with thermalized neutrinos. 
This effect leads to a significant increase of the diffusion timescale.
\item $\alpha_{\nue,\nueb{\rm blk}} = \alpha_{\rm blk} = 0.55$, while $\alpha_{\numt,{\rm blk}} = 0 $.
\item $\tau_{\rm cut} = 20$.
\end{itemize}
We will refer to this set of values as our {\it standard set} (second row in Table~\ref{tab:parameter and runs}).

\subsubsection{Collapse and bounce phase}

In the following, we describe the results we obtain for our 15~$\msun$ model and how they compare with
our reference solution.
The stellar iron core collapses until densities in excess of nuclear saturation density are reached 
in the center and a shock wave forms. Free streaming 
neutrinos reduce the total lepton number, while the electron fraction is further decreased by the conversion
of electrons into $\nue$.
The core reaches core bounce on a time $t_{\rm bounce} = 248 \, {\rm ms}$ after the beginning of the simulation. 
The central density at bounce is $\rho_{\rm bounce} = 3.36 \times 10^{14} \, {\rm g \, cm^{-3}}$ and the 
enclosed mass at the shock formation point is 
$M_{\rm enc,bounce} = 0.75~\msun$.
The central density and the initial PNS mass compare closely with the corresponding values obtained by \aboltz
($\rho_{\rm bounce, AB} = 3.34 \times 10^{14} \, {\rm g \, cm^{-3}}$ and $ M_{\rm enc,bounce,AB} = 0.74~\msun $, respectively),
while the collapse time of the reference model is shorter, $t_{\rm bounce, AB} = 205 \, {\rm ms}$.
In Figure~\ref{fig1}, we compare the central values obtained for \ye and $Y_{\nu_e}$ (left panel), and for 
the entropy and temperature (right panel), as a function of the central density during the collapse.
Moreover, in Figure~\ref{fig2}, we plot radial profiles of \ye, temperature and radial velocity, as a function 
of the enclosed mass, for three different times during the collapse (labeled by their central density).
In all cases, we have obtained a good agreement with the reference solution. 
The decrease of \ye is well reproduced during the deleptonization process, while neutrino trapping occurs when the central 
density reaches $\rho_{\rm c} \approx 2 \times 10^{12} \, {\rm g \, cm^{-3}}$. After that, the further decrease
of \ye is compensated by the growth in $Y_{\nu_e}$, keeping the total lepton number and the entropy roughly constant.
In the ASL model, the entropy per baryon and the electron fraction stay almost constant after neutrino trapping 
up to core bounce, inside the innermost 0.8~$\msun$.
In the \boltz reference solution, the detailed treatment of the equilibrium approach 
and diffusion process slightly reduces the total lepton number ($\Delta Y_{\rm l} \approx 0.02$), 
compared with the ASL solution.
Similarly, the entropy per baryon rises just before the formation of the shock wave 
($\Delta s \approx 0.2 \, k_{\rm B }$). This difference is due to full 
thermal and weak equilibrium with matter assumed in the ASL treatment.
However, this equilibrium is only approximated and deviations from it 
leads to a slightly larger matter entropy \citep{Cooperstein.etal:1986,Cooperstein.etal:1987a}.
Finally, in the radial velocity profile, we notice a wiggle appearing after neutrino-trapping sets in, around
$M_{\rm enc} \approx 1~\msun$.
This is due to the neutrino stress, computed according to Equation~(\ref{eqn: vdot opaque regime}),
which can overestimate the stress at the interface between the trapped and the free-streaming regime,
and neglect additional momentum transfer in optically thin conditions.

\subsubsection{Post bounce phase}

After core bounce, the shock wave propagates outwards and iron group nuclei falling into the shock are photo-dissociated
into neutrons and protons. Once the shock reaches the relevant neutrinospheres, electron capture on free protons in almost free-streaming conditions
causes a peak in $\nue$ luminosity and a fast neutronization of the shocked matter.
Later, the absorption of electrons and positrons on free nucleons, together with neutrino pair processes, 
produces an intense radiation emission of neutrinos of all flavors.
The combined effect of the nuclei photo-dissociation and neutrino emission causes the prompt
shock expansion to stop and the shock itself to stall within a few tens of milliseconds. The absorption of neutrinos
inside the so-called gain region increases the shock radius during several tens of milliseconds after the stalling. 
However, this energy deposition is not enough to revive the shock and lead to an explosion of the star. 

\begin{figure*}
\begin{minipage}{0.33 \linewidth}
\centering
\includegraphics[width = 0.65 \linewidth,angle=-90]{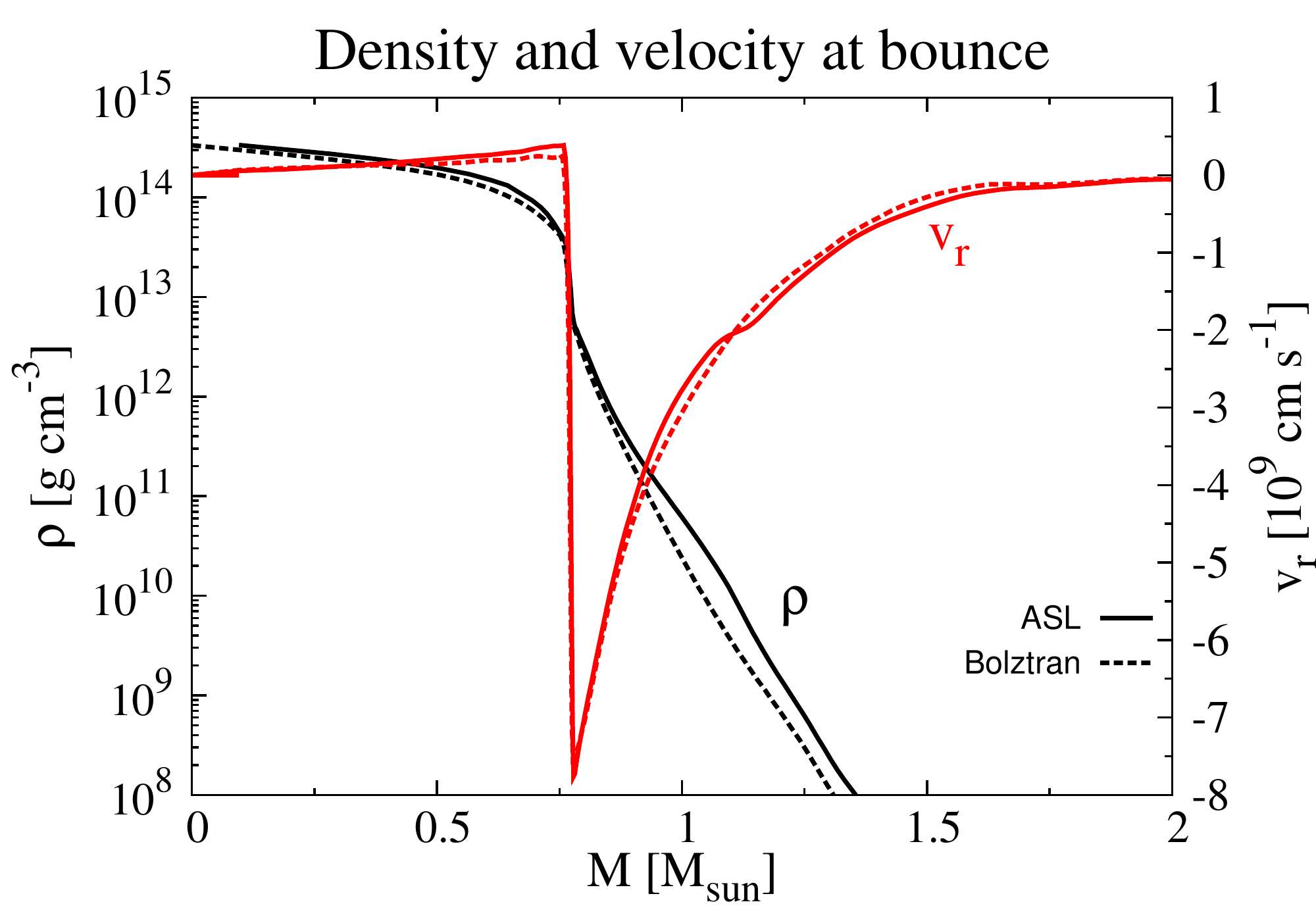}
\end{minipage}
\begin{minipage}{0.33 \linewidth}
\centering
\includegraphics[width = 0.65 \linewidth,angle=-90]{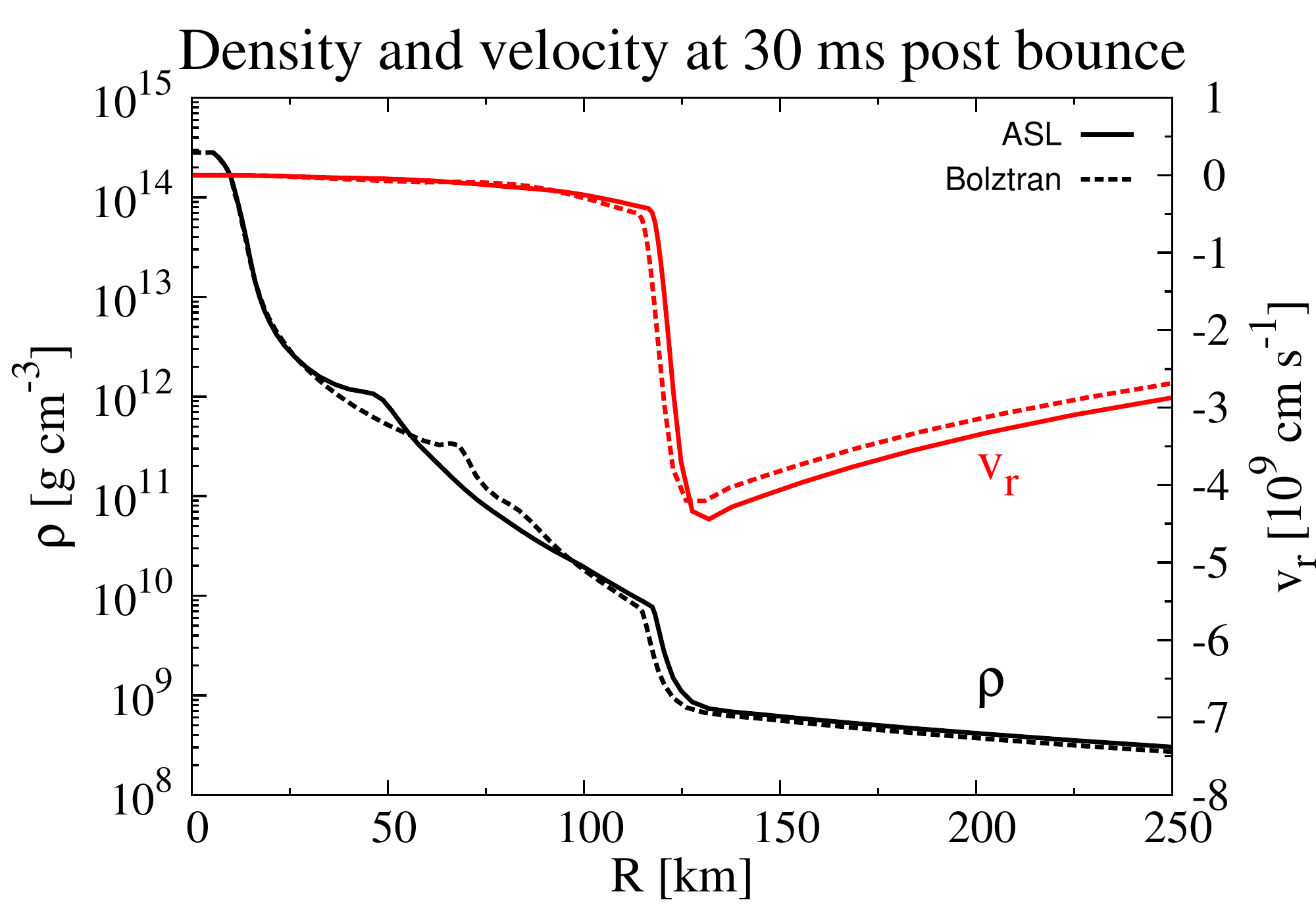}
\end{minipage}
\begin{minipage}{0.33 \linewidth}
\centering
\includegraphics[width = 0.65 \linewidth,angle=-90]{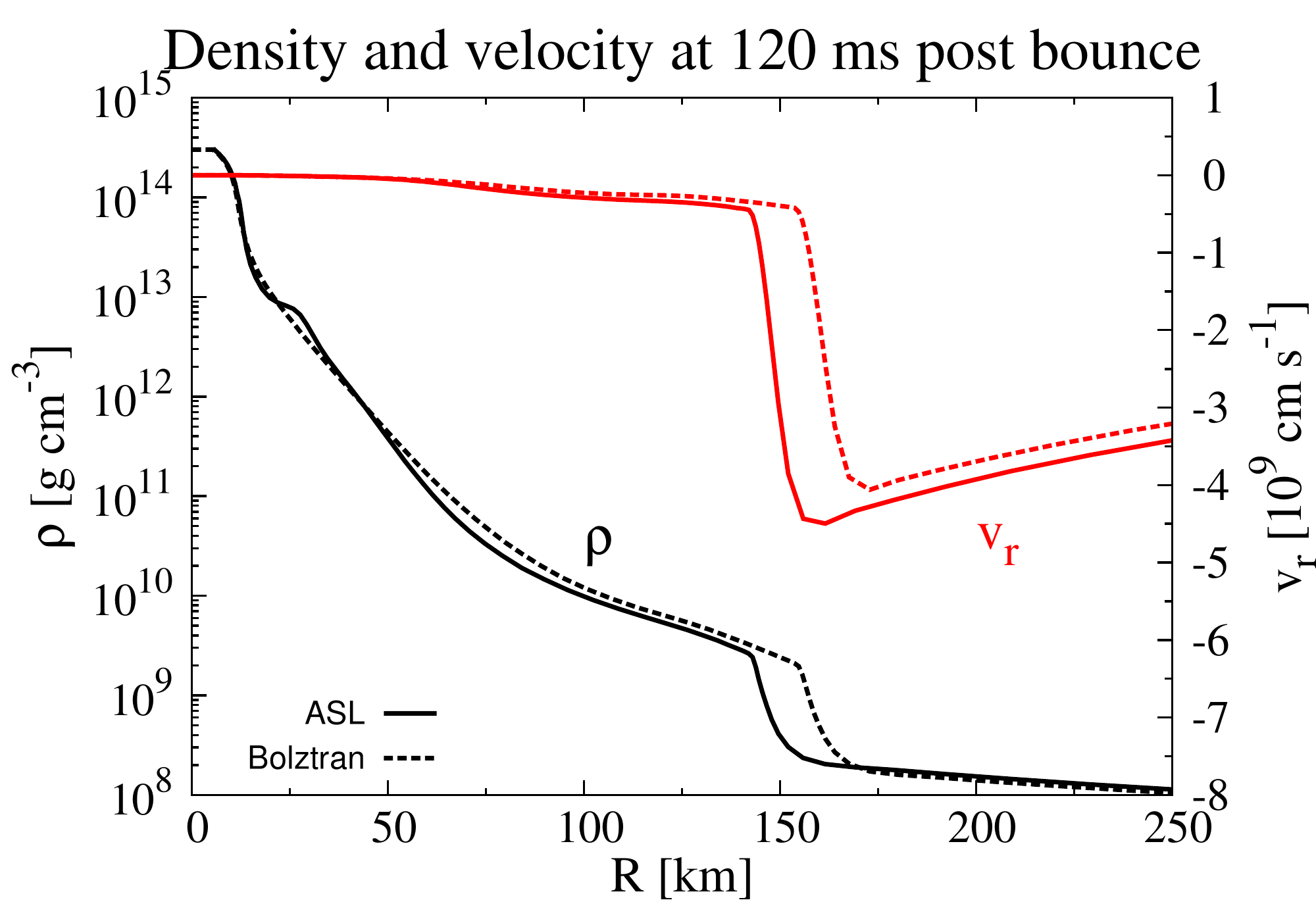}
\end{minipage}\\
\begin{minipage}{0.33 \linewidth}
\centering
\includegraphics[width = 0.65 \linewidth,angle=-90]{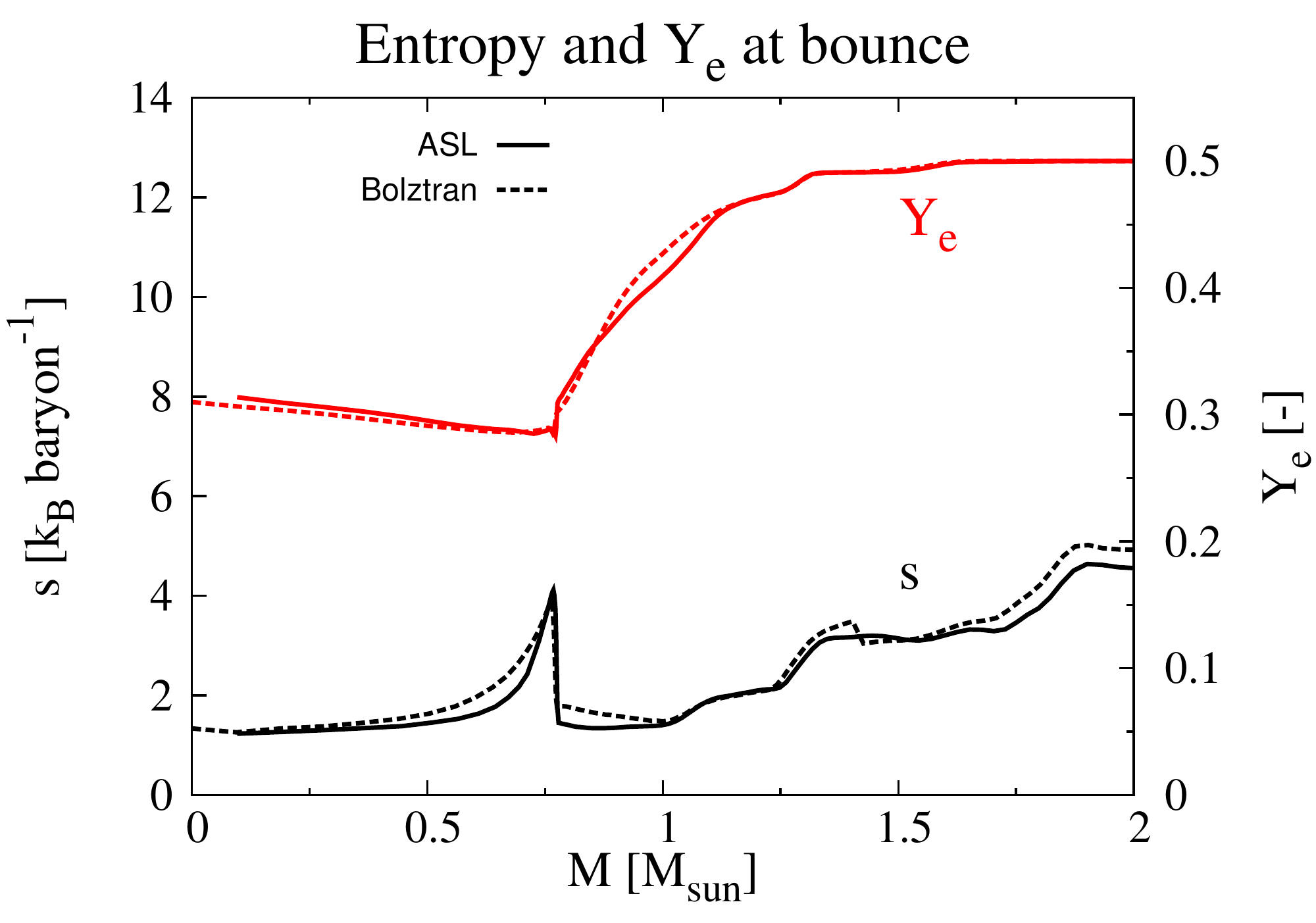}
\end{minipage}
\begin{minipage}{0.33 \linewidth}
\centering
\includegraphics[width = 0.65 \linewidth,angle=-90]{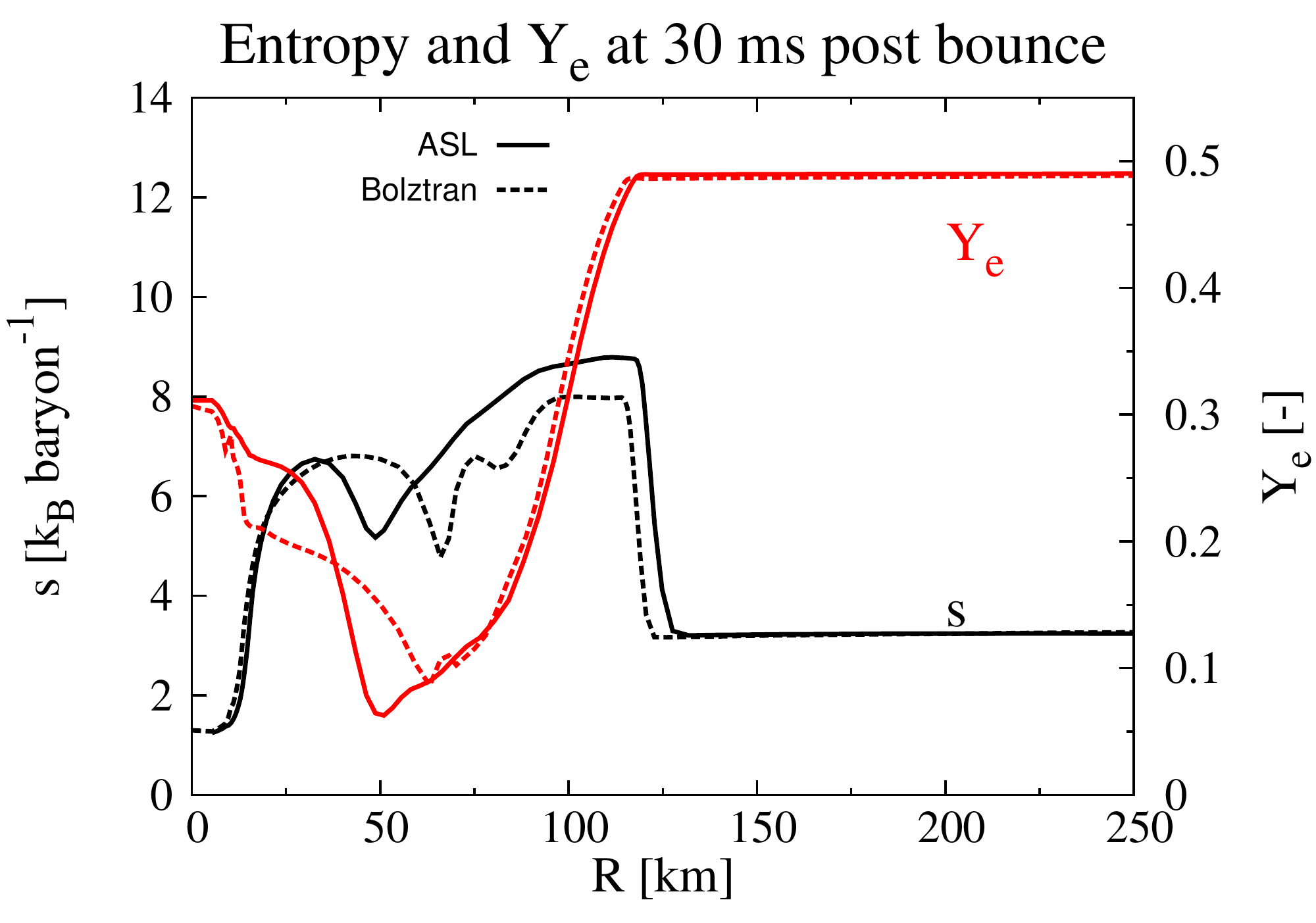} 
\end{minipage}
\begin{minipage}{0.33 \linewidth}
\begin{center}
\includegraphics[width = 0.65 \linewidth,angle=-90]{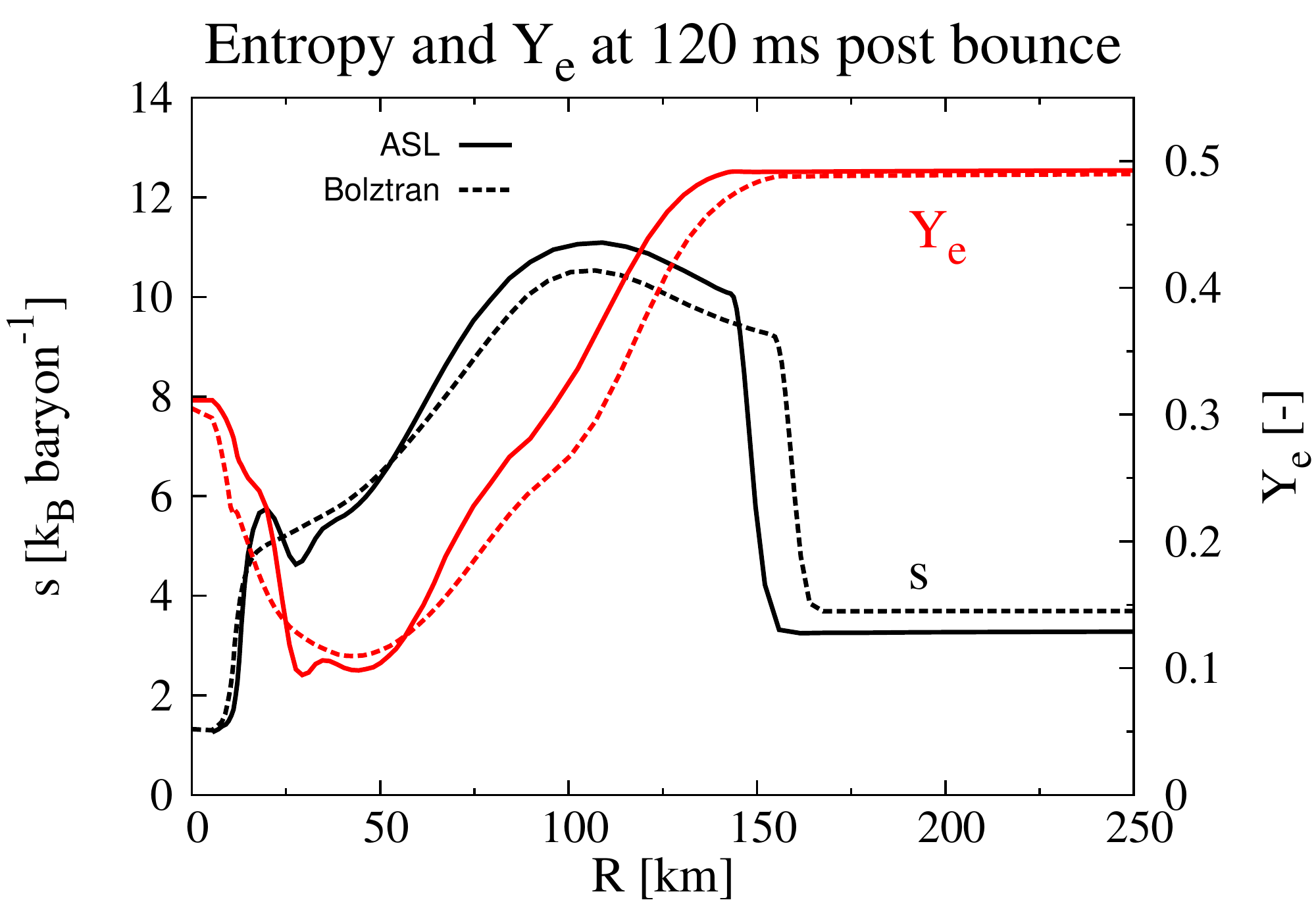}
\end{center}
\end{minipage}
\caption{Profiles of the density and radial velocity (top row), and of the entropy and electron fraction (bottom row),
at three different times for the 15~$\msun$ model: at core bounce (left panels, using the enclosed mass 
as independent variable), at 30 ms and 120 ms after core bounce
(central and right panels, respectively, using the radial distance as independent variable).
The solid lines represent the run obtained with the ASL scheme, the dashed to the run obtained 
with \boltz.}
\label{fig3}
\end{figure*}

Figure~\ref{fig3} shows radial profiles of several quantities 
at core bounce, as a function of the enclosed mass, as well as at two different times after core bounce 
($t \approx 30 \,{\rm ms}$ and $t \approx 120 \, {\rm ms}$), as a function of the radial 
distance from the center. 
The results obtained with the ASL scheme
show the most relevant features and the expected typical evolution. 
We find a good agreement for the location
of the shock during the different phases. We recognize the effect of the passage of the shock wave in 
the electron fraction profile as well as the result of the neutrino emission and absorption 
on the entropy profile (especially around 120 ms, where the increase of the entropy inside the 
gain region can be seen).
A detailed comparison with the reference solution shows several quantitative differences between the two models.
They originate from our approximate treatment, compared to a detailed neutrino transport scheme.
However, the overall qualitative (and also a partial quantitative) agreement between the two models is preserved
during the entire simulation time.

In the left panel of Figure~\ref{fig4}, we present the temporal evolution of the shock and PNS radii 
(defined as $\rho(R_{\rm PNS}) = 10^{11} \, {\rm g \, cm^{-3}}$).
Overall, $R_{\rm shock}$ and $R_{\rm PNS}$ evolutions are in good agreement with the reference model.
In the \aasl run, the shock expansion reaches its maximum, 
$\approx 145 \, {\rm km}$, around at 140 ms. This maximum extension is $\sim 10 \%$ smaller than the maximum $R_{\rm shock}$ obtained
by the \aboltz simulation. We also notice that the latter reaches its maximum earlier ($\approx 105 \,{\rm ms}$).
In general, the shock evolution is more pessimistic in the ASL model than in the reference one.
On one hand, this is a consequence of the smaller PNS radius \citep[e.g.][]{Marek.Janka:2009}, which in turn
is due to an overestimated neutrino cooling happening in the semi-transparent regime.
On the other hand, the comparison with a \boltz model including also neutrino-electron scattering (thin long-dashed lines) 
suggests that the smaller shock radius obtained by the ASL during the first 30ms after core bounce could also be
the result of the effective inclusion of the neutrino thermalization provided by this inelastic process.
This effect, together with the enhanced $\nue$ luminosity around neutrino burst, compensates for the larger
enclosed mass at core bounce. Finally, despite the slightly smaller extension, 
the PNS contraction rate within the first 300 ms is similar to the one of the reference solution.

\subsubsection{Neutrino quantities}

\begin{figure*}
\centering
\includegraphics[width = 0.23 \linewidth,angle=-90]{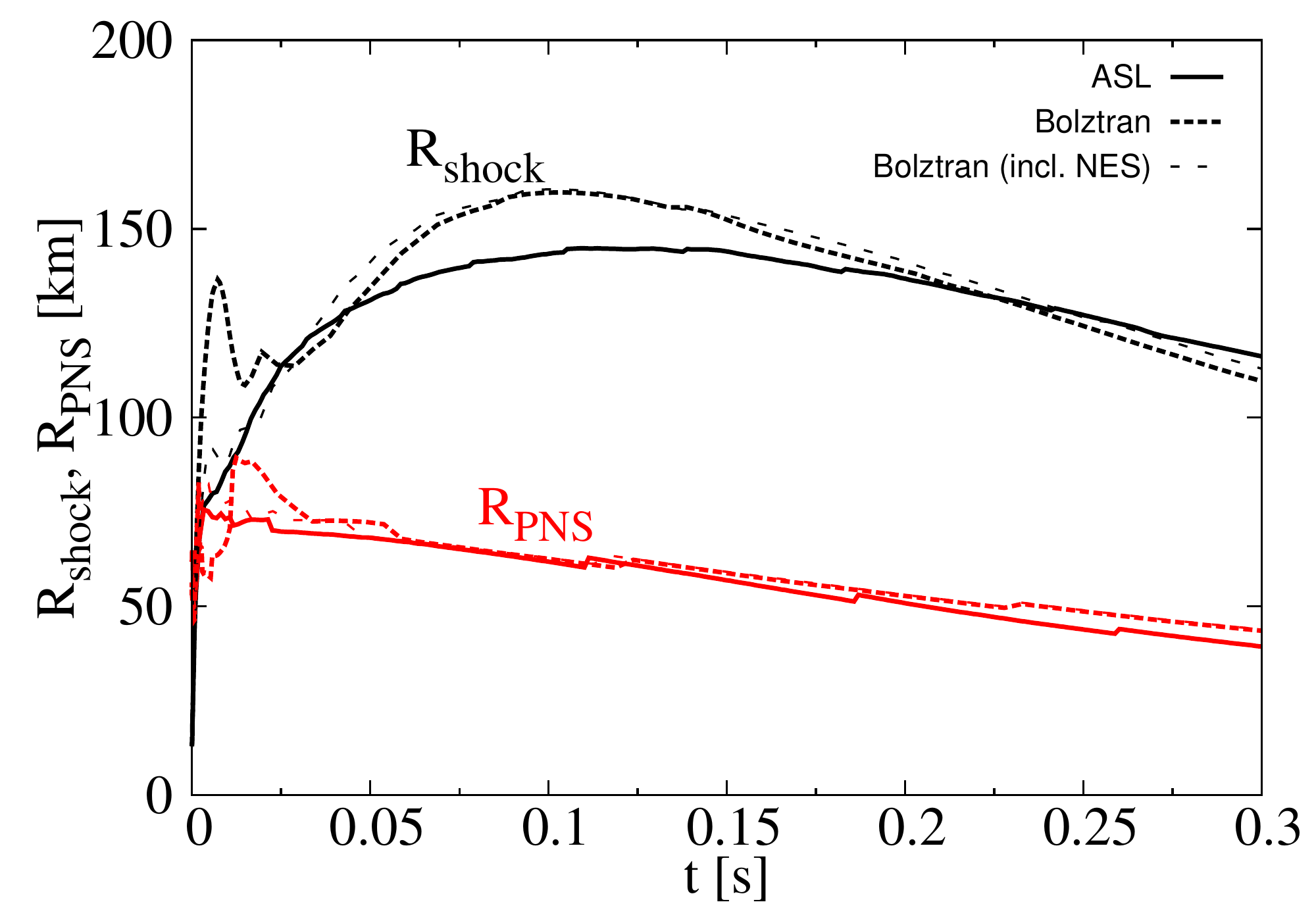}
\includegraphics[width = 0.23 \linewidth,angle=-90]{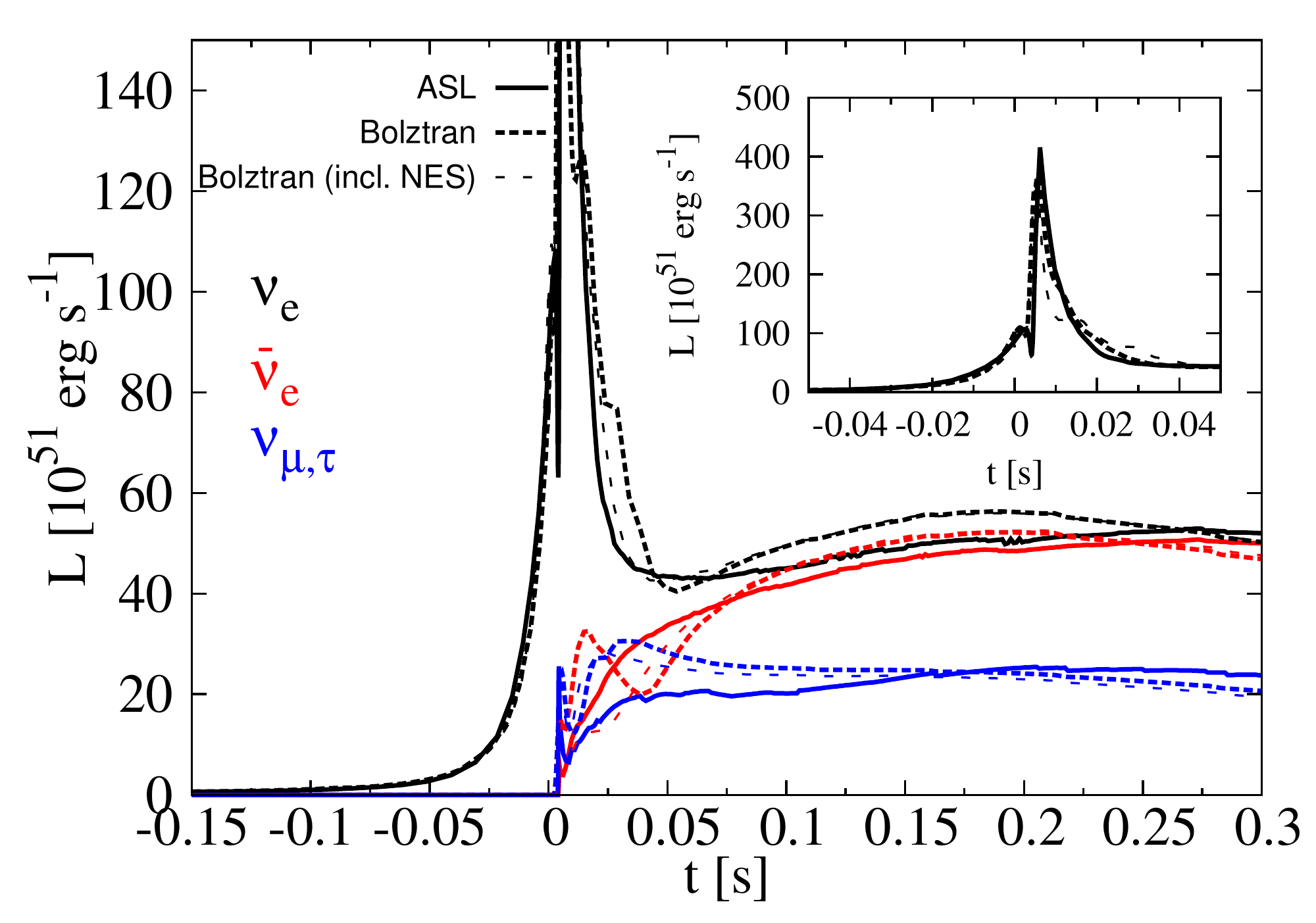}
\includegraphics[width = 0.23 \linewidth,angle=-90]{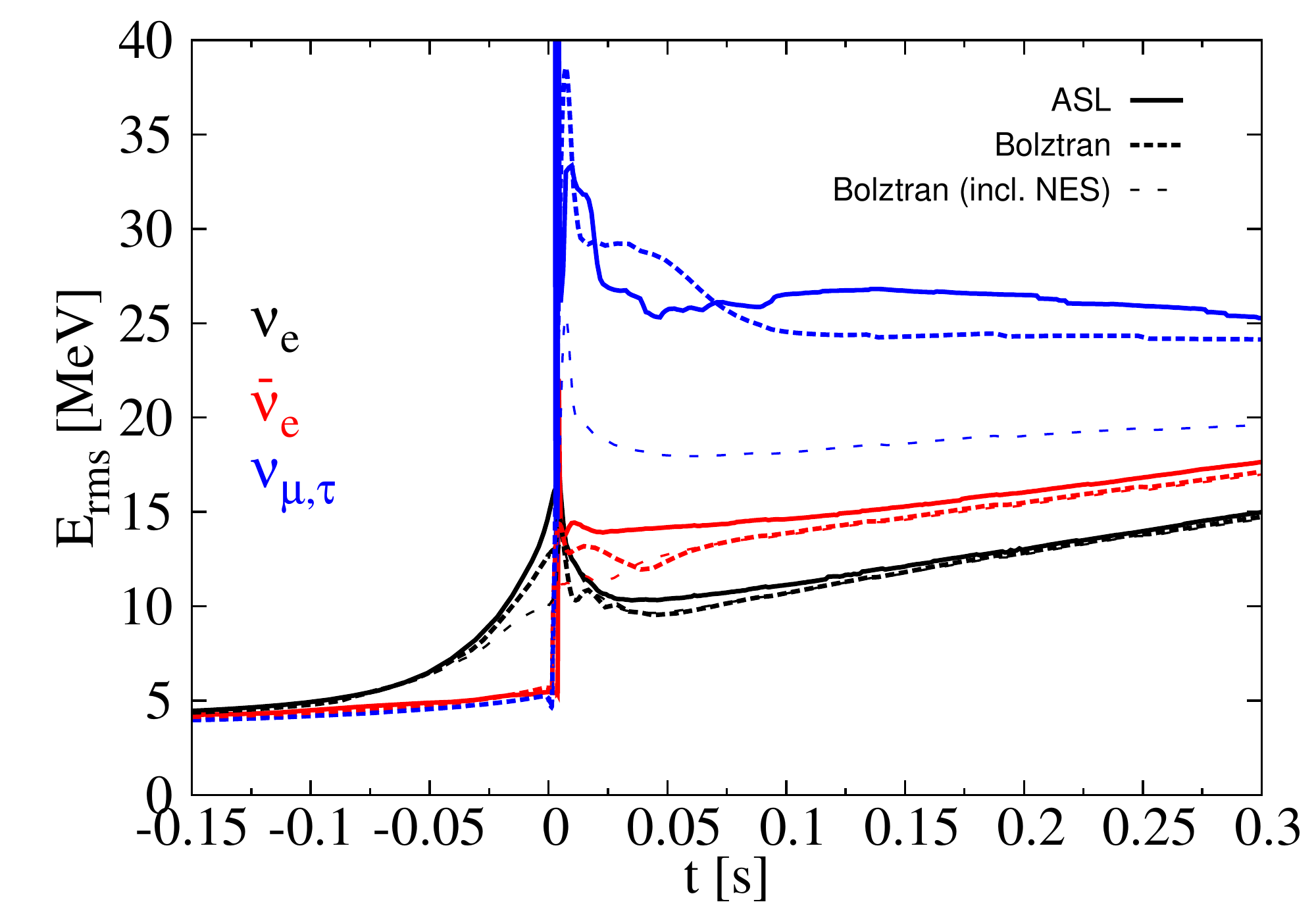}
\caption{Temporal evolution of the shock and PNS radii (left panel), of the neutrino luminosities (middle panel) and of the neutrino RMS energies
(right panel) for the calibration run of a 15~$\msun$ progenitor. The solid lines represent the run obtained with the ASL scheme, 
the thick-short dashed lines the run obtained with \boltz. For comparison purposes, we also plot results obtained with \boltz
including neutrino-electron scattering (thin-long dashed lines).}
\label{fig4}
\end{figure*}

The neutrino luminosities and RMS energies are displayed in the middle and right panels of Figure~\ref{fig4}.

During the collapse phase, electron capture on nuclei causes the $\nue$ luminosity to rise. This increase proceeds monotonically up to
neutrino trapping ($L_{\nue} \approx 10.6 \times 10^{52} \, {\rm erg \, s^{-1}}$), when the enhanced core opacity and the fast decreasing 
collapse timescale halt (and even slightly decrease) the $\nue$ luminosity. Once the core has bounced and 
the shock has passed through 
the relevant neutrinospheres, $L_{\nue}$ shows a burst and peaks at $4.1 \times 10^{53} \, {\rm erg \, s^{-1}}$. 
The same behavior appears in the reference solution obtained by \aboltz, even if the luminosities 
are a bit smaller ($9.5 \times  10^{52} \, {\rm erg \, s^{-1}}$ at
trapping and $3.5 \times  10^{53} \, {\rm erg \, s^{-1}}$ at burst).
After the neutrino burst, $L_{\nue}$ decreases and stabilizes during the accretion phase, $L_{\nue} \approx 5.0 \times 10^{52} \, {\rm erg \, s^{-1}}$.
At the same time, the $\nueb$ luminosity rises and becomes almost equal to $L_{\nue}$ during the whole accretion phase.
The same trend is observed in the reference solution and the differences in the absolute values are usually within 10\%.
The smaller values obtained by the ASL scheme are related with the smaller neutrinosphere radii.
Also the rise of $L_{\numt}$ proceeds after core bounce, but a few milliseconds before $L_{\nueb}$. 
This is expected because of the negative $\nueb$ degeneracy parameter and to their larger opacity. 
In the rising phase (i.e., within the first $\sim 75 \, {\rm ms}$), the ASL scheme 
underestimates the $\numt$ luminosities, while the agreement increases during the stationary accretion phase. This discrepancy is due
to the difficulty by the ASL scheme in modeling the dynamical rise of the $\numt$'s, which are characterized by an extended scattering atmosphere 
above the radius where neutrino bremsstrahlung and pair production freeze out.

The RMS energies obtained by the ASL scheme show trends in agreement with the reference solution. 
A harder spectrum is obtained for $\nue$ during the collapse phase and for $\nueb$ in the first tens 
of milliseconds after core bounce.
We notice that the RMS energies for $\numt$'s are consistent with the values obtained by the reference solution,
i.e. without including neutrino-electron scattering. However, the inclusion
of this process in \aboltz leads to significantly smaller energies (by $\sim 20-25\%$). 
This indicates that this process is responsible for efficiently down-scattering high energy $\numt$'s, 
while they diffuse out from the core.
The same behavior is visible for $\nue$'s during the latest phases of the collapse.

\subsubsection{Parameter variations}
\label{sec: parameter variations}

\begin{figure*}
\centering
\includegraphics[width = 0.23 \linewidth,angle=-90]{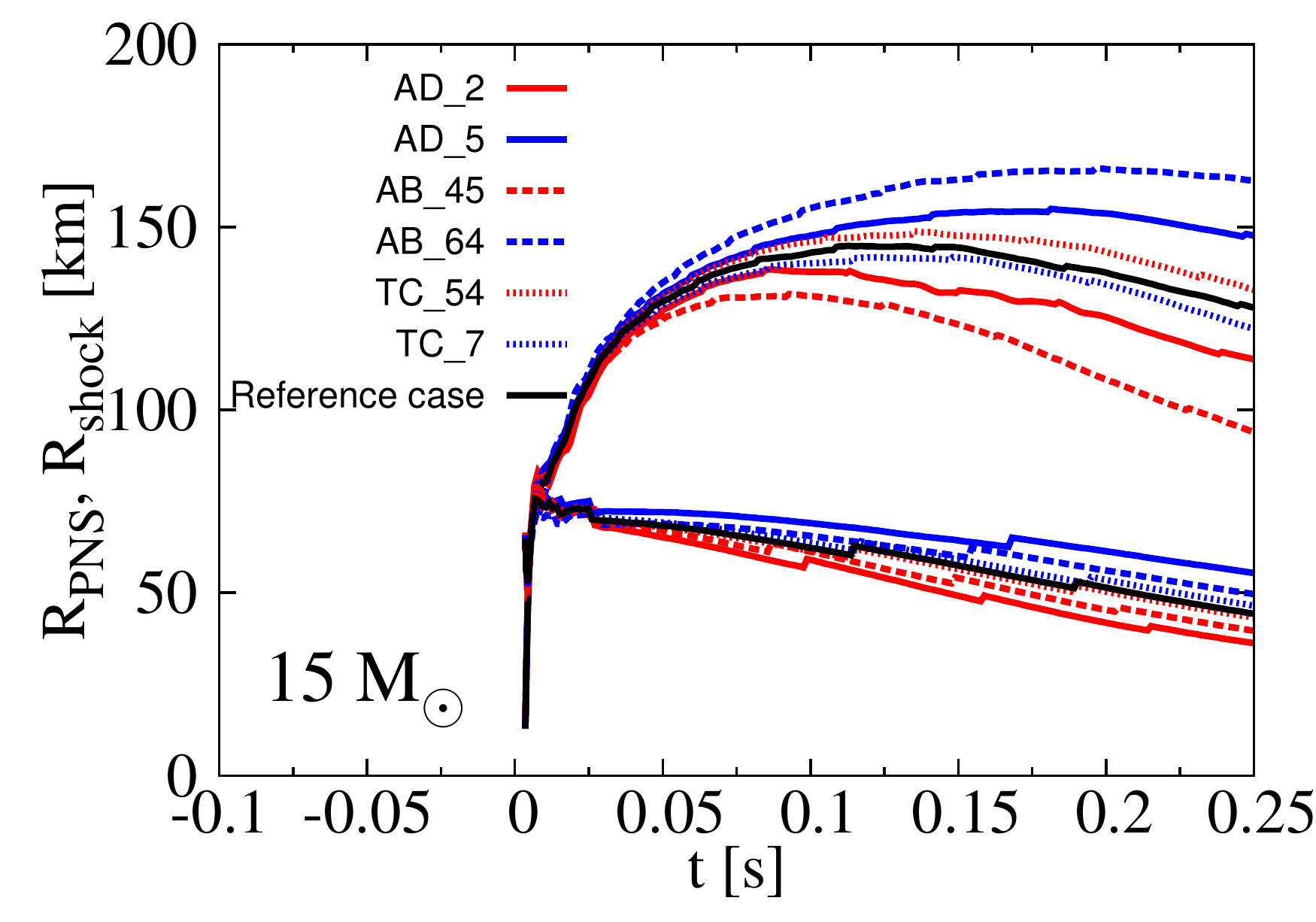}
\includegraphics[width = 0.23 \linewidth,angle=-90]{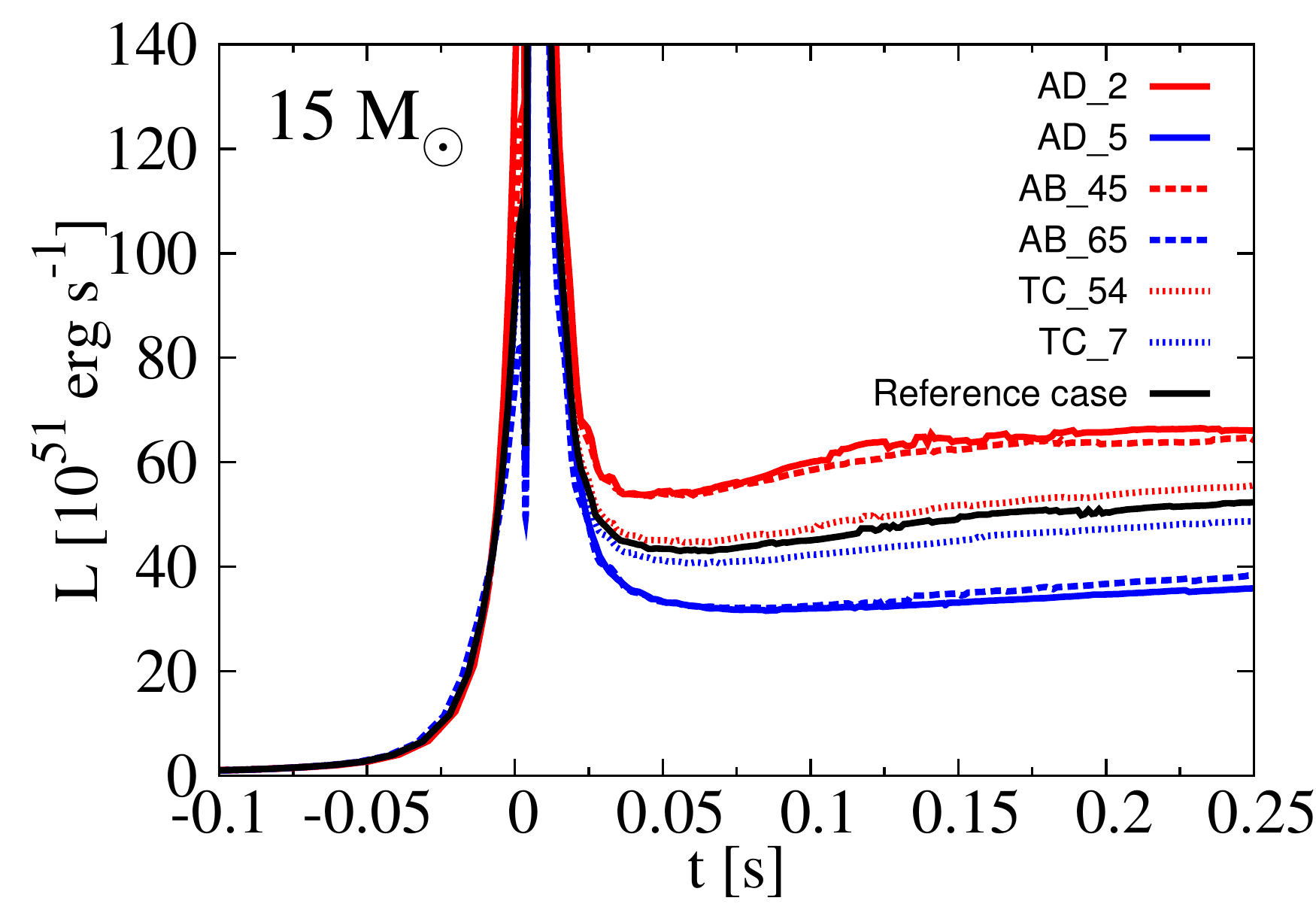}
\includegraphics[width = 0.23 \linewidth,angle=-90]{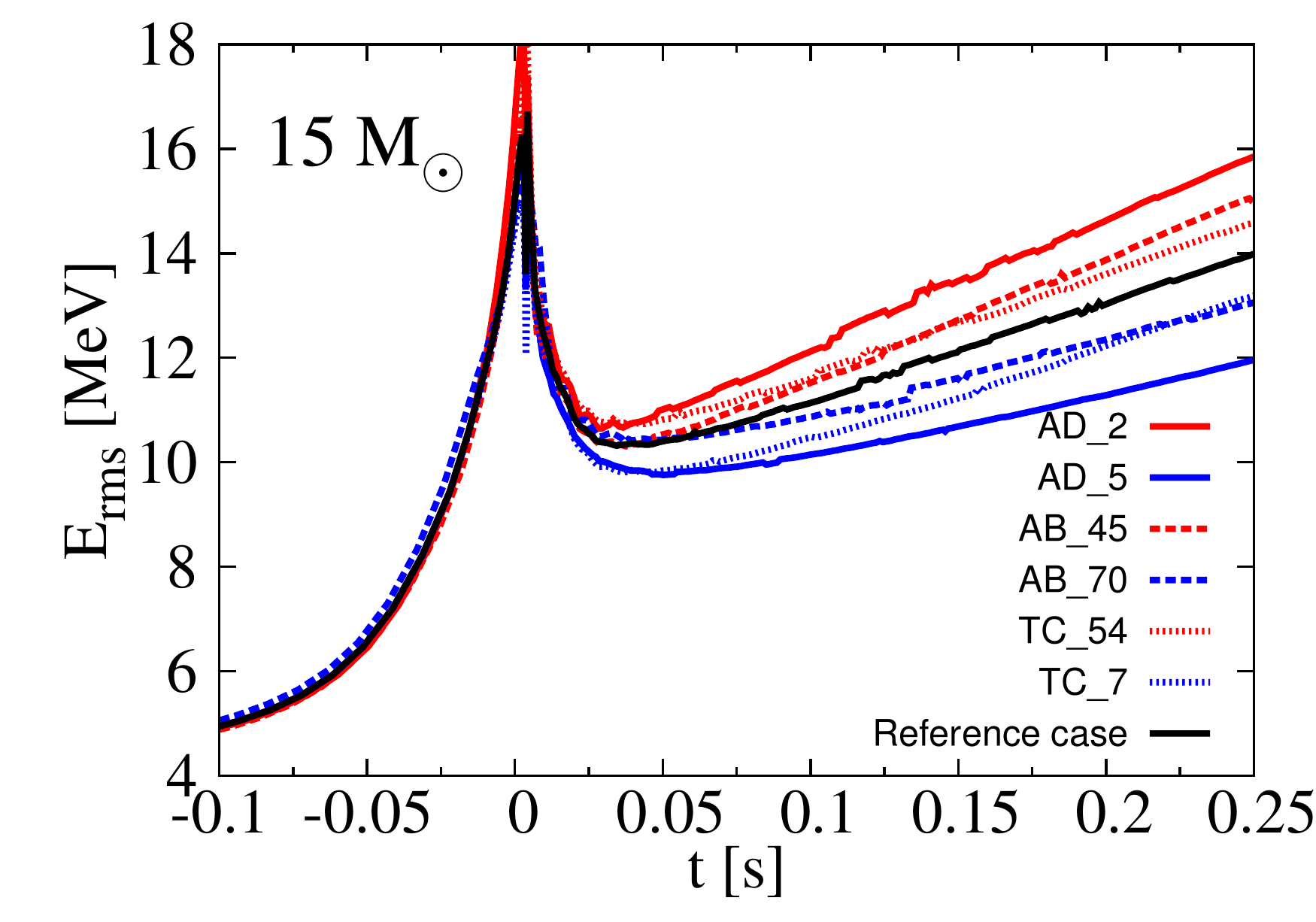}
\caption{Comparison of the temporal evolution of the shock and PNS radii (left panel), of 
the neutrino luminosities (middle panel) and of the neutrino RMS energies (right panel)
between our reference case and six runs obtained varying the three free parameters, $\alpha_{\rm blk},
\alpha_{\rm diff}$ and $\tau_{\rm cut}$ (see the text for details).}
\label{fig5}
\end{figure*}

We briefly explore the sensitivity of the ASL scheme with respect to variations of its
free parameters ($\alpha_{\rm diff}, \alpha_{\rm blk}, \tau_{\rm cut}$) around 
the calibrated values. To do this, we consider the 15~$\msun$ model of Section \ref{subsec: 15msun test},
with its standard set of parameters, and we vary independently each of them. We choose 
$\alpha_{\rm diff} = 2$ (AD\_2 model) and $\alpha_{\rm diff} = 5$ (AD\_5 model),
$\alpha_{\rm blk} = 0.45$ (AB\_45 model) and $\alpha_{\rm blk} = 0.65$ (AB\_65 model),
and $\tau_{\rm cut} = 7$ (TC\_7 model) and $\tau_{\rm cut} = 54$ (TC\_54 model),
representing six independent tests (see Table~\ref{tab:parameter and runs}).
We notice that the diffusion rates are proportional to 
$1 / \alpha_{\rm diff}$, and the chosen diffusion parameters could be also expressed as
$1 / \alpha_{\rm diff} = 0.5$ (AD\_2 model) and $1 / \alpha_{\rm diff} = 0.2$ (AD\_5 model).
Thus, the variations of $\alpha_{\rm diff}$ and $\alpha_{\rm blk}$ span an effective interval
of roughly $\pm 30 \%$ around the calibrated values. For $\tau_{\rm cut}$, we choose two values such
that $ \ln{\left( \tau_{\rm cut} \right)} / \ln{\left( 20 \right)} \approx 1 \pm \left( 1/3 \right) $

In Figure~\ref{fig5}, we show
the shock and PNS radius,
the $\nue$ luminosity, 
and the $\nue$ RMS energy (the corresponding curves for $\nueb$ and $\numt$ present analogous trends)
for each of the six tests, in comparison with the reference case. 
The parameter $\alpha_{\rm blk}$ alters the emission
rates everywhere inside the core (cf. Equation~(\ref{eqn: r_eff last version})), causing a variation 
of the total neutrino luminosity roughly equal to the variation of the parameter itself. 
Since it applies equally to all neutrino 
energies, it does not affect directly the neutrino spectrum and it modifies only marginally 
the evolution of the neutrino mean energies. The variations of the RMS energies that we observe 
in AB\_45 and AB\_65 are mainly due to the different evolution of the radial profiles of the 
thermodynamical quantities. A variation of $\pm 30 \%$ of $\alpha_{\rm blk}$ does not change 
the qualitative behavior of the simulations, but it changes the shock radius 
significantly, by a few tens of kilometers. Variations of the PNS radius are more restrained.

Since the diffusion rates affect mainly the behavior of the deep interior 
of the collapsing core, the radius of the PNS is more sensitively affected by 
variations of the diffusion parameter $\alpha_{\rm diff}$.
In particular, an increase of $\alpha_{\rm diff}$ (AD\_5)
causes a decrease of the diffusion rates. It also moves outwards the
transition region between the diffusion and the production rates (cf. Equation~(\ref{eqn: r_eff old version})).
The combined results of these effects are a decrease
of the neutrino luminosities (also in this case, roughly equal to the relative variation
of the parameter) and a significant softening of the neutrino spectrum.
Lower neutrino mean energies and luminosities translate into a less efficient neutrino
heating. Nevertheless the consequent reduced energy deposition behind the shock 
is partially compensated by the slower contraction of the PNS and of the shock radius, 
which is expected in the case of reduced neutrino luminosities. 
A decrease of $\alpha_{\rm diff}$ (AD\_2) provides parallel, but opposite effects.

The variations of the parameter $\tau_{\rm cut}$ have the smallest impact on the ASL scheme 
results. In particular, a large increase of $\tau_{\rm cut}$ from 20 to 54 (TC\_54) provides 
larger neutrino mean energies, but it also increases the emission rates associated with
high energy neutrinos, Equation~(\ref{eqn: r_eff last version}). 
The more intense energy emission coming
from the optically thick region is still not enough to modify significantly the evolution of the PNS radius. 
However, it almost compensates the more efficient absorption provided in optically thin conditions, 
and limits the differences in the radial profiles and in the shock conditions. 
Analogous, but opposite, considerations apply to the TC\_7 run.

Our brief parametric study has shown that even significant variations of the free parameters
of the model (of the order of $\pm 30 \%$) around the calibrated values do not change qualitatively
the results of the simulations for the tested 15~$\msun$ progenitor. On the other hand,
quantitative differences are present: variations of the parameters $\alpha_{\rm blk}$ and 
$\alpha_{\rm diff}$ have the largest impact, since they modify significantly the neutrino luminosities
and mean energies. The diffusion parameter regulates also the contraction rate of the PNS, while
it has a less pronounced effect on the neutrino mean energies, compared with $\alpha_{\rm blk}$.
Variations of $\tau_{\rm cut}$ affect mainly the neutrino mean energies, but have a reduced impact on
the overall dynamics.

\subsection{12~$\msun$ and 40~$\msun$ progenitors}

\begin{figure*}
\centering
\includegraphics[width = 0.23 \linewidth,angle=-90]{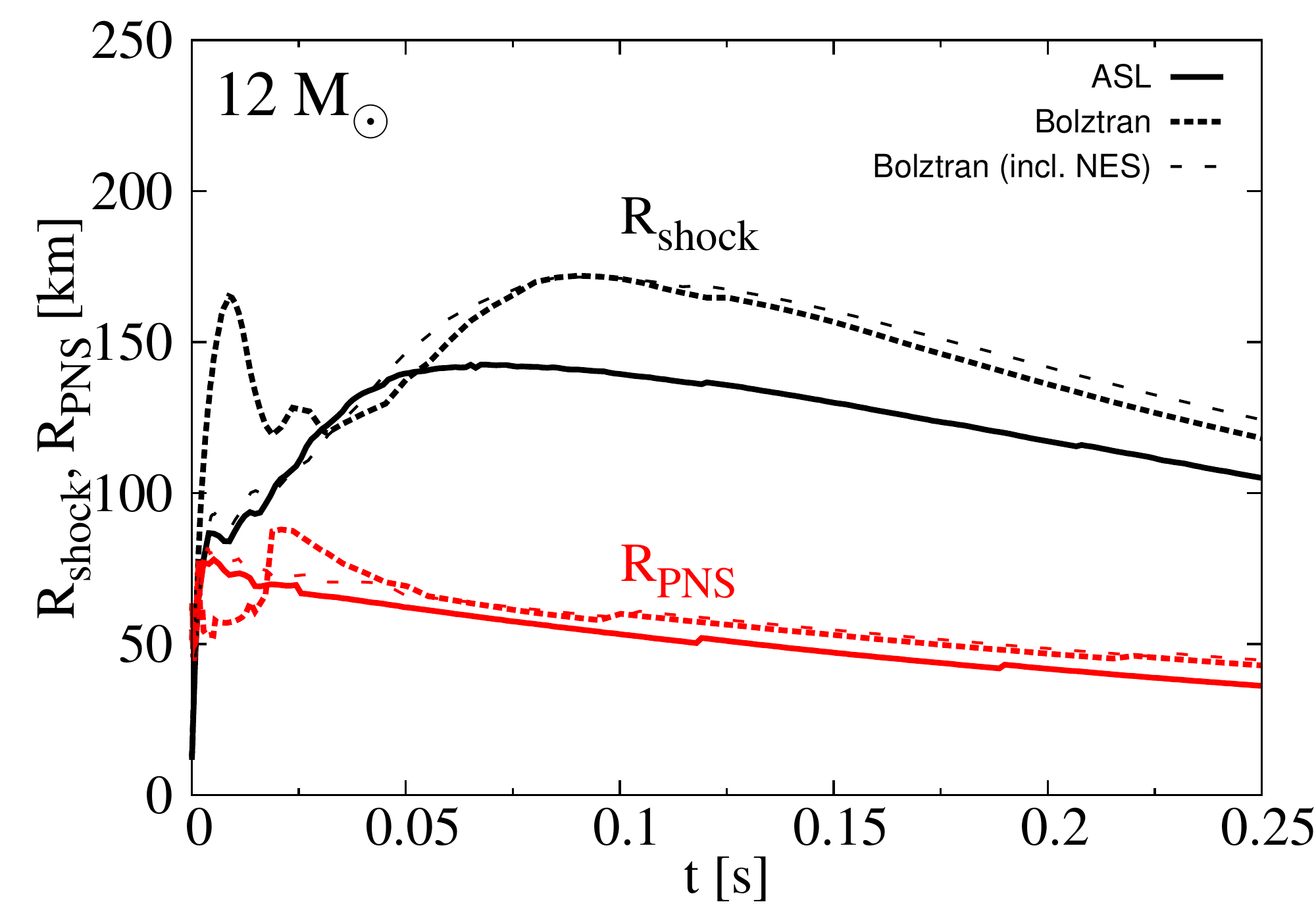}
\includegraphics[width = 0.23 \linewidth,angle=-90]{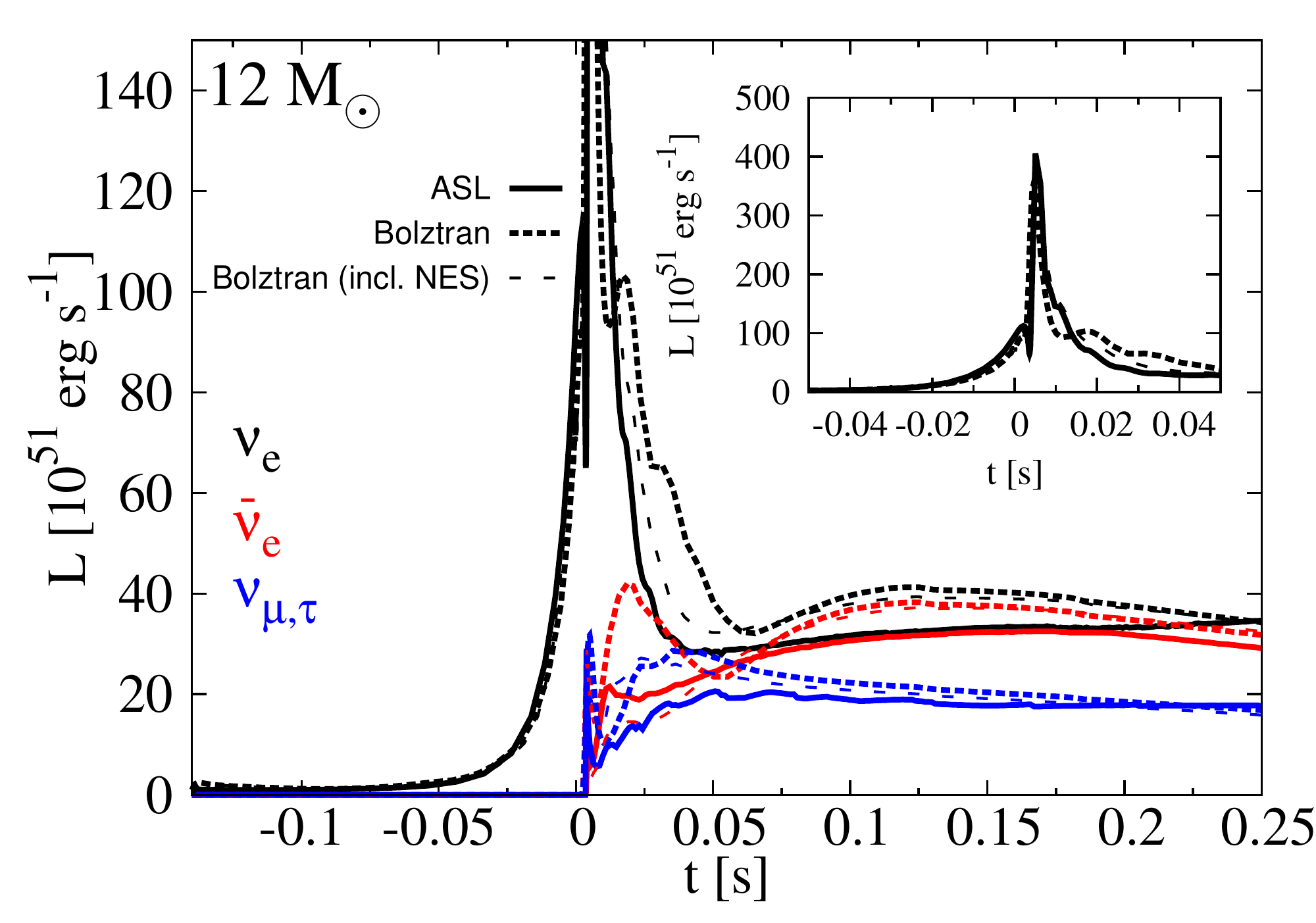}
\includegraphics[width = 0.23 \linewidth,angle=-90]{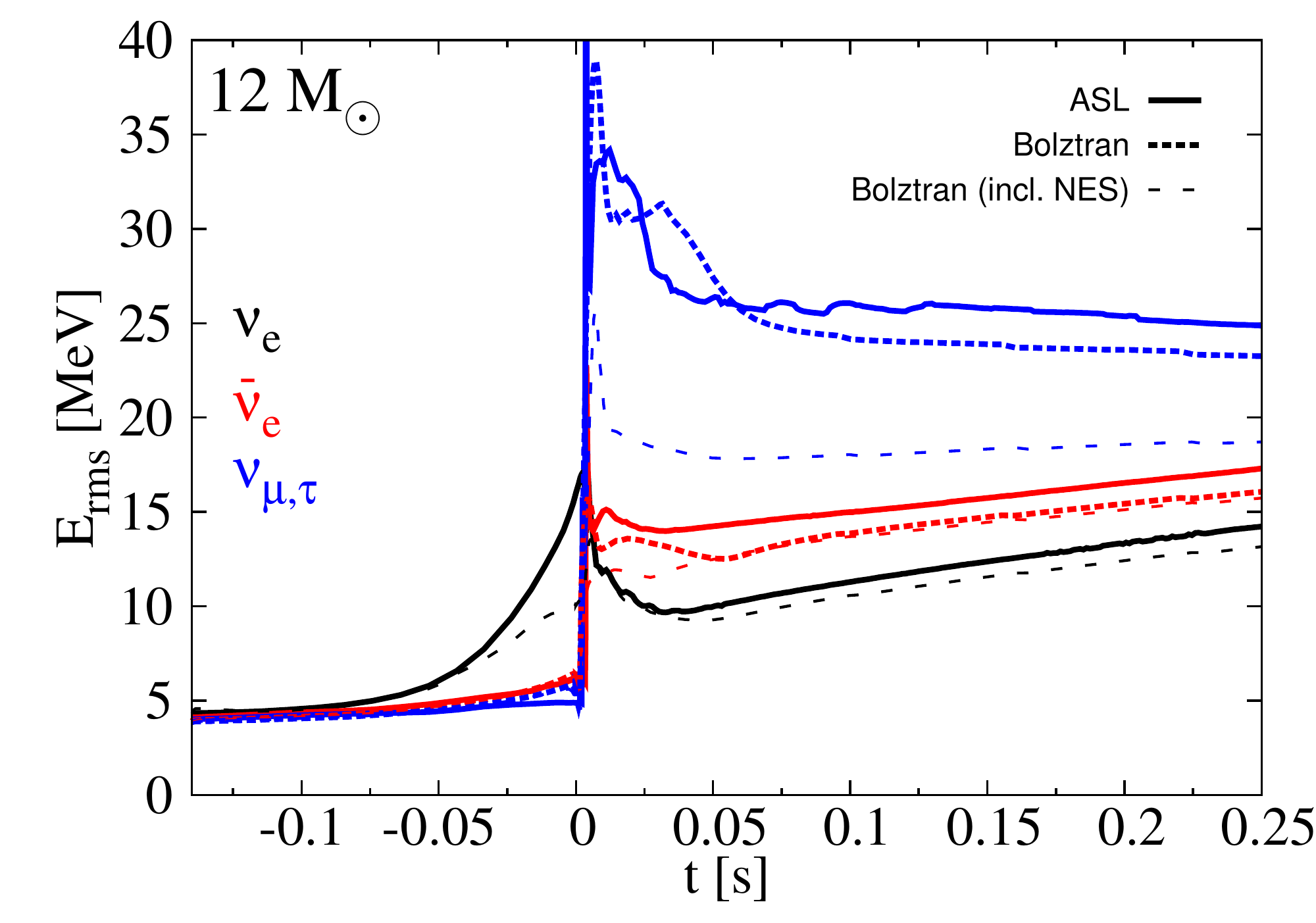}\\
\includegraphics[width = 0.23 \linewidth,angle=-90]{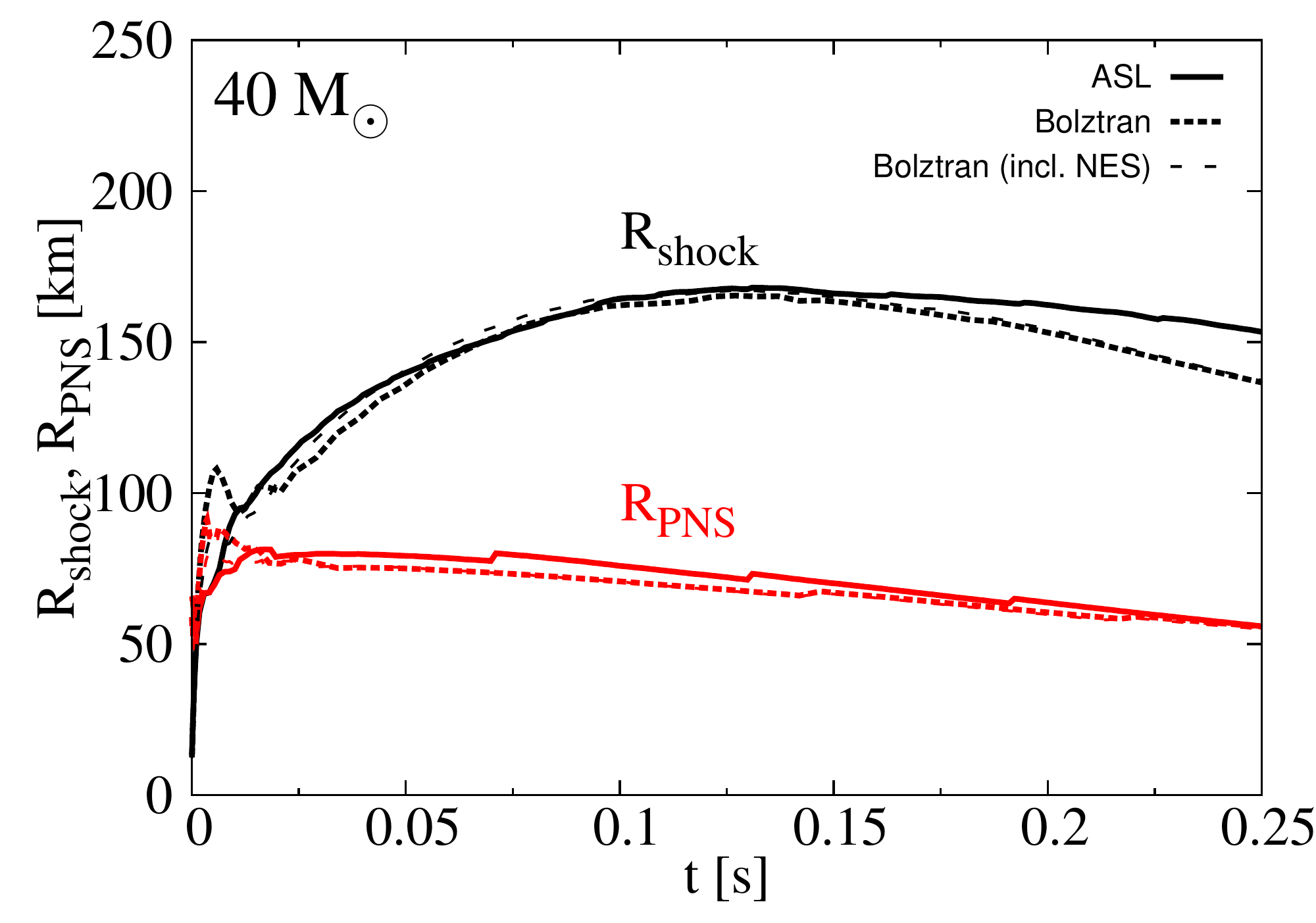}
\includegraphics[width = 0.23 \linewidth,angle=-90]{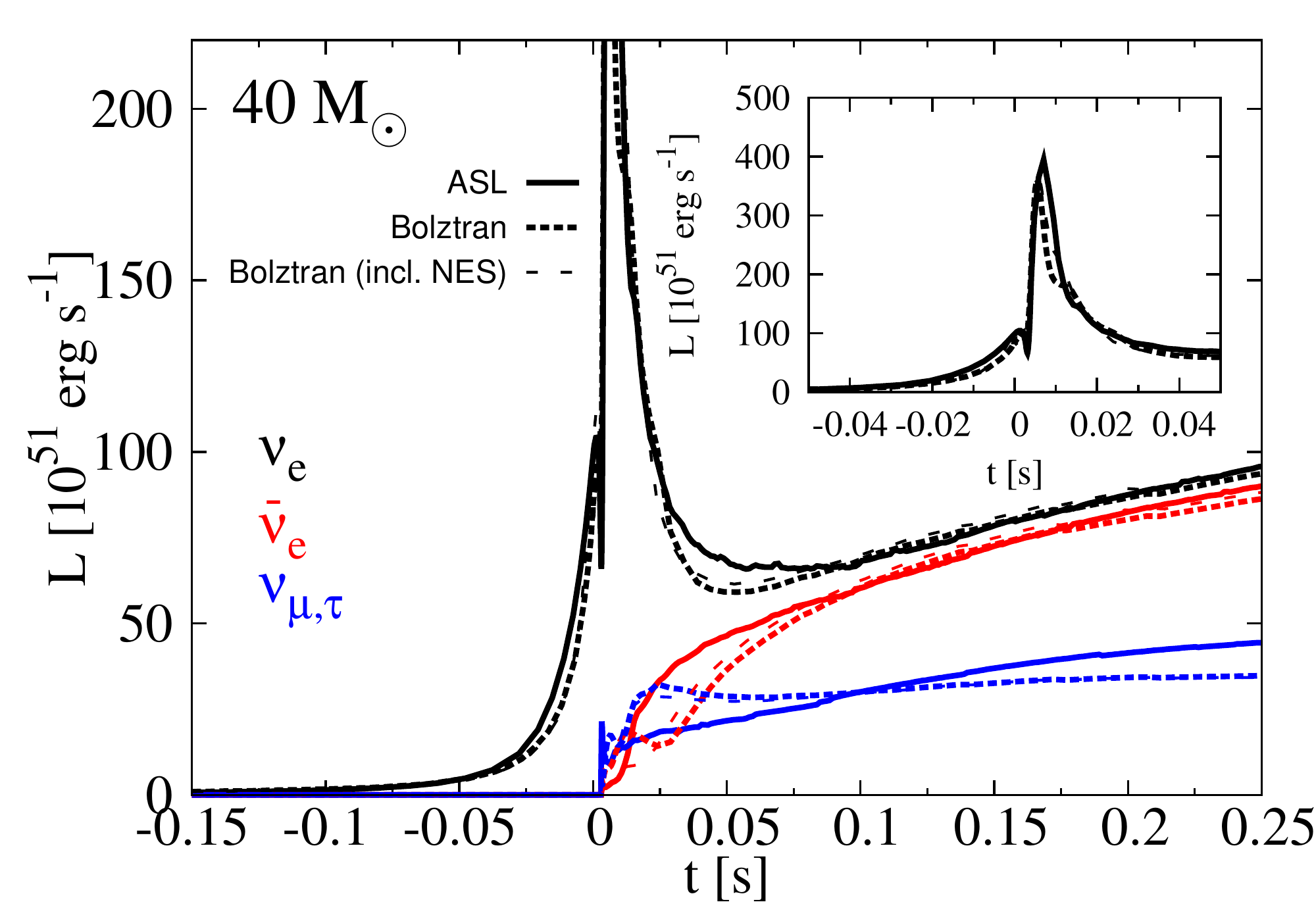}
\includegraphics[width = 0.23 \linewidth,angle=-90]{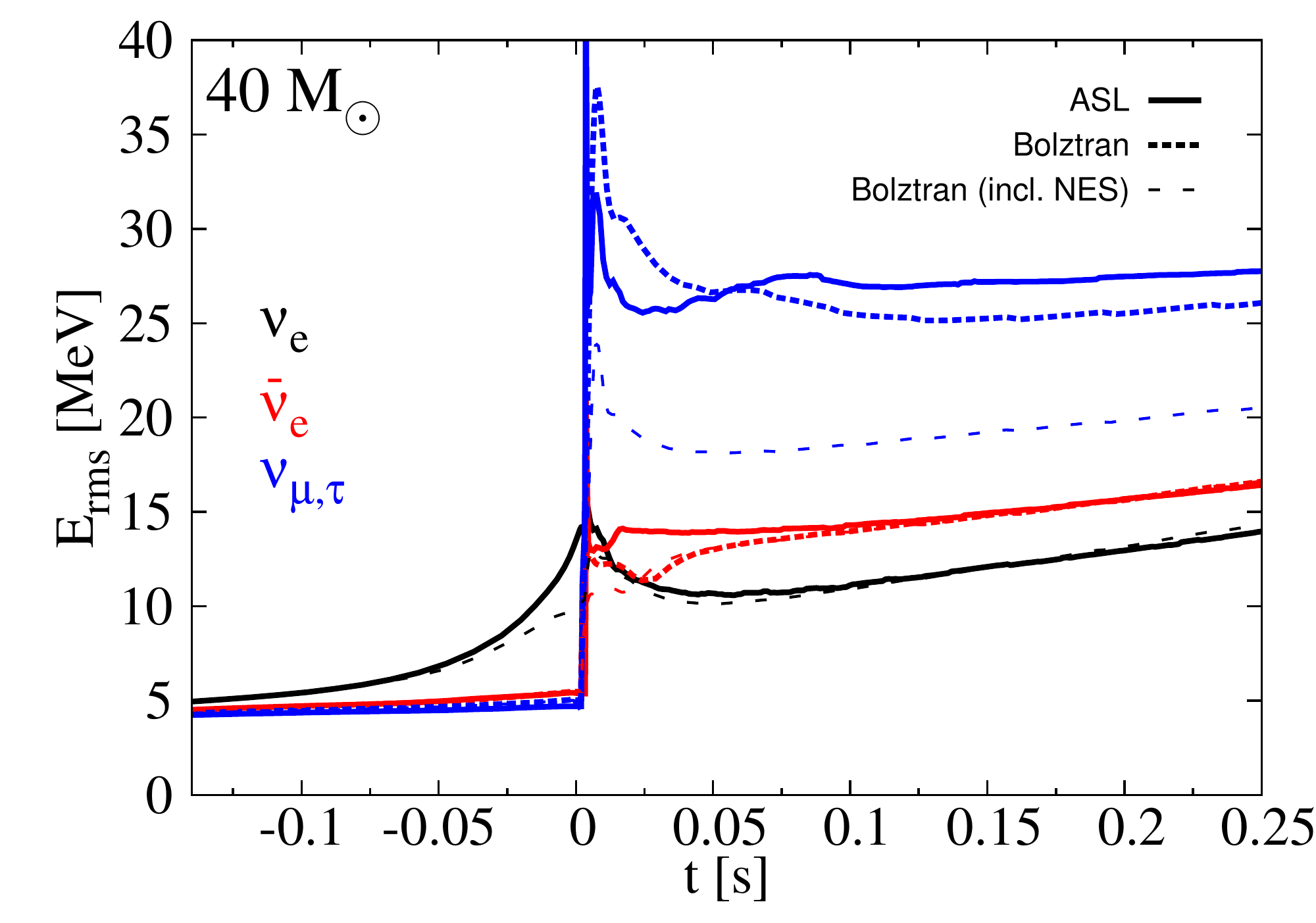}
\caption{Same as in Figure~\ref{fig4}, but for 12~$\msun$ (top panels) and 40~$\msun$ (bottom panels) progenitors.}
\label{fig6}
\end{figure*}

After having presented the calibration of the ASL free parameters for the core-collapse 
of a 15~$\msun$ ZAMS star, we test them with two different progenitors: 
12~$\msun$ \citep{Woosley2002} and 40~$\msun$ \citep{Woosley2007} ZAMS stars.
Also for these cases, we compare the results obtained in Newtonian simulations performed
with \aasl with the ones of \aboltz.
For the 12~$\msun$ case, we include 1.67~$\msun$ from the initial progenitor, distributed over 103 radial zones.
The initial radius extends up to $6800 \, {\rm km}$. For the 40~$\msun$ case, we include 2.60~$\msun$ from the 
initial progenitor, distributed over 135 radial zones, ranging initially from 0 up to $5100 \, {\rm km}$ from the center.

The comparison of the collapse profiles between the \aasl and the \aboltz results
shows a very good agreement, for example in the evolution of the central entropy and electron fraction, 
similar to the one we have observed for the 15~$\msun$ calibration model. 
The time necessary to reach core bounce is again larger by 15-20\% in the ASL case: 169 ms and 354 ms, to be compared 
with 142 ms and 313 ms obtained in the detailed neutrino transport run, for the 12~$\msun$ 
and 40~$\msun$ case, respectively.
The enclosed mass where the shock forms is larger in the ASL models by only a few percents (2-3 \%).
In Figure~\ref{fig6}, we present the evolution of some key quantities during 250 ms after core bounce, for both
the 12~$\msun$ (upper panels) and the 40~$\msun$ (lower panels) progenitor models.
For comparison purposes, we plot also the results obtained by the \aboltz code.
Overall, the results obtained with the ASL model reproduce qualitatively, and partially quantitatively, the 
results obtained by the detailed Boltzmann neutrino transport. The agreement is better for the 40~$\msun$ case.
This is due to the fact that, despite the large difference in the ZAMS mass, the core properties of the 15~$\msun$
progenitor show similarities with more massive progenitors and differences with lighter progenitors.
In the 12~$\msun$ case, the results obtained by the ASL scheme
look more pessimistic, due to a faster and more intense energy loss above the neutrinospheres during the 
first tens of milliseconds after core bounce. The lower $\nue$ and $\nueb$ luminosities, and the larger RMS energies, 
observed for $t \gtrsim 0.75 \, {\rm ms}$, are a consequence of the more compact PNS and shock.
Also for these two progenitor models (and especially for the 12~$\msun$ one), the $\nue$ luminosity and 
shock radius evolutions within the first tens of milliseconds after core bounce follow more closely 
the results obtained with \boltz, once the neutrino scattering on electrons and positrons 
has been included. The reasons are analog to the 15~$\msun$ case.
According to the analysis reported in Section~\ref{sec: parameter variations}, the more pessimistic 
results obtained for the 12~$\msun$ case can also indicate that light progenitors
would require a slightly different parameter choice to better match results from the reference model 
(in particular, larger $\alpha_{\rm diff}$ and $\alpha_{\rm blk}$).

\section{Examples in multidimensional simulations}
\label{sec: multiD models}

In Section \ref{sec: tests and validity}, we have compared results from the ASL scheme against
Boltzmann transport in spherically symmetric models. To do that, we used the same hydrodynamics code, \agile.
In order to show the possibility for the scheme to be implemented in multidimensional contexts, 
we report the following two tests, performed with two different hydrodynamical codes. 
In the first one, we apply the ASL scheme in a multidimensional setting by coupling it
to an axisymmetric Eulerian and non-relativistic hydrodynamics solver, to model the core-collapse of
a stellar iron core.
In the second one, we couple our algorithm with a Lagrangian hydrodynamics code,
and we simulate the same stellar core-collapse in 3D using smoothed particle hydrodynamics (SPH).
In both cases we consider a 15~$\msun$ progenitor from \citet{Woosley2002}.
We use the \cite{Lattimer1991} EoS with nuclear compressibility
$K = 220$ MeV and the ASL standard parameter set, as described in Section
\ref{sec: tests and validity}, with 20 geometrically increasing energy bins
between 3 MeV and 300 MeV.

The different dimensions, implementations and numerical techniques are expected to introduce
differences among the multidimensional tests and compared with 1D results (see Section~\ref{sec: tests and validity}). 
Nevertheless, during the collapse phase and in the first tens of milliseconds after core bounce, 
the profiles and the shock shape are expected to behave similarly to spherically symmetric models
\citep[e.g.,][]{Marek.Janka:2009,Mueller.etal:2012b,Bruenn.etal:2013}, even if deviations due to PNS and prompt 
convection can appear \citep[e.g.,][]{Buras.etal:2006a,Mueller.etal:2012b}. 
At later times, multidimensional effects change significantly the dynamics of the system. 
Several multidimensional CCSN results,
employing more sophisticated neutrino treatments, have been published. 
In the case of axisymmetric models of a 15~$\msun$ model, see e.g. 
\cite{Scheck.etal:2006,Buras.etal:2006b,Mueller.etal:2012,Suwa.etal:2013,Zhang.etal:2013,Bruenn.etal:2013,Bruenn.etal:2014,
Mueller.Janka:2014,Pan.etal:2015,Dolence.etal:2015}.
For three-dimensional SPH models, see e.g. \cite{Fryer.Warren:2004}. They provide reference cases 
to check the qualitative behavior of our simulations and of the neutrino quantities within
a few hundreds milliseconds after core bounce.
However, our focus is to demonstrate the versatility and portability 
of the ASL algorithm, along with the validation of the results it provides 
with 1D detailed simulations, but not to compare to other hydrodynamical
codes and among different dimensions.
% However, detailed quantitative analysis 
% and comparisons of our results go beyond the scope of this paper.

\subsection{2D grid-CCSN model}
\label{subsec:2D_CCSN}

\begin{figure*}
\begin{minipage}{0.33 \linewidth}
\centering
\includegraphics[width = 0.7 \linewidth,angle=-90]{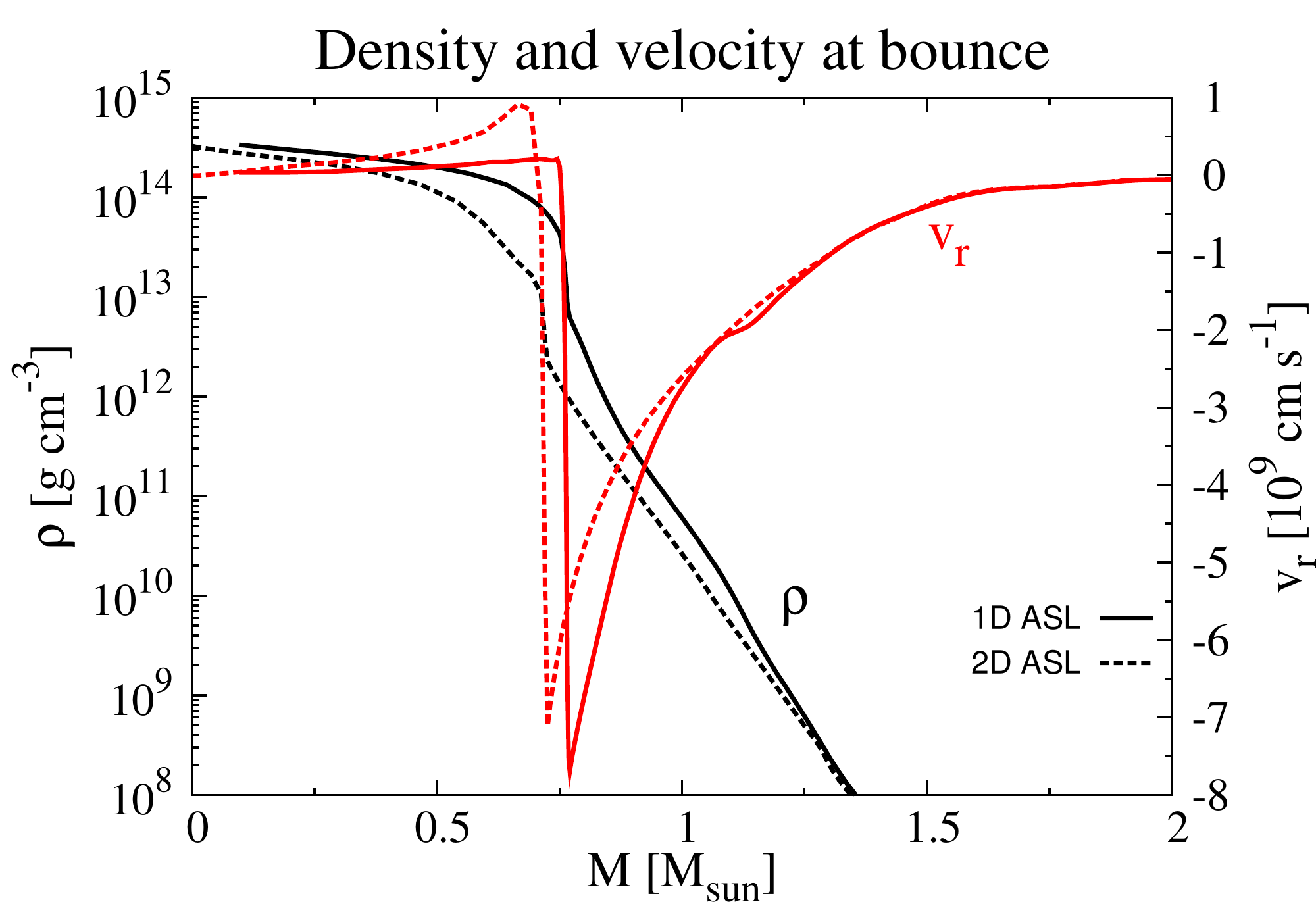}
\end{minipage}
\begin{minipage}{0.33 \linewidth}
\centering
\includegraphics[width = 0.7 \linewidth,angle=-90]{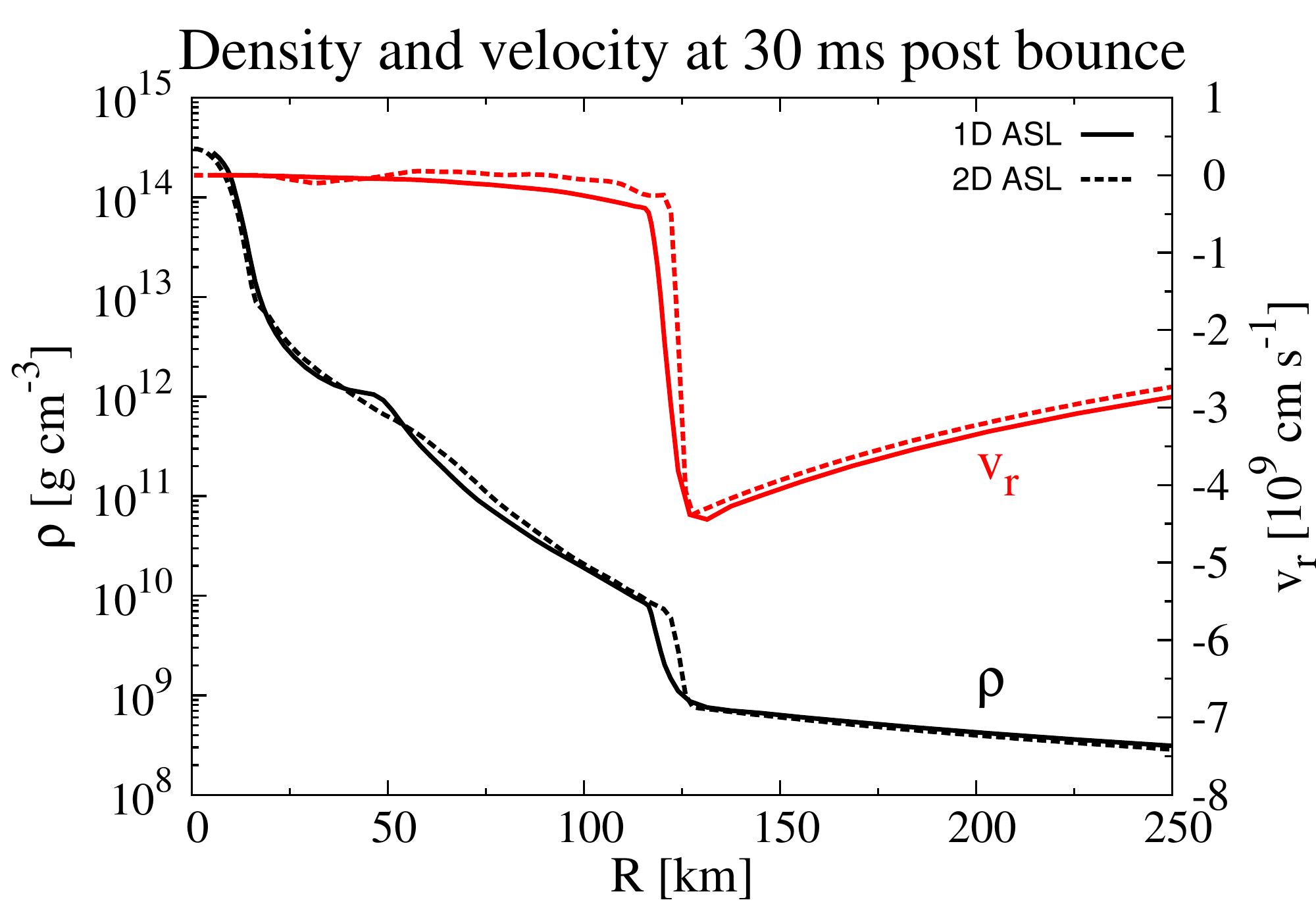}
\end{minipage}
\begin{minipage}{0.33 \linewidth}
\centering
\includegraphics[width = 0.7 \linewidth,angle=-90]{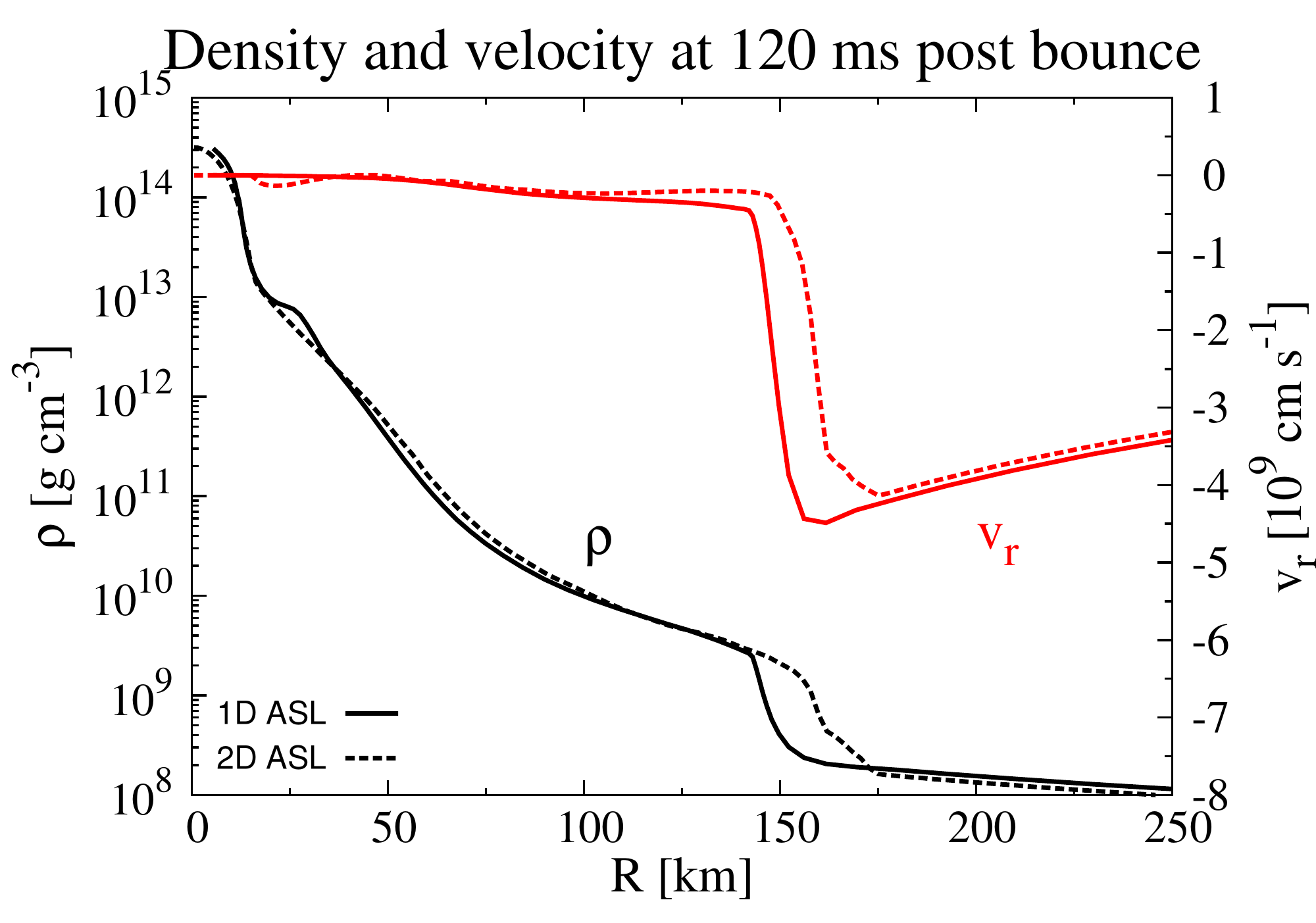}
\end{minipage}\\
\begin{minipage}{0.33 \linewidth}
\centering
\includegraphics[width = 0.7 \linewidth,angle=-90]{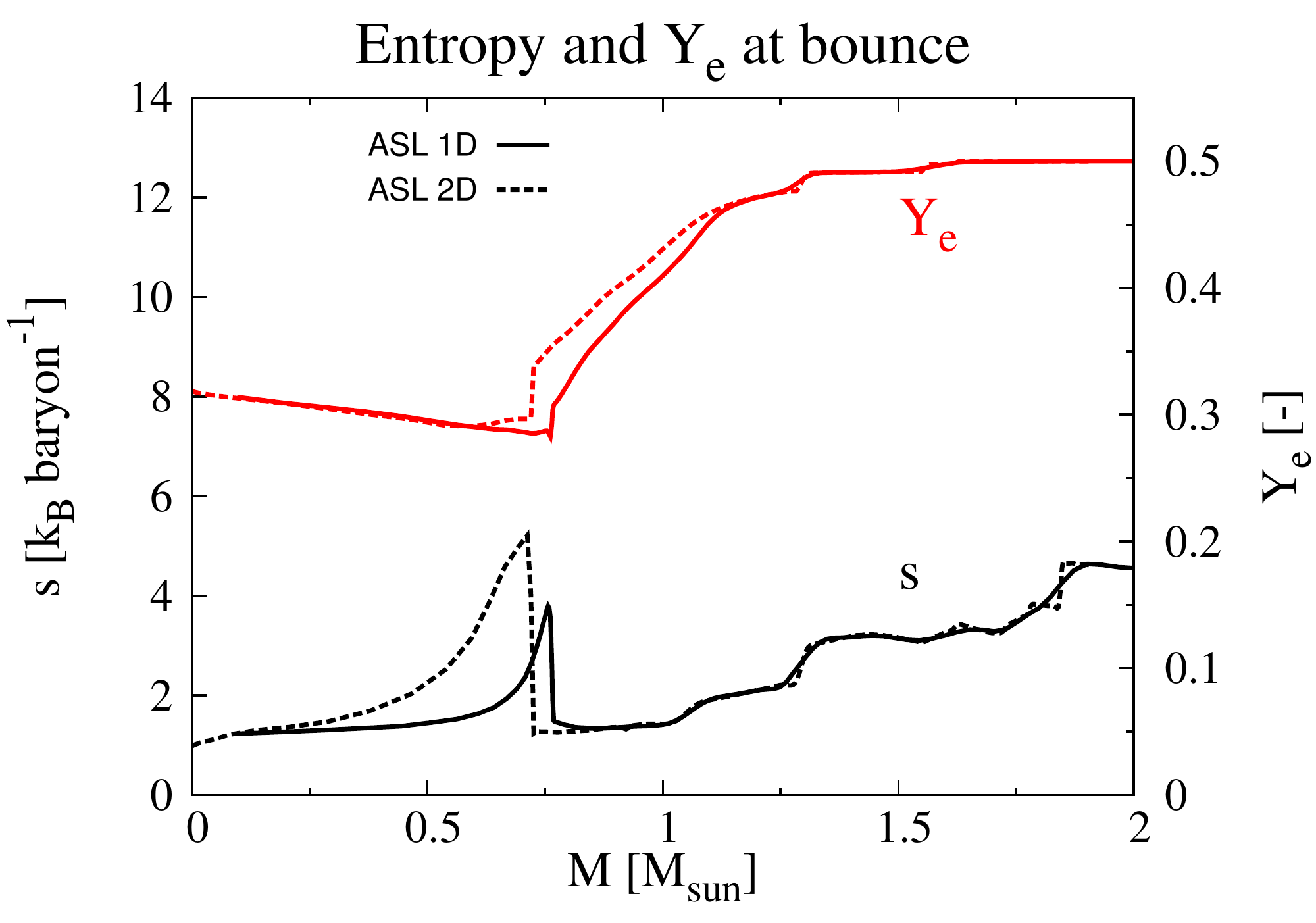}
\end{minipage}
\begin{minipage}{0.33 \linewidth}
\centering
\includegraphics[width = 0.7 \linewidth,angle=-90]{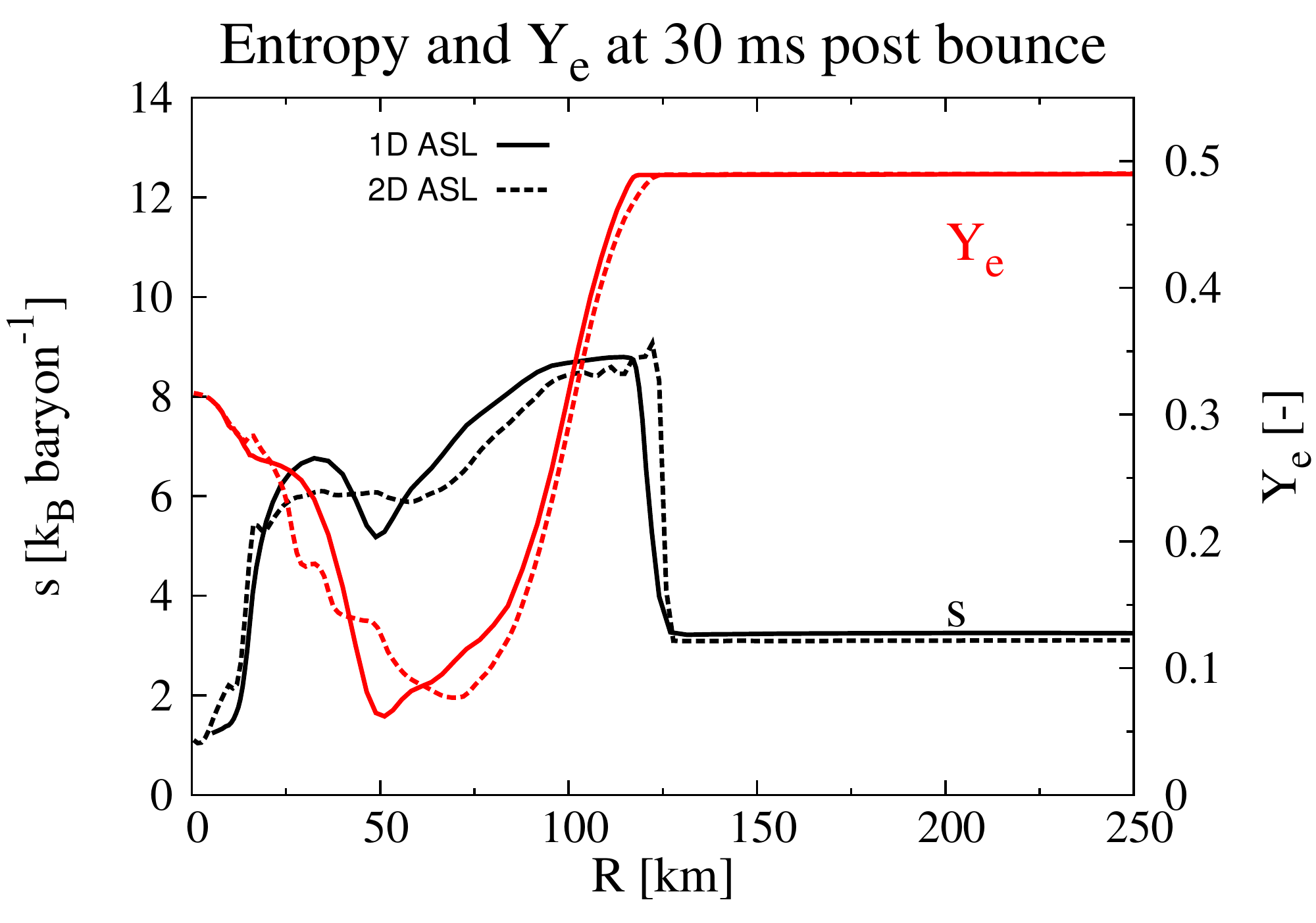}
\end{minipage}
\begin{minipage}{0.33 \linewidth}
\begin{center}
\includegraphics[width = 0.7 \linewidth,angle=-90]{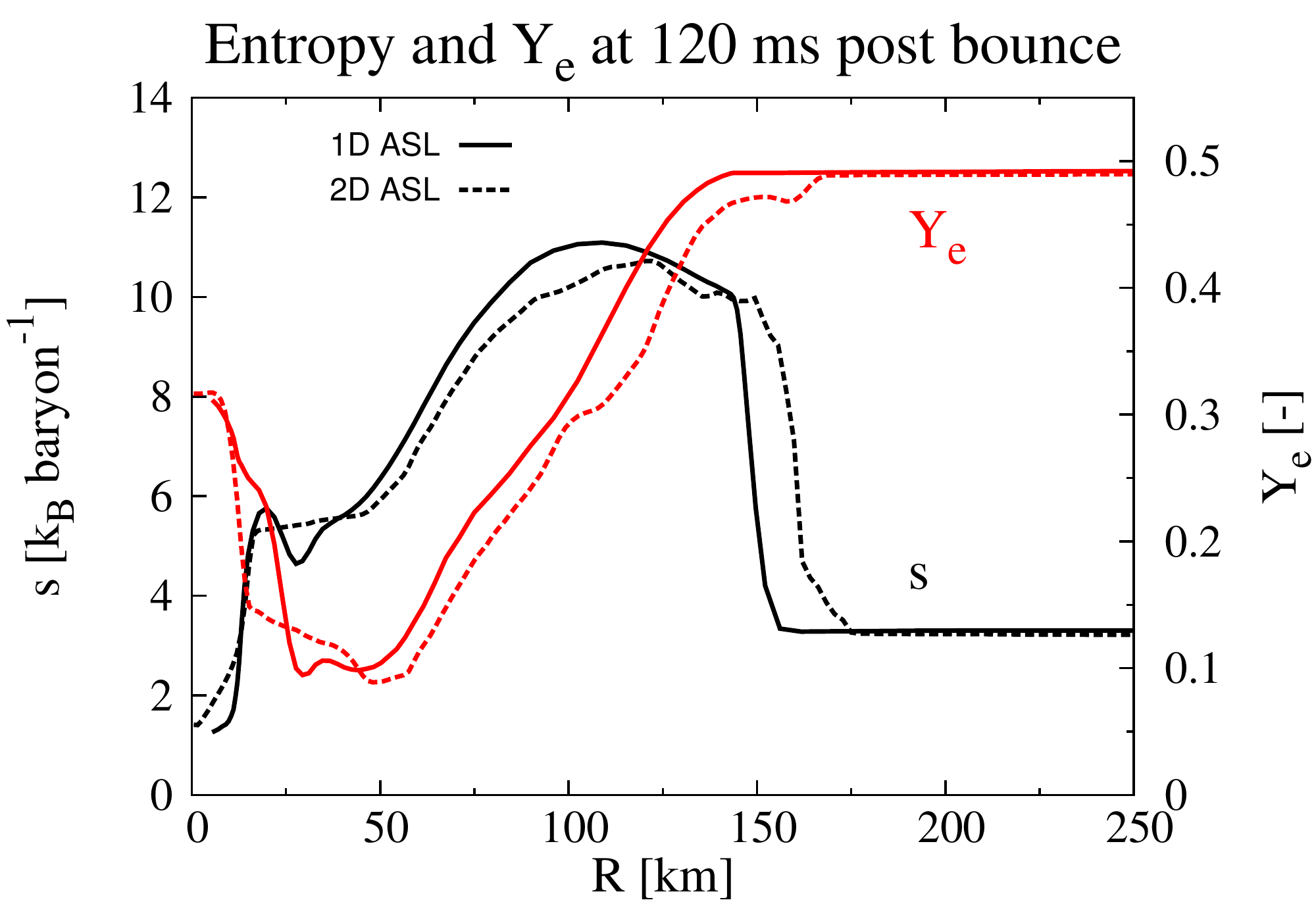}
\end{center}
\end{minipage}
\caption{Profiles of the density and radial velocity (top row), and of
         the entropy and electron fraction (bottom row) for the 1D ASL
         reference (solid lines) and the spherically averaged 2D ASL model
         (dashed lines).
         The left, middle and right panels refer to 0, 30 and 120 ms after bounce,
         respectively.}
\label{fig7}         
\end{figure*}

\begin{figure*}
 \begin{minipage}{0.33 \linewidth}
  \includegraphics[width= \linewidth]{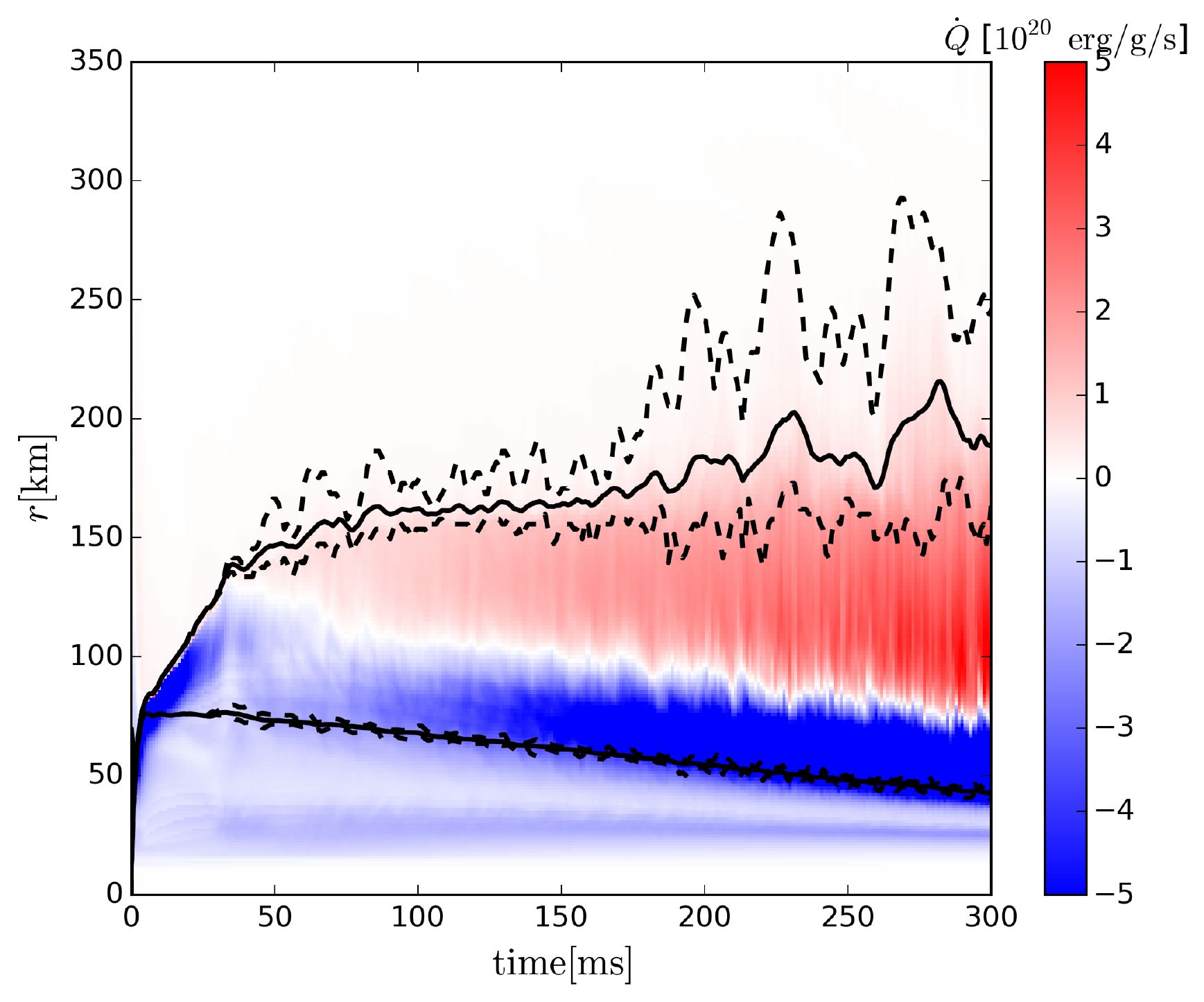}
 \end{minipage}
 \begin{minipage}{0.33 \linewidth}
  \includegraphics[width=0.7  \linewidth,angle=-90]{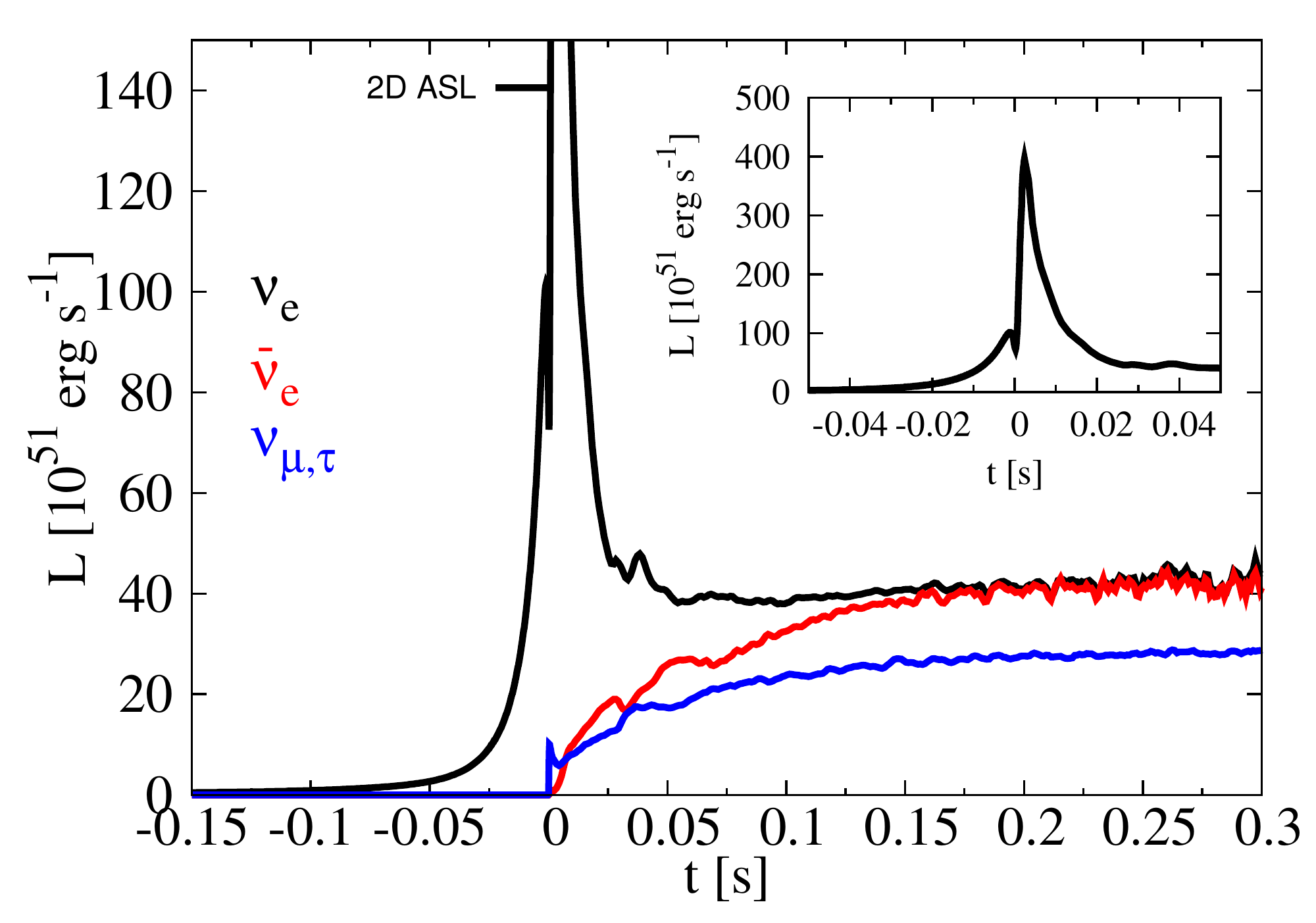}
 \end{minipage}
 \begin{minipage}{0.33 \linewidth}
  \includegraphics[width=0.7  \linewidth,angle=-90]{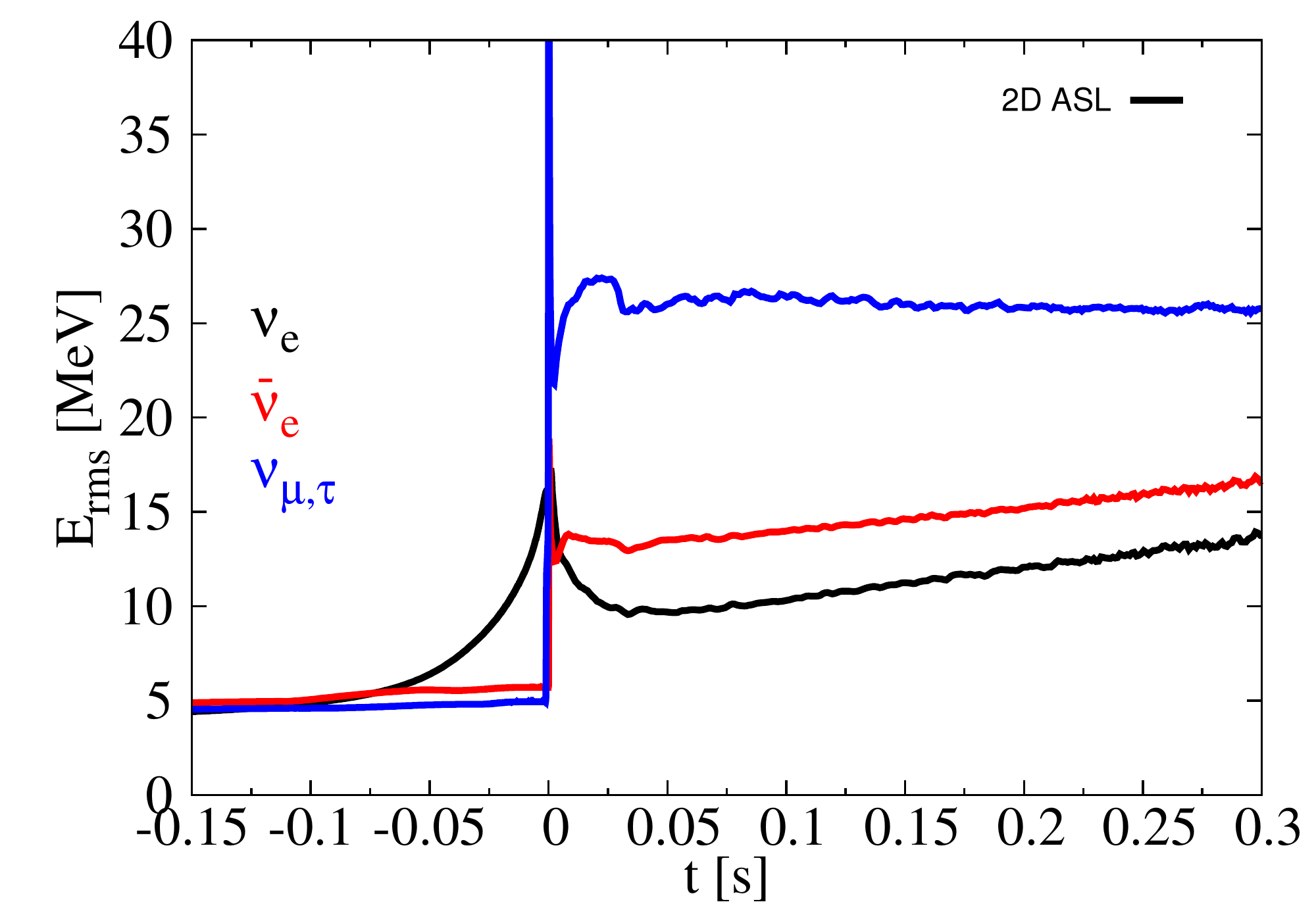}
 \end{minipage}  
  \caption{Left: evolution of the PNS and shock radii for the 2D CCSN model of a 15~$\msun$ progenitor.
           The solid lines represent the average radius while the dashed
           lines represent the minimal and maximal radii, respectively.
           Furthermore, color coded is the net energy deposition by the
           electron flavor neutrinos.
           Central: temporal evolution of the neutrino luminosities for the 2D CCSN model. 
           The black, red and blue lines represent electron neutrino, electron 
           antineutrino and heavy neutrinos, respectively.
           Right: same as in the central panel, but for the neutrino RMS energies.}
  \label{fig8}
\end{figure*}

\begin{figure*}
  \begin{minipage}{0.48\linewidth}
    \centering
    \includegraphics[width=\linewidth]{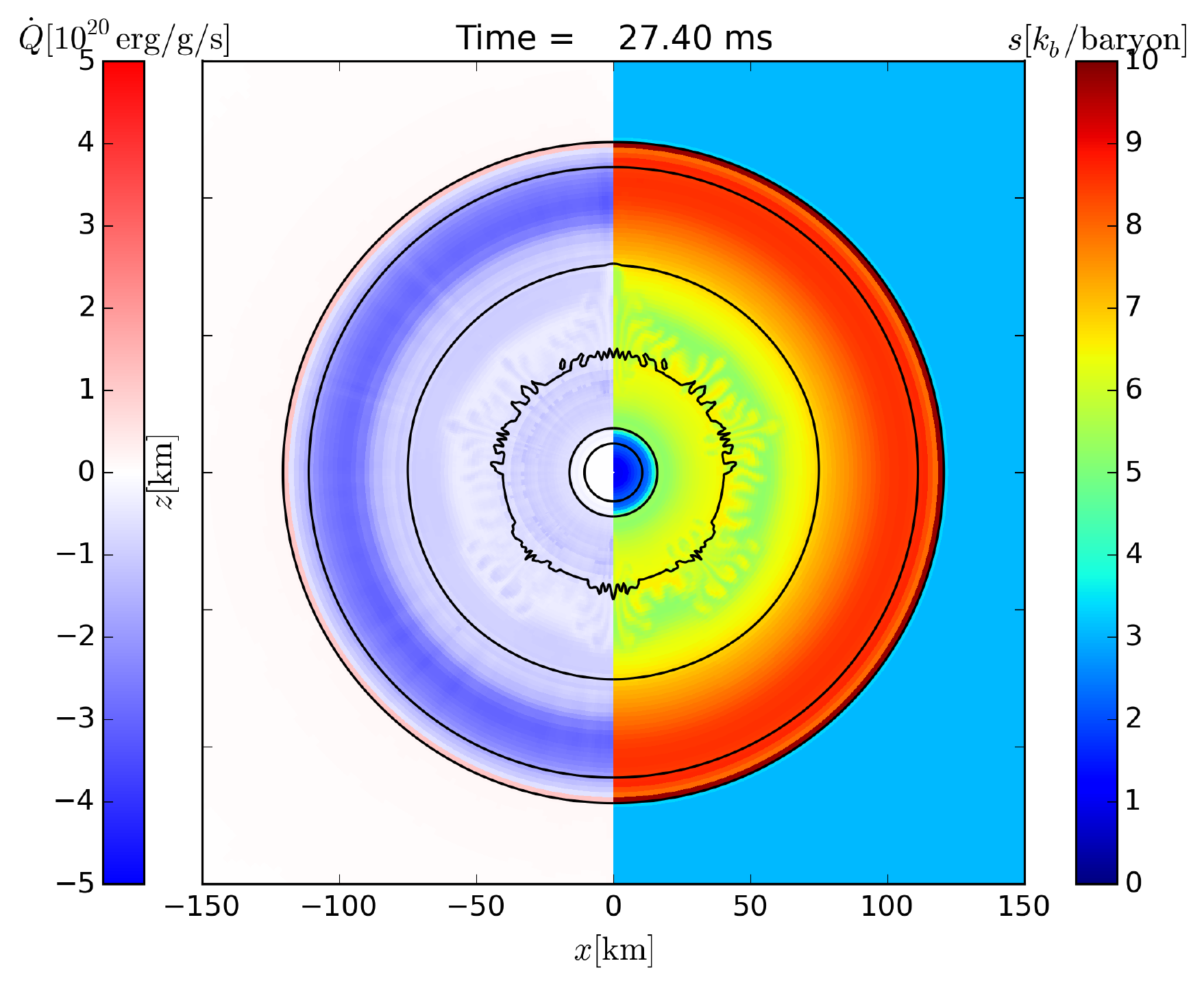}
  \end{minipage}
  \begin{minipage}{0.48\linewidth}
    \centering
    \includegraphics[width=\linewidth]{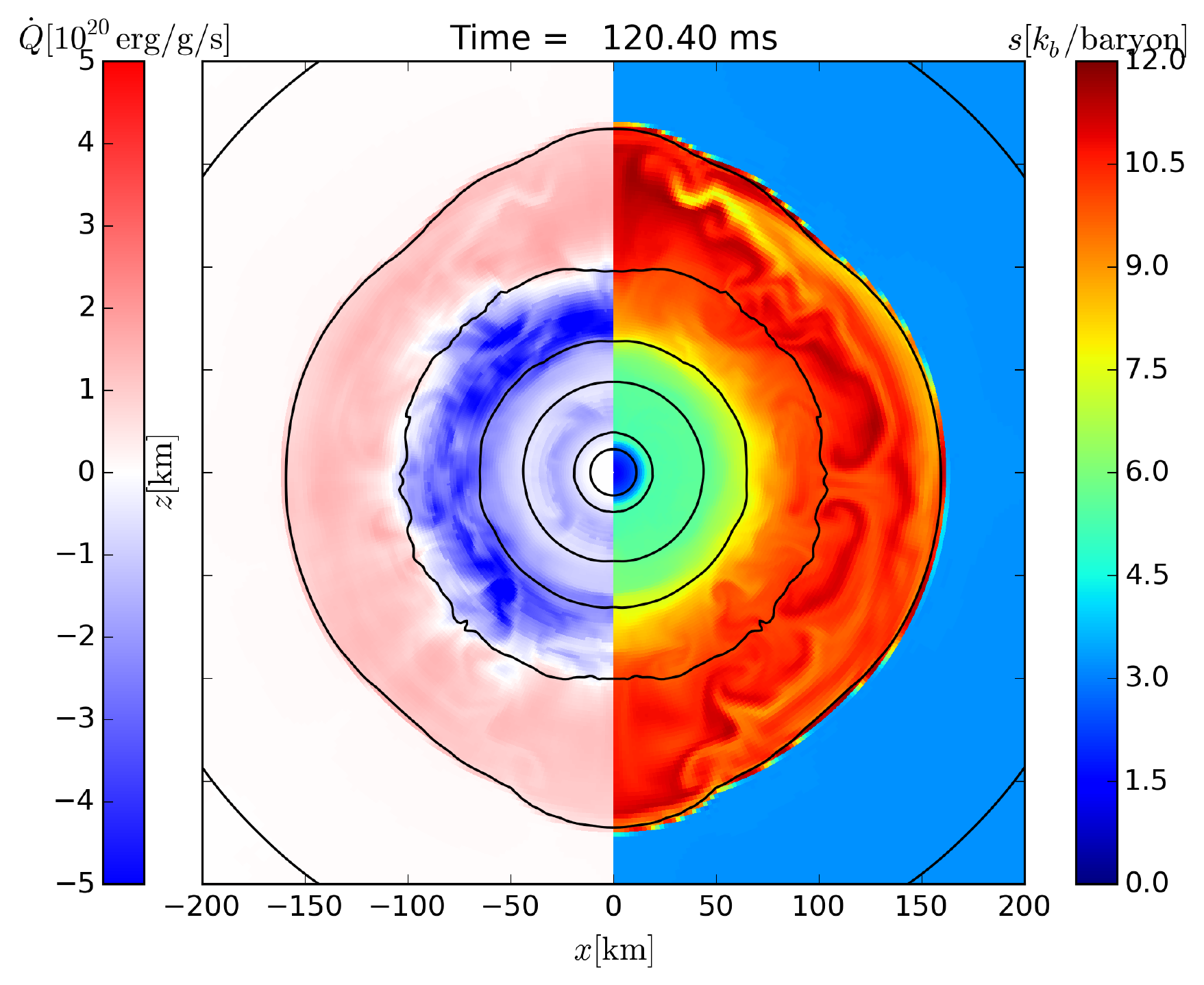}
  \end{minipage}\\
  \begin{minipage}{0.48\linewidth}
    \centering
    \includegraphics[width=\linewidth]{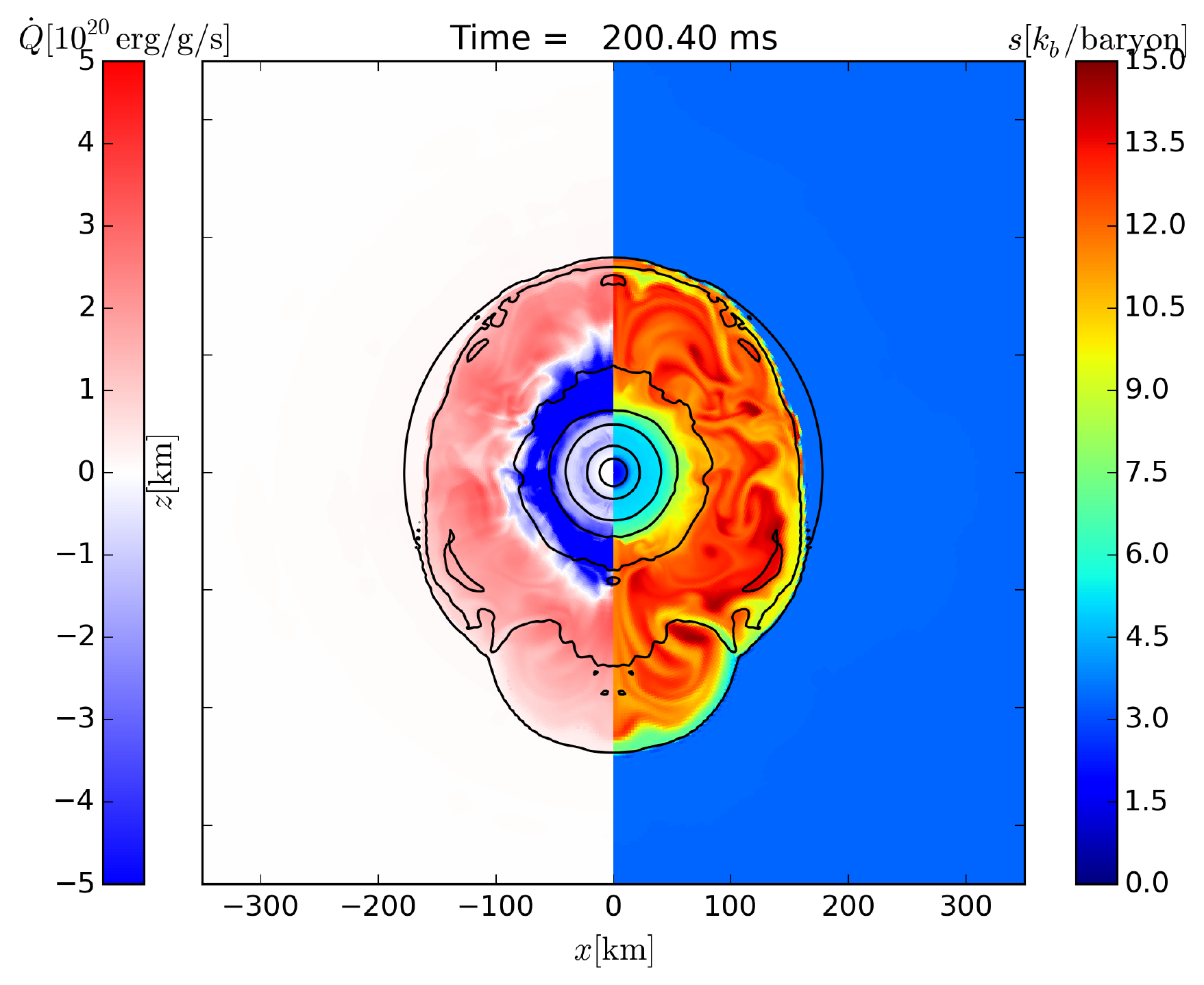}
  \end{minipage}
  \begin{minipage}{0.48\linewidth}
    \centering
    \includegraphics[width=\linewidth]{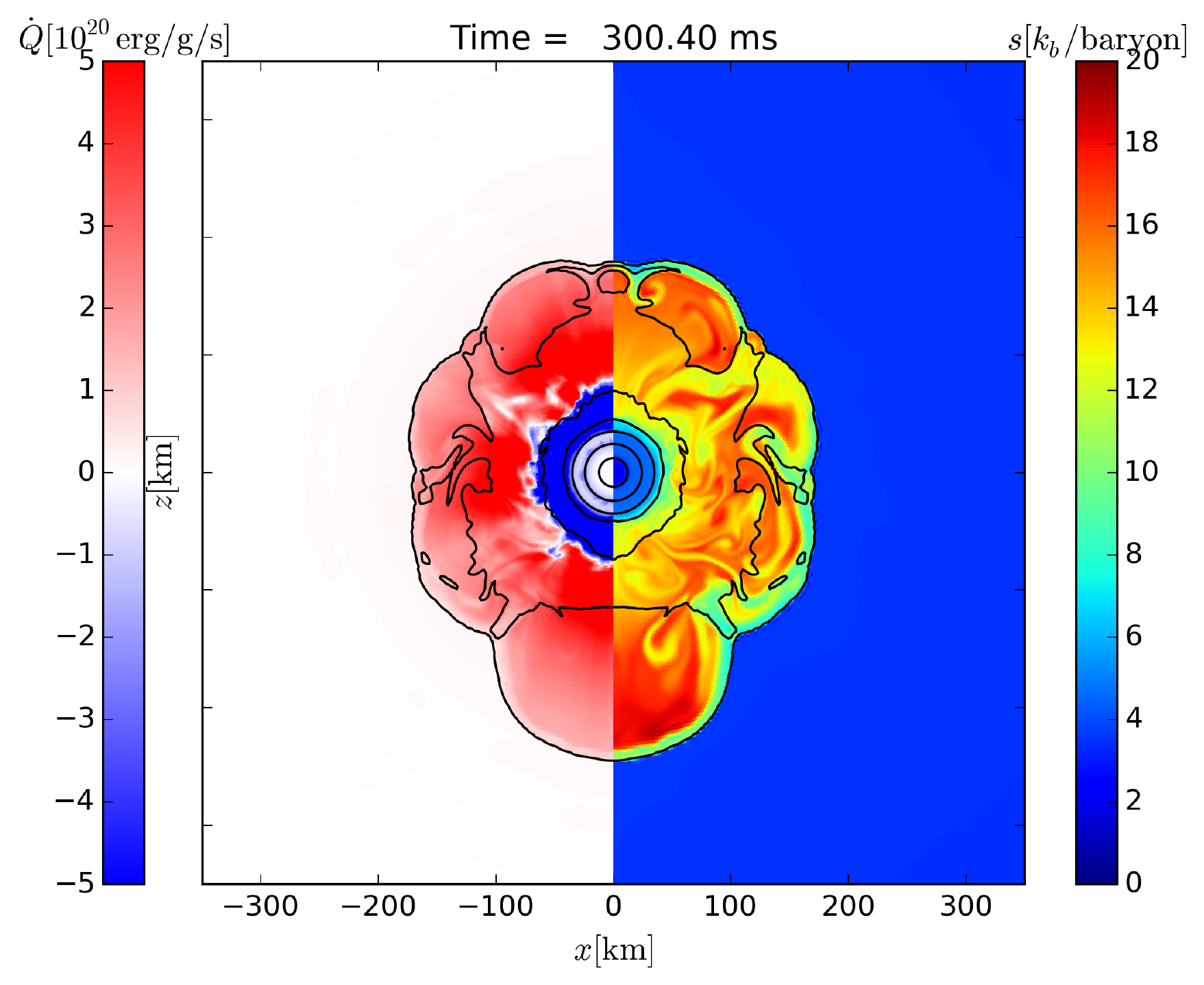}
  \end{minipage}
  \caption{Snapshots of the post-bounce evolution of the 2D CCSN model at
           different times (27 ms, 120 ms, 200 ms and 300 ms).
           In each snapshot the data is mirrored along the symmetry axis
           displaying on the left the net neutrino heating rate and on the
           right the specific entropy.
           }
  \label{fig9}
\end{figure*}

The equations of hydrodynamics in spherical coordinates and azimuthal
symmetry are evolved with a directionally unsplit finite-volume scheme.
The scheme is of Godunov type using a second-order in space well-balanced
reconstruction \citep{Kaeppeli.Mishra:2014,Kaeppeli.Mishra:2014b}
with characteristic limiting, a HLLC
approximate Riemann solver \citep{Toro1997} and a second-order in time
Strong Stability Preserving Runge-Kutta (SSP-RK2) \citep{gottlieb:89} time
integration.
Spurious solutions near strong (grid aligned) shocks are avoided by the
use the H-correction method by \cite{Sanders.etal:1998}.

Newtonian self-gravity is approximated by a spatially second-order accurate
discretization of the monopole term (i.e., by spherically averaging the mass
density and integrating the resulting one-dimensional profile) and a
spatially second-order five-point discretization of the Poisson equation
for the deviation from the monopole term.

The ASL scheme is coupled to the hydrodynamics by a "ray-by-ray"
approach: apart from the trapped neutrino components ($Y_\nu, Z_\nu$), which
we evolve according the corresponding multidimensional advection equations 
(see Equations~(\ref{eqn:hydro equations})-(\ref{eqn: variables and fluxes})),
the ASL scheme is applied as described in Section \ref{sec: the new ASL} along each radial "ray".
However, in the present implementation we
have neglected the neutrino stress in the momentum equation.

The computational domain includes the innermost 5000 km and the full
[0,$\pi$] polar realm.
The radial direction is discretized by $N_r = 512$ logarithmically spaced
cells: $\Delta r_i = \Delta r_ 1 a^i$, $i = 1, ..., N_r$,
$a - 1 = 5.659 \times 10^{-3} $ and $\Delta r_ 1 = 1$ km.
The polar direction is uniformly discretized by $N_\theta = 256$ cells.
The progenitor is then mapped (without adding any rotation and perturbations)
onto the computational domain and evolved numerically through collapse,
bounce and until $300$ ms post-bounce.

The collapse proceeds without any noticeable deviations from spherical
symmetry until it is halted by parts of the inner core bouncing at
$t_{\mathrm{bounce}} = 222$ ms after the start of the simulation 
due to the stiffening of the EoS.
At that time the central density reaches
$\rho_{\mathrm{bounce}} = 3.29 \times 10^{14}$ g cm$^{-3}$.
The enclosed mass at the shock formation radius is
$M_{\rm enc,bounce} = 0.66~\msun$, which is slightly
lower than in the one-dimensional reference simulation.
In Figure~\ref{fig7}, we present a detailed comparison of the
profiles from the axisymmetric simulation with the one-dimensional reference.
At bounce (left panels), we observe that all the quantities are in good agreement.
The 2D simulation has a slightly more compact PNS, which we attribute
to the fact that we have neglected the neutrino stress.
At 30 ms after bounce (middle panels) the agreement is still very good up
to the negative entropy gradient, which was washed out by prompt convection.
At later times, the two-dimensional simulation deviates from the
one-dimensional reference due to multidimensional effects, e.g., convection
and shock instabilities.
This is illustrated in the right panels of Figure~\ref{fig7},
where we compare the profiles at 120 ms after bounce.

In the left panel of Figure~\ref{fig8}, we show the minimum, 
average and maximum shock and PNS radii.
In the same panel, the spherically averaged net heating rate by
electron flavor neutrinos is also shown.
The shock and PNS radii evolution can be separated in three distinct phases.
The first lasts from bounce up to $\approx 30$ ms after bounce.
This phase features the initial very strong acceleration of the shock wave.
When the shock passes the neutrino spheres, the neutronization burst induces
strong cooling as indicated by the negative heating rate.
During this phase the evolution is almost perfectly spherically symmetric.
The second phase starts at $\approx 30$ ms after bounce and lasts up to
$\approx 170$ ms.
During this phase the unstable entropy profile left behind by the shock wave
triggers strong convective motions, i.e., the so-called prompt-convection.
This induces anisotropic shock movements, which are visible in the minimum
and maximum shock radii.
The effect is also visible in the minimum and maximum PNS radii, but to a
much lesser extent.
However, the anisotropic shock movements remain mild, i.e., the difference
between minimum and maximum shock radius does not exceed $\approx 40$ km.
In the upper-left panel of Figure~\ref{fig9}, we show a
snapshot at $27$ ms post-bounce when prompt-convection just sets in.
During this second phase, one observes the appearance and progressive growth
of regions below the shock with net neutrino heating.
In the right-upper panel we show a snapshot at 120 ms post-bounce.
It can be seen, that there is an extended heating region below the shock.
At this time, convective motions due to neutrino energy deposition start
to set in.
The third and last phase starts $\approx 170$ ms and lasts until we stopped
the simulation.
During this phase, the energy deposited by the neutrinos triggers strong
convective motions.
Plumes rise by buoyancy against the continuous accretion flow.
This in turn triggers increasingly strong shock movements and
it is illustrated in the two lower panels of Figure~\ref{fig9}.

In the central and right panels of Figure~\ref{fig8}, we show the neutrino luminosities and
RMS energies from the axisymmetric simulation.
Both quantities agree well with the one-dimensional reference.
Moreover, the multidimensional one shows some oscillations in all the
quantities after $\approx$ 50 ms post bounce.
These oscillations become stronger after $\approx 150$ ms, especially in
the luminosities.
We attribute these differences again to multidimensional effects.

\subsection{3D SPH-CCSN models}

The calculation is performed using our SPH code, \texttt{SPHYNX}, that solves the 
Euler equations derived from a variational principle \citep[see, e.g.,][and references therein]{Rosswog2009}. 
\texttt{SPHYNX} uses high order interpolating kernels, namely the $sinc$ kernels with $n=5$ \citep{Cabezon2008, Garcia.etal:2014}, 
and an improved gradients evaluation based in the integral approach
IAD$_0$ \citep{Garcia2012,Cabezon2012,Rosswog:2015b}. 
Three dimensional gravity is calculated with a hierarchical tree structure created using the 
Barnes-Hut algorithm \citep{Hernquist1989}, and the neutrino treatment is handled with a three dimensional 
version of the ASL treatment presented in Section~\ref{sec: the new ASL}.

\subsubsection{SPH + ASL coupling}

\begin{figure*}
\includegraphics[width = \linewidth,angle=-90]{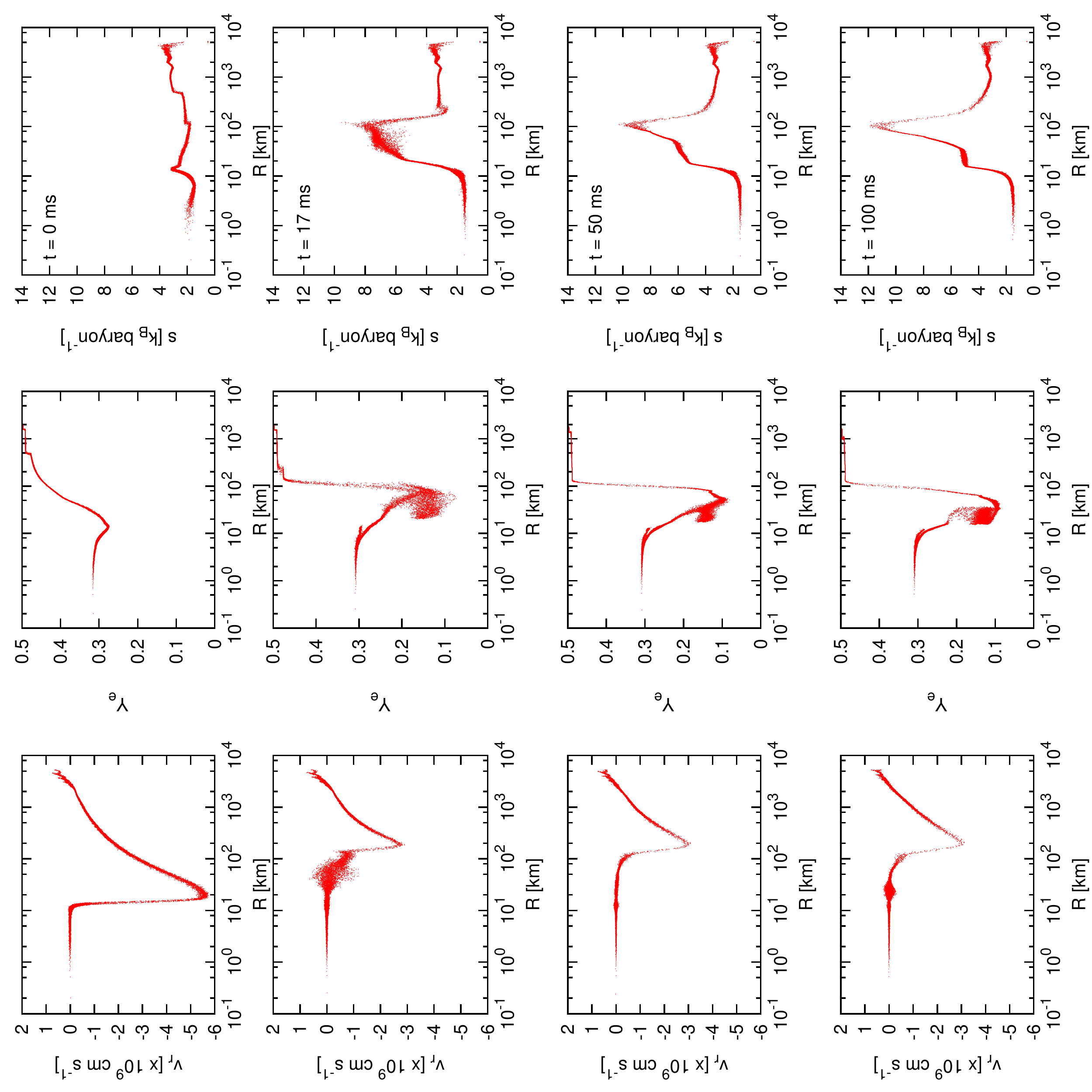}
\caption{Snapshots of the radial profiles of radial velocity (left), \ye (middle), and entropy (right) at 
different times (bounce, 17~ms, 50~ms and 100~ms, from the top to the bottom)
from our SPH model of a 15 $\msun$ CCSN. 
Each point represents one SPH particle, and only 1 every 10 are plotted.}
\label{fig10}
\end{figure*}

\begin{figure*}
\centering
\includegraphics[width = 0.85 \linewidth]{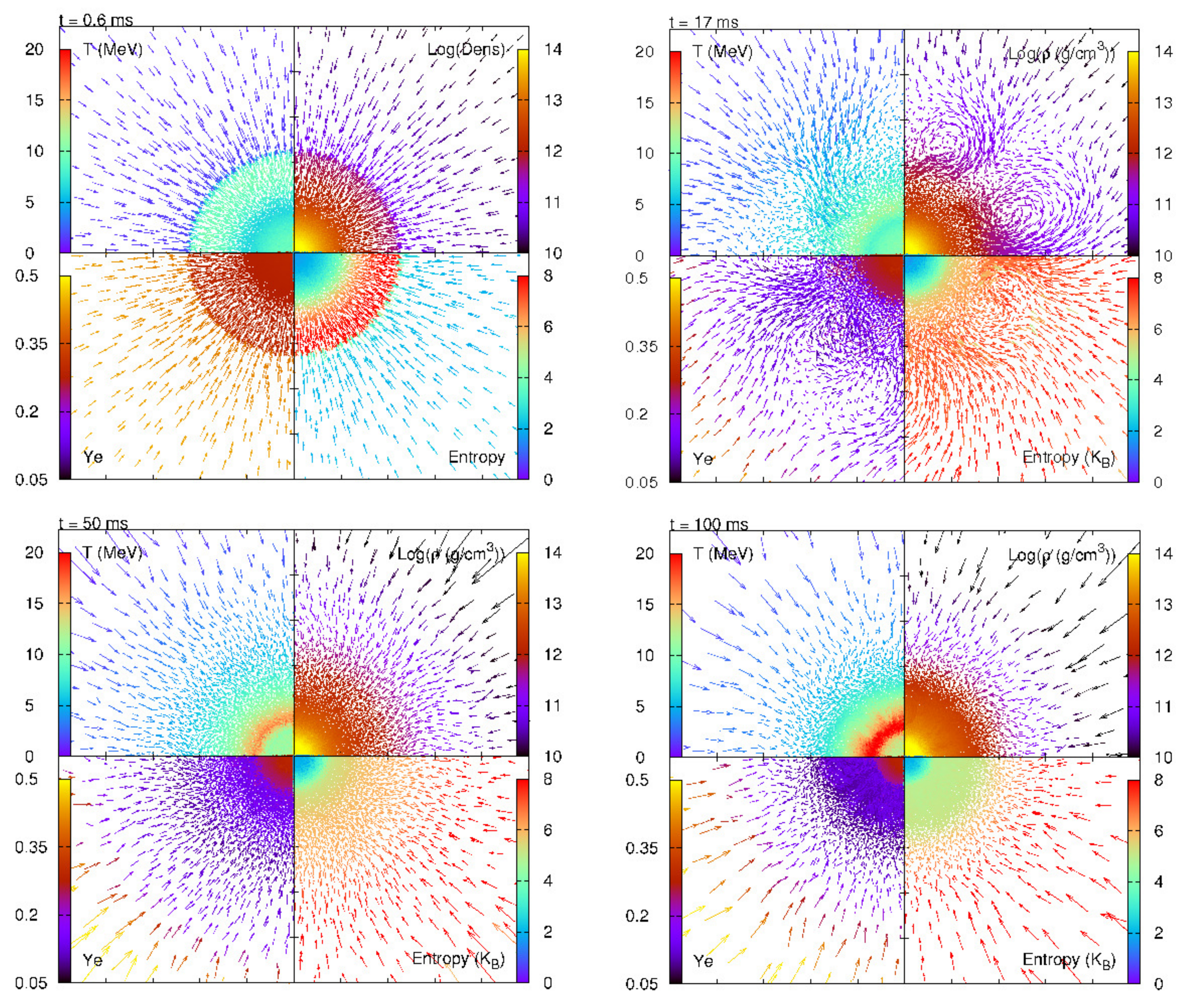}
\caption{Snapshots of the post-bounce evolution at different times (0.6~ms, 17~ms, 50~ms and 100~ms)
from our SPH model of a 15 $\msun$ CCSN. 
We show here only a thin slice of the 3D domain on the $xy$ plane and each box is 200~km wide, i.e. $-100~{\rm km} < x,y < 100~{\rm km}$. 
Each arrow represents one SPH particle and 
shows its projected velocity. Temperature, density, \ye, and entropy are color coded in each snapshot.}
\label{fig11}
\end{figure*}

To evolve the system with \texttt{SPHYNX} we solve the hydrodynamical equations in Lagrangian form,
including a gravitational and a neutrino source term.
In the momentum equation, we add the total neutrino pressure, Equation~(\ref{pnu1}), to the plasma pressure.
Therefore, the stress provided by the trapped neutrinos is directly taken into account.
The evolution of each $Z_{\nu}$ is provided by an energy equation (similar to the equation
for the plasma internal energy $e$) which consistently uses the neutrino pressure $P_{\nu}$.

From the position of the SPH particles and the EoS, we compute the
local density, gradient of pressure, gravitational potential and internal energy.
Moreover, each particle carries information regarding the electron fraction, \ye and
the neutrino trapped components, $(Y_{\nu}, Z_{\nu}$). 
Similarly to the method presented in Section~\ref{sec: the new ASL}, the ASL scheme ultimately provides 
the rates of change for these quantities ($\dot{Y}_{e}$, $\dot{Y}_{\nu}$, $\dot{Z}_{\nu}$) and for the 
internal energy ($\dot{e}_{\nu}$).

The abundances of electrons and neutrinos are evolved explicitly, 
while the implementation of the energy equation for $Z_{\nu}$
is described in Appendix~\ref{appB}.

Most of the quantities which are needed to compute the previous terms are local, so the implementation 
of the ASL scheme in the SPH structure is straightforward and directly done in 3D with very 
few modifications of the hydrodynamical part of the code. 
The only non-local quantities are the spectral optical depths,
$\tau_{\nu,{\rm tot}}(\spc,E)$ and $\tau_{\nu,{\rm en}}(\spc,E)$, and the spectral neutrino densities, 
$n_{\nu}(E,\spc)$, used in the calculation of the non-local absorption rates, 
$h_{\nu}(E,\spc)$, Equation~(\ref{eqn: heating term outside})). 

To compute the optical depths we use the expected 
quasi-spherical symmetry of a collapsing stellar core by defining a one dimensional radial grid. 
On this grid we calculate the spherical averages of the neutrino spectral mean free paths 
(which are computed locally in 3D, at each SPH particle position). 
Then, we integrate $1/\lambda(R,E)$ radially, from the 
external edge up to each radial position to obtain the radial optical depth.
Finally, the spherically symmetric optical depth is mapped back on the three dimensional 
SPH particle distribution, interpolating with respect to the distance from the center 
of mass. Using the 3D density as interpolation variable led to no significant differences.
% for performing this remapping led to no significant differences.
Regarding the neutrino densities, we also consider them to be 
spherically symmetric, and to evaluate them we calculate the spectral number luminosity 
via performing a radial integration over all the particles sorted from lower to higher radius.

\subsubsection{Setup and results}

\begin{figure*}
\begin{minipage}{0.33 \linewidth}
\includegraphics[width = \linewidth]{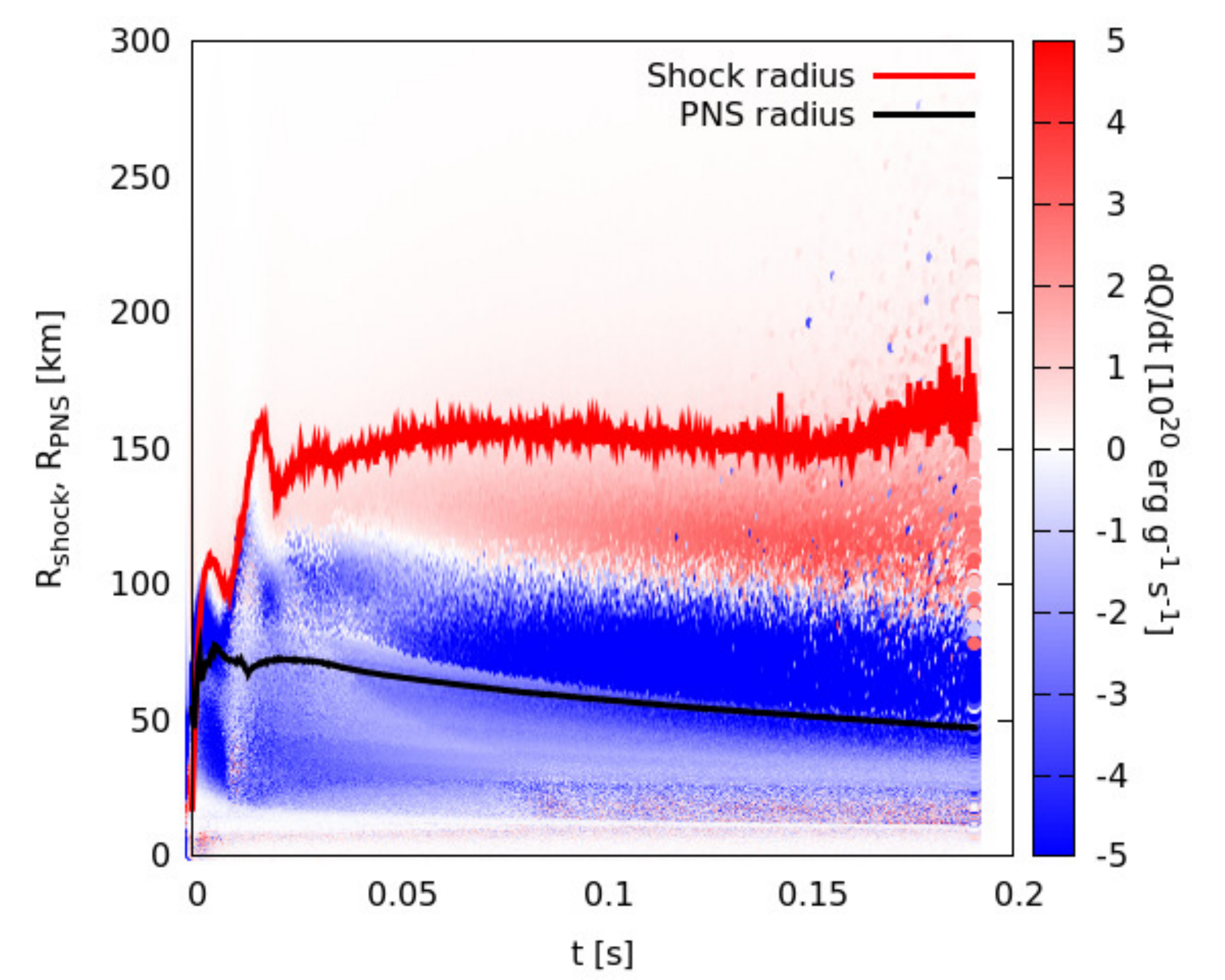} 
\end{minipage}
\begin{minipage}{0.33 \linewidth}
\includegraphics[width = 0.7 \linewidth,angle = -90]{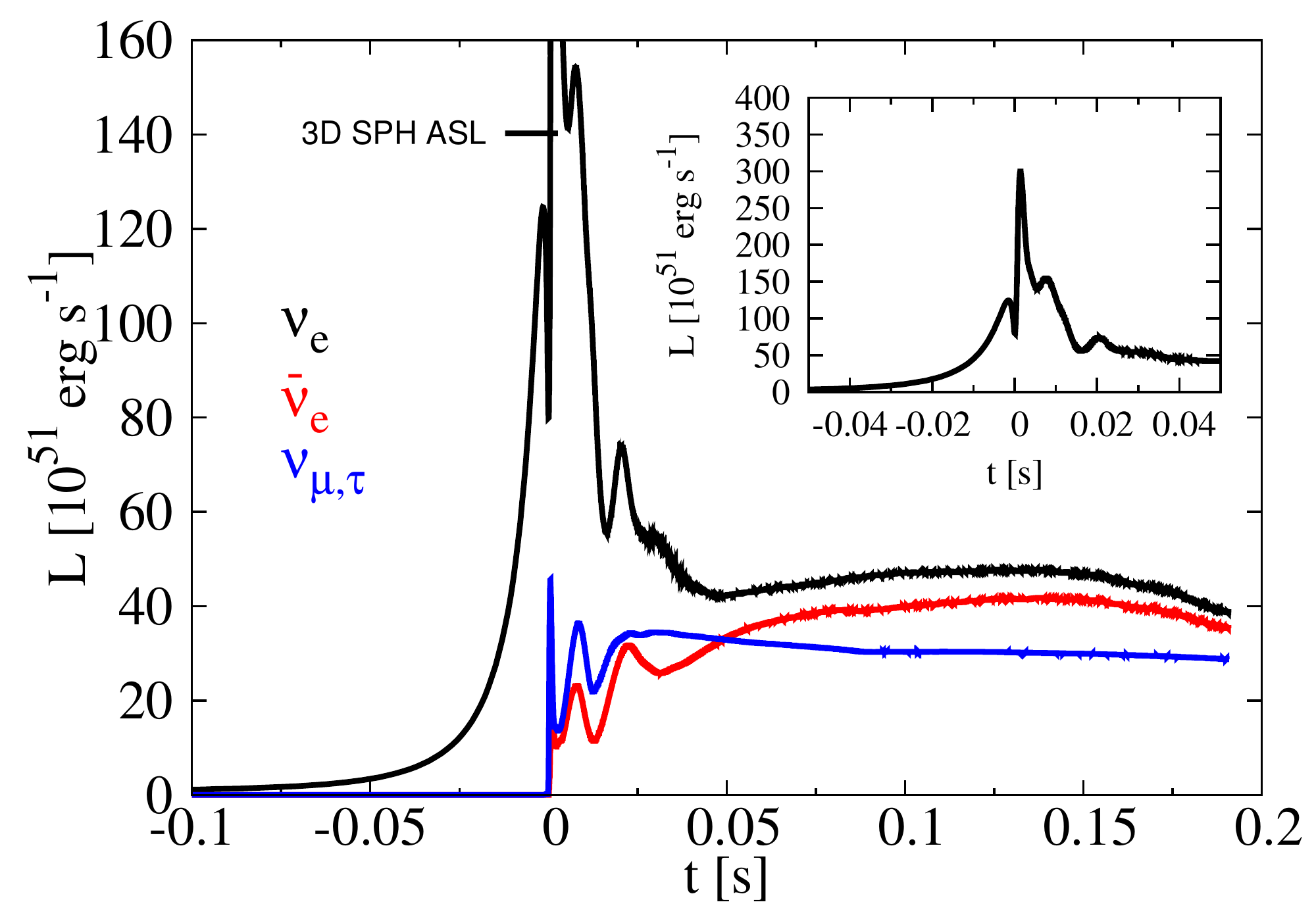}
\end{minipage}
\begin{minipage}{0.33 \linewidth}
\includegraphics[width = 0.7 \linewidth,angle = -90]{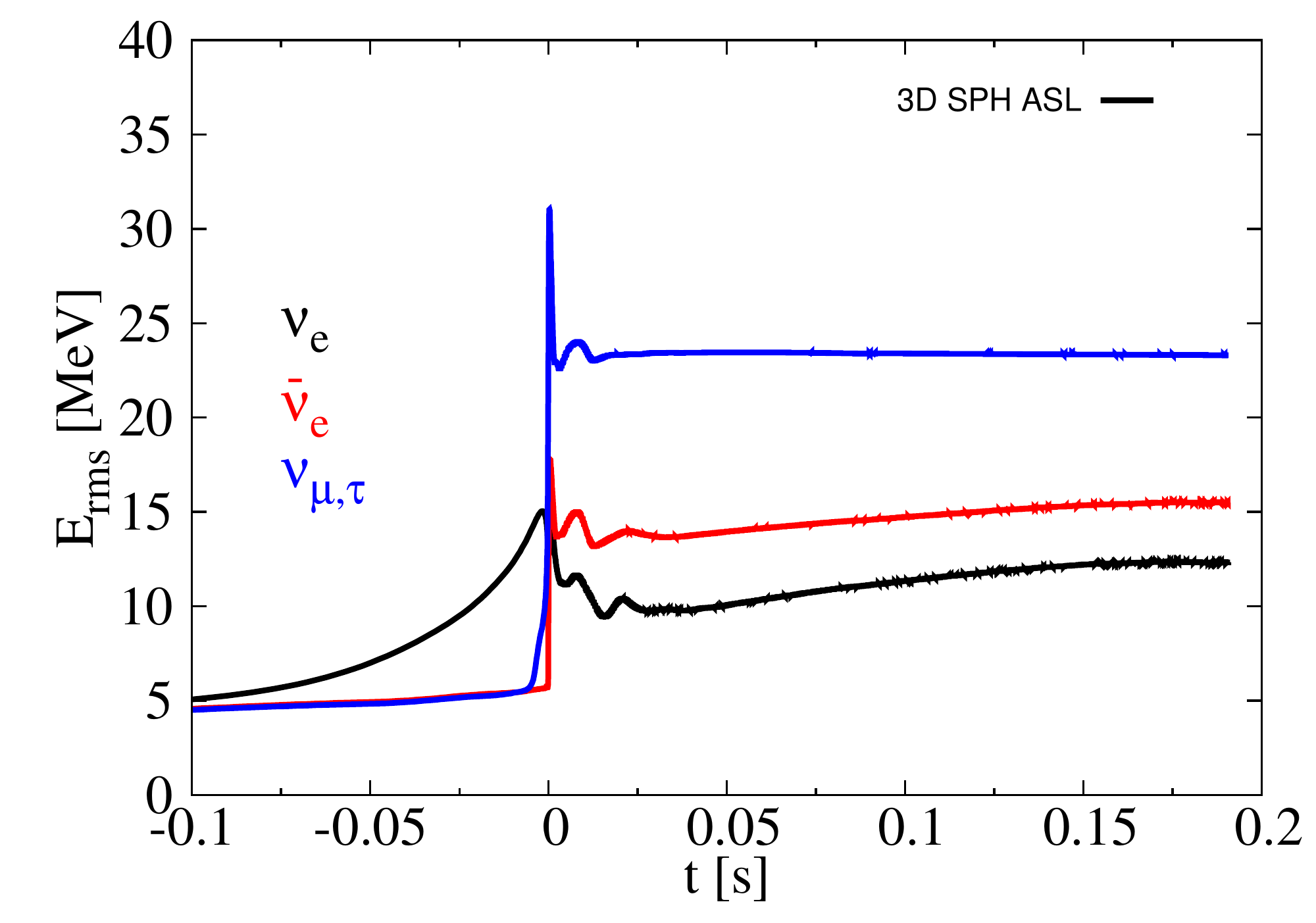}
\end{minipage}
\caption{Same as in Figure~\ref{fig8}, but for our 3D SPH simulation of a 15~$\msun$ CCSN.}
\label{fig12}
\end{figure*}

We set up a CCSN SPH simulation with 200,000 particles, using Newtonian gravity. 
We map the spherically symmetric progenitor model into a three dimensional 
quasi-random Sobol distribution of equal mass particles. 
We simulate $\sim 1.8~\msun$ of mass and 
up to $3,800$ km in radius.
Next we perform an angular relaxation of the system by allowing the particles to move, 
but with fixed radius. In this way we erase artificial gradients of pressure formed by 
random clumps of particles, and obtain clean radial profiles that adjust to the initial 1D model. 

The dynamics of the collapse and of the early post-bounce phase are in very good agreement with the previous results.
The electron fraction in the center of the PNS decreases until it reaches $Y_{\rm e} \sim 0.31$. When the central density becomes 
$\rho_{\rm bounce} = 3.3 \cdot 10^{14}$~g~cm$^{ -3}$, the core bounces and a shock wave forms at the surface of the newly born PNS.
More specifically, the shock is formed at $ \approx 14$~km, which corresponds to an enclosed mass of $M_{\rm enc,bounce} = 0.76~\msun$ 
of unshocked material. 

In Figure~\ref{fig10}, we present radial profiles of radial velocity (left column), 
electron fraction (central column) and entropy (right column), at four different times from
core bounce up to 100ms.
Each point in the panels represents one of the SPH particles. 
From the low dispersion of the profiles at bounce, we can see that the collapse preserves 
spherical symmetry, as expected. We notice the position of the shock wave from the entropy spike and 
from the corresponding starting deleptonization in the \ye profile. 
As the shock proceeds, multidimensional effects and inhomogeneities appear and grow 
inside the shocked material.
This effect can be seen from the spread of particles in the profiles of all represented quantities 
in the second row. After 50 ms the initial prompt convection has been quenched and the evolution 
proceeds more steadily. Later on, convection slowly settles again in the shocked material, but 
with a longer timescale, which can be seen in the profiles of radial velocity and \ye 
as a increased scatter of the particles (last row).
The shape and the evolution of multidimensional instabilities can be better seen in 
Figure~\ref{fig11}. In this series of snapshots, we project onto the $xy$ plane 
all the particles included in a thin layer of 20~km around the equatorial plane ($\text{Z}=0$),
for the same time steps used in Figure~\ref{fig10}.
The size of the box is 200~km side.
Material processed by the shock is slowed down, deleptonized, and accreted onto the hot PNS. 
At 17~ms we show the development of a transitory violent convection that comes right after the shock launch, 
which in this snapshot has reached the outer edge of the plot, at $R\simeq 100$~km. 
At 50~ms the convection modes have been quenched and the PNS accretes with a steady flow. 
This leads to the more evolved image at 100~ms, where accretion occurs smoothly on a slowly 
compactifying PNS.
As time proceeds, the accretion and compression of matter increases 
the temperature above the surface where the shock was launched.
The low numerical diffusion of SPH helps keeping the heat located in this region. 
The strong deleptonization via neutrino emission can be seen at work in all snapshots, 
even at 17~ms, where evident 3D features occur, showing that the ASL is working correctly 
in multidimensional simulations.

In the left panel of Figure~\ref{fig12}, we show the evolution of the PNS radius (black line)
and of the shock radius (red line) as a function of time after bounce.
The PNS shows a stable evolution with a slowly decreasing radius, corresponding to a 
more compactified configuration due to accretion and cooling.
The shock position is determined within the local resolution 
as the radius at which the artificial viscosity (AV) peaks. The AV is specifically 
designed to dissipate energy only when shocks are present. We saw in all of our 
simulations that, once the shock is formed and launched, the AV is approximately 
two orders of magnitude higher at the shock position than in the surrounding matter.
The overall evolution of the shock is similar to the spherical results, 
with a fast expansion reaching 100~km within the first 5~ms, and after 
some oscillations, triggered by 3D convection, it settles around 150~km. 
In the same figure, we also present the time evolution 
of the local cooling and heating rates color-coded. 
The thin white region between the cooling (blue) and the heating (red) 
regions is the location of the gain radius, which settles around 100~km
at 50~ms after bounce and slowly recedes, while the neutrino heating sets 
in behind the shock.

In the central and right panels of Figure~\ref{fig12}, we show the temporal evolution of the neutrino 
luminosities and RMS energies, respectively,
calculated during our SPH simulation with the ASL scheme.
The obtained results agree well with the 1D simulation performed with the ASL scheme.
In particular, in the luminosity evolutions we distinguish 
all relevant features expected in the collapse, burst and accretion phases.
Also the RMS energies have an evolution that corresponds very 
well with the results obtained with the reference 1D model, and their hierarchy is 
preserved through the simulation.
Another feature which looks interesting 
is the oscillation in the luminosities within the first milliseconds after neutrino burst: 
they appear in all three curves and they are related to violent convection inside the PNS during 
the very early post-bounce phase. Prompt, violent convection inside the PNS was also observed, 
for example, by \cite{Herant.etal:1994} in 2D SPH simulations, and
more recently by \cite{Dolence.etal:2015} and \cite{Pan.etal:2015}).
This feature deserves a deeper investigation (see, e.g., the analysis performed
by \cite{Bruenn.etal:2004} and \cite{Buras.etal:2006a}). 

\section{Conclusions}
\label{sec: conclusions}
We have presented the Advanced Spectral Leakage (ASL) scheme.
This provides an approximate treatment for the neutrino transport problem in astrophysical contexts,
like the core-collapse of massive star or the merger of compact objects.
The goal of the scheme is to provide an efficient and physically motivated treatment that contains all the 
major aspects of neutrino emission and absorption, with a level of accuracy lower than other more sophisticated
(multidimensional) neutrino transports (like, for example, M1 schemes, MGFLD schemes or IDSA), 
but higher than the classical gray leakage schemes.
It allows the application to different astrophysical contexts, 
codes, and geometrical dimensions, with a reduced computational cost. 
Due to its effective nature, details of the neutrino transport can not always be reproduced.
Nevertheless, it is optimal:
1) to study problems where a spectral neutrino treatment is required, but the details of the neutrino behavior 
are of secondary importance; 
2) to perform extensive parametric or high-resolution studies, which are still computationally too 
costly in multidimensional simulations with detailed neutrino transports;
3) to accomplish preparatory and exploratory tests; 
4) during the developing and testing of a hydrodynamic code, when the 
usage of an easily verifiable, but still reliable neutrino treatment could be useful; 
5) to study complex and very dynamical systems in which, due to the lack of symmetries, other more sophisticated neutrino 
treatments are still not available.

Due to its approximate nature, it is not well suited to investigate aspects where 
the details of the neutrino transport are crucial
\citep[e.g.,][]{Lentz.etal:2012a,Lentz.etal:2012b,Mueller.etal:2012b,Melson.etal:2015b}.
Moreover, since it avoids the solution of the transport problem in the diffusive
regime by estimates of the diffusion timescales, it is not designed to study the detailed
cooling of compact objects, especially over long timescales \citep{Hudepohl.etal:2010,Fischer2012,Roberts.etal:2012a,Suwa:2014}.

We have developed and tested the scheme against reference models provided by numerical solutions of the Boltzmann
equation. We have explored three progenitors with 12,15 and 40~$\msun$ ZAMS masses.
The 15~$\msun$ case has been more extensively studied to calibrate the free parameters of the 
scheme. We have also investigated the impact of the variation of their values on the obtained results.
Usually, the changes we have tested produced differences qualitatively 
in agreement with what we expected. Large quantitative discrepancies are observed when the parameter values
differ significantly from the calibrated ones.
The 12~$\msun$ and 40~$\msun$ cases have been used to show the robustness of the calibrated parameters with
respect to the progenitor model.

Overall, the radial profiles of several hydrodynamical and thermodynamical quantities 
obtained by the ASL scheme show a good agreement with
the reference solutions during the whole simulated period (from the collapse to the neutrino heating phase).
Small differences are present during the collapse phase and in the prompt shock expansion, while in
the neutrino heating phase, where a detailed treatment would be required to model with accuracy 
both the neutrino emission and absorption, differences tend to grow.
Usually, the shock position is well reproduced, 
with typical differences not larger than $10-15~{\rm km}$ even at later times (larger discrepancies have been
observed only for the 12~$\msun$ case).
The profiles of electron fraction and entropy are the ones that present the most notable differences,
even though some differences can be interpreted as radial or temporal shifts.
We have also compared the temporal evolution of the neutrino luminosities and mean energies, for all neutrino flavors.
Again, the most relevant features 
are present also in the approximate results, especially for $\nue$'s. 
Quantitative differences are nevertheless visible, especially in the rapid growth of
$\nueb$'s and $\numt$'s in the early post-bounce phase.
The behaviors and the values we have obtained for the RMS energies of $\nue$ and $\nueb$ 
are consistent with the reference solution. This is true especially in the neutrino heating phase, where
describing correctly the mean neutrino energies is crucial to model the neutrino 
absorption in the optically thin region .

We have also shown that the scheme can be applied, without conceptual changes, to different types of codes 
and to different spatial dimensions. In this respect, we have implemented and tested it in multidimensional 
core-collapse models using an axisymmetric Eulerian grid code and a three-dimensional Lagrangian SPH code.
These models agrees with our spherically symmetric solution during the collapse
and in the early post-bounce phase. Multidimensional features appear in the post-bounce phase and they
agree (at least, at a qualitative level) with published results
obtained in multidimensional models employing more detailed neutrino transport.
In fact, the ASL treatment have already been applied to multidimensional CCSN models 
\citep{Winteler.etal:2012,Perego.etal:2015} 
and to the study of the aftermath of neutron star mergers \citep{Perego.etal:2014b}. 

The scheme presents a modular structure which allows the inclusion of new neutrino reactions and opacities, as well as the possibility
to include more sophisticated treatments (for example, for the reconstruction of the trapped distribution functions or for
the neutrino thermalization process), without changing its basic features.
The inclusion of additional neutrino reactions, like neutrino-electron scattering, or of some relevant relativistic and 
Doppler effects \citep[e.g.,][and references therein]{Lentz.etal:2012b} in the ASL scheme will be carried out in the nearby future.

\section*{Acknowledgments}
The authors thank A. Arcones, E. Gafton, M. Liebend\"orfer, S. Rosswog and
F-K. Thielemann for useful discussions and comments about this work.
AP is supported by the Helmholtz-University Investigator grant No. VH-NG-825.
He also thanks the University of Basel and ETH Z\"urich for their hospitality.
RC acknowledges the support from the HP2C Supernova project, the European Research Council (FP7) 
under the ERC Advanced Grant Agreement N° 321263 - FISH, and the DIAPHANE project 
within the Platform for Advanced Scientific Computing (PASC).
RC and RK thank TU Darmstadt for its hospitality.
AP, RC and RK acknowledge the use of computational resources provided by 
the Swiss SuperComputing Center (CSCS), under the allocation grant s414.
The authors thank also M. Liebend\"orfer for the access to \boltz runs, and acknowledge 
the support of sciCORE (http://scicore.unibas.ch/) scientific computing core facility 
at University of Basel, where some of the calculations were performed.
RK acknowledges the computing resources provided by the Brutus and Euler clusters
at ETHZ.

\appendix

\section{Implementation of the pair processes in the ASL scheme}
\label{appA}

In this appendix, we present our implementation of the neutrino pair processes, Eqs. (\ref{eqn: ep annihilation}) 
and (\ref{eqn: bremsstrahlung}), in the context of the ASL scheme.
Our goal is the computation of the associated emissivities, $\left( j_{\nu} \right)_{\rm pair}$
and absorptivities, $\left( \chi_{\nu} \right)_{\rm pair}$.
These quantities are necessary to compute the local
mean free paths, Eqs. (\ref{eqn: lambda tot}) and (\ref{eqn: lambda eff}) , the production and diffusion timescales
Equations~(\ref{eqn: tprod}) and (\ref{eqn: diffusion timescale}), and the production rates, Equation~(\ref{eqn: r_prod definition}).
We start from the expression of the collision integral for pair processes 
in the Boltzmann equation for the neutrino species $\nu$ (see, e.g., \cite{Bruenn1985} or \cite{Hannestad1998}):
\begin{align}
         \left. \dot{f}_{\nu}(\mathbf{k}_\nu) \right|_{\rm coll,pair} =  
          \left( 1-f_{\nu}(\mathbf{k}_{\nu}) \right) \, 
          \frac{1}{c \left( 2 \pi \hbar c \right)^3} 
          \int \: {\rm d}^3 \mathbf{k}_{\nub} \,
           \left( 1-f_{\nub}(\mathbf{k}_{\nub} ) \right) 
         S^{\rm pr}_{\rm pair}(\mathbf{k}_{\nu},\mathbf{k}_{\nub}) \nonumber \\
             - \, f_{\nu}(\mathbf{k}_{\nu}) \frac{1}{c \left( 2 \pi \hbar c \right)^3} 
          \int \: {\rm d}^3 \mathbf{k}_{\nub} \, 
          f_{\nub}(\mathbf{k}_{\nub}) \,
         S^{\rm ab}_{\rm pair}(\mathbf{k}_{\nu},\mathbf{k}_{\nub}) 
         \label{eqn: pair collision term}
\end{align}
where $\nub$ denotes the antiparticle of $\nu$, $\mathbf{k}_{\nu}$ and $\mathbf{k}_{\nub}$ the neutrino momenta, $f_{\nu}$ and $f_{\nub}$
the neutrino distribution functions, and $S^{\rm pr}_{\rm pair}$ and $S^{\rm ab}_{\rm pair}$ the kernel of the pair reactions.

To compute the local emissivities, we consider the first term in the integral of Equation~(\ref{eqn: pair collision term}) and we perform
the integral over the $\nub$ phase space, neglecting Pauli blocking factors for $\nub$ in the final state, since the production rate 
is mainly relevant in the optically thin region:
\begin{equation}
          j_{\rm em,pair}(E_{\nu}) =  
          \frac{1}{c \left( 2 \pi \hbar c \right)^3} 
          \int \: {\rm d}^3 \mathbf{k}_{\nub} \, 
         S^{\rm pr}_{\rm pair}(\mathbf{k}_{\nu},\mathbf{k}_{\nub}) \, .
         \label{eqn: pair emission}
\end{equation}
For the absorptivities, we consider the right term of the integral in Equation~(\ref{eqn: pair collision term}),
and we integrate it over the phase space of $\nub$,
\begin{equation}
          \chi_{\rm ab,pair}(E_{\nu}) =  
          \frac{1}{c \left( 2 \pi \hbar c \right)^3} 
          \int \: {\rm d}^3 \mathbf{k}_{\nub} \,
          f_{\nub}(\mathbf{k}_{\nub}) \,  
         S^{\rm ab}_{\rm pair}(\mathbf{k}_{\nu},\mathbf{k}_{\nub}) \, ,
         \label{eqn: pair absorption}
\end{equation}
assuming that $f_{\nub}$ is described by Fermi-Dirac distribution functions in weak and thermal equilibrium.
Within this assumption, we recover the correct limit in the diffusive regime, where the calculation of
the mean free path and of the optical depth are more relevant. In the optically thin limit, where the
actual distribution functions are expected to differ significantly from Fermi-Dirac distributions,
our approach is expected to overestimate the absorptivity due to pair processes. This would lead to a smaller 
$\lambda_{\rm pair}$. 
However, we have tested that, for the corresponding
relevant thermodynamical conditions, $\lambda_{\rm pair}$ is still significantly larger than the mean free path
due to assorption or scattering on nucleons. Moreover, $\lambda_{\rm pair}$ is always much larger than the 
linear dimension of the system. Thus, our overestimated absorptivities in optically thin conditions do not
affect critically the location of the neutrino surfaces.

The reaction kernels, $S^{\rm ab}_{\rm pair}$ and $S^{\rm pr}_{\rm pair}$, are calculated following \cite{Hannestad1998} 
for the bremsstrahlung process, and \cite{Bruenn1985,Mezzacappa.Messer:1999} for the pair annihilation process.
The calculation of the integrals (\ref{eqn: pair emission}) and (\ref{eqn: pair absorption}) during runtime would be 
by far the most expensive part of the rate computation. Thus, we decide to tabulate these emission and absorption rates, 
and to interpolate them during the program execution. We compute them as a function of the matter density, electron fraction
and temperature. For our tables we consider a three-dimensional grid, where we uniformly sample: 
i) the logarithm of the matter density, with 104 points between $10^{5.1} \, {\rm g \, cm^{-3}}$ and $10^{15.4} \, {\rm g \, cm^{-3}}$;
ii) the electron fraction, with 72 points between 0.0 and 0.56; 
iii) the logarithm of the matter temperature, with 31 points between 0.1 MeV and 100 MeV.
We perform the integral over the neutrino energy by splitting the $\left[ 0,+\infty \right)$ integration interval
into two segments, $\left[ 0,E_{\rm ref}\right)$ and $\left[ E_{\rm ref},+\infty \right)$, with $E_{\rm ref} = 3 \, T /2 $.
For the second, improper integral, we perform the change of variable $E \rightarrow 1/E$. Then, we integrate
each of the two integrals using the Gauss-Legendre quadrature with 16 points. The interpolation at runtime is accomplished by
a trilinear interpolation method.

\section{Calculation of the energy density of neutrinos in SPHYNX}
\label{appB}

In this appendix, we present our implementation of the equation that evolves $Z_{\nu}$ in the 
SPH implementation of the ASL.
In the following, all SPH equations use IAD$_0$ for calculating derivatives \citep{Garcia2012,Cabezon2012,Rosswog:2015b}. 
To make a conversion to the traditional SPH prescription simply substitute all $\mathcal{A}_{ij}$ by $\nabla_i W_{ij}$.

The trapped neutrino energy $Z_{\nu}$ is evolved according to
\begin{equation}
\frac{ {\rm d}Z_{\nu}}{\rm dt} =-\frac{P_{\nu}}{\rho}\nabla \cdot \mathbf{v} + \dot{Z}_{\nu} 
\label{sphznu}
\end{equation}
From Equation~(\ref{sphznu}) it is clear that the variation of $Z_{\nu}$ consists of two contributions: 
The first term is is due to the $P{\rm d}V$ work and can be calculated using a 
SPH equation similar to the SPH (baryonic) energy equation, which takes into account 
the neutrino pressure (instead of the baryonic pressure) and the density changes of the fluid.
The second one is the source term provided by the ASL scheme and takes into account the rate of change of energy 
of the trapped neutrinos due to production and diffusion. 
In overall, evolving independent equations for $\dot{Z}_{\nu}$ is equivalent to split 
the equation for the total internal energy $u$ in two components: baryonic and neutrinos. 
The baryonic part is accounted with a regular SPH equation for specific internal energy $e$, 
while the neutrino component is calculated with:

\begin{equation}
 \left( \frac{{\rm d}{Z}_{\nu}}{{\rm d}t} \right)_i = \frac{P_{\nu,i}}{\Omega_i\rho_i^2}\sum_j m_j
 (\mathbf{v}_i-\mathbf{v}_j)\cdot\mathcal{A}_{ij}+\dot{Z}_{\nu,i} \, ,
 \label{eqn:energysph2}	
\end{equation}
where $i$ is the particle index, $\mathbf{v}$ is the velocity vector, and $\Omega_i$ is the grad-h term. 
Then, $Z^{n+1}_{\nu}$ could be calculated using the same integration method used for the specific internal energy. 

Although this scheme is quite straightforward, in order to calculate the new $Z_{\nu}$ we evaluate 
the rate of change $\dot{Z}_{\nu}$, which in fact depends on $Z_{\nu}$ itself via $P_{\nu}$ (Equation~\ref{pnu1}).
Therefore, we opted for developing a semi-implicit scheme that preserves the consistency 
between both magnitudes at a very low computational cost.
Noting that our objective is
\begin{equation}
P^{n+1}_{\nu}=\frac{1}{3}\rho Z^{n+1}_{\nu}\,
\label{pnu}
\end{equation}
we can now substitute Equation~(\ref{pnu}) into Equation~(\ref{eqn:energysph2}) to explicitly show its dependence on 
$Z^{n+1}_{\nu}$. Taking into account all sources we can write:

\begin{equation}
Z^{n+1}_{\nu,i}=Z^n_{\nu,i}+\dot{Z}^n_{\nu,i} \, \Delta t+\frac{1}{3}\frac{Z^{n+1}_{\nu,i}}{\Omega_i\rho_i}\sum_j m_j(\mathbf{v}_i-\mathbf{v}_j)\cdot\mathcal{A}_{ij}\Delta t.
\end{equation}

Regrouping and isolating $Z^{n+1}_{\nu,i}$ on the left hand side, we obtain the final version for the evolution of $Z_{\nu,i}$ for each neutrino species:

\begin{equation}
Z^{n+1}_{\nu,i}=\frac{Z^n_{\nu,i}+\dot{Z}^n_{\nu,i} \, \Delta t}{1-\frac{1}{3}\frac{1}{\Omega_i\rho_i}\sum_j m_j(\mathbf{v}_i-\mathbf{v}_j)\cdot\mathcal{A}_{ij}\Delta t}.
\label{znueq}
\end{equation}

Equation~(\ref{znueq}) shows a very stable evolution for $Z_{\nu}$ in all the simulations and it provides consistent 
values for the neutrino pressure and the neutrino energy by construction. After its evaluation, we use the new 
$Z^{n+1}_{\nu}$ to calculate, via Equation~(\ref{pnu}), the neutrino pressure $P_{\nu}$ that is afterwards included 
in the momentum equation via adding it to the baryonic pressure.

\bibliographystyle{apj}
% \bibliography{references}

\end{document}